\documentclass[prd,aps,a4paper,superscriptaddress,twocolumn,footinbib]{revtex4}
\usepackage{graphicx}
\usepackage{color}
\usepackage{dcolumn}
\usepackage{bm}
\usepackage{slashed}
\usepackage{amsmath}
\usepackage{latexsym}
\usepackage{amssymb}
\usepackage{amsfonts}
\allowdisplaybreaks
\begin{document}
\title{Binary black hole simulation with an adaptive finite element method \uppercase\expandafter{\romannumeral2}: Application of local discontinuous Galerkin method to Einstein equations}

\author{Zhoujian Cao}
\email{zjcao@amt.ac.cn} \affiliation{Institute of Applied
Mathematics and LSEC, Academy of Mathematics and Systems Science,
Chinese Academy of Sciences, Beijing 100190, China}
\author{Pei Fu} \affiliation{School of Mathematical Sciences, University
of Science and Technology of China, Hefei, Anhui 230026, P.R. China}
\author{Li-Wei Ji} \affiliation{CAS Key Laboratory of Theoretical Physics, Institute of Theoretical Physics,
Chinese Academy of Sciences, Beijing 100190 and School of Physical Sciences, University of Chinese Academy of Sciences, Beijing 100049, China}
\author{Yinhua Xia} \affiliation{School of Mathematical Sciences, University
of Science and Technology of China, Hefei, Anhui 230026, P.R. China}
\begin{abstract}
Finite difference method and pseudo-spectral method have been widely used in the numerical relativity to solve the Einstein equations. As the third major category method to solve partial differential equations, finite element method is much less used in numerical relativity. In this paper we design a finite element algorithm to solve the evolution part of the Einstein equations. This paper is the second one of a systematic investigation of applying adaptive finite element method to the Einstein equations, especially aim for binary compact objects simulations. The first paper of this series has been contributed to the constrained part of the Einstein equations for initial data. Since applying finite element method to the Einstein equations is a big project, we mainly propose the theoretical framework of a finite element algorithm together with local discontinuous Galerkin method for the Einstein equations in the current work. In addition, we have tested our algorithm based on the spherical symmetric spacetime evolution. In order to simplify our numerical tests, we have reduced the problem to a one-dimensional space problem by taking the advantage of the spherical symmetry. Our reduced equation system is a new formalism for spherical symmetric spacetime simulation. Based on our test results, we find that our finite element method can capture the shock formation which is introduced by numerical error. In contrast, such shock is smoothed out by numerical dissipation within the finite difference method. We suspect this is the part reason for that the accuracy of finite element method is higher than finite difference method. At the same time kinds of formulation parameters setting are also discussed.
\end{abstract}


\maketitle


\section{Introduction}
Regarding to the problem of solving partial differential equations, there are three categories including finite difference
method, spectral method and finite element method. Finite difference method is the oldest one. This is also true to numerical relativity. Among the existing numerical relativity codes, most of them use finite difference method \cite{AylBakBog09,PhysRevD.78.124011,YamShiTan08,0264-9381-32-24-245011}. Besides finite difference method, pseudo-spectral method has also been successfully used in numerical relativity \cite{PhysRevD.93.063006,0264-9381-23-16-S09,PhysRevD.79.024003}. But unfortunately finite element code for numerical relativity is still missing (but see \cite{SopLag06,SopSunLagXu06,Zum09}). Previously we have developed a new finite element code for numerical relativity, iPHG \cite{PhysRevD.91.044033}, but it is only for constraint part of the Einstein equations. It only works for solving initial data.

On the one hand, numerical relativity has been mature in the sense that it can be applied to study kinds of physical phenomena related to strong gravitational field and highly dynamical spacetime \cite{Cardoso2015,0034-4885-80-9-096901}. On the other hand, numerical relativity still faces some challenges. For example, the current gravitational wave form model such as effective one body numerical relativity (EOBNR) model \cite{PhysRevD.96.044028} is valid only for mass ratio between 1:1 and 1:20. Such mass ratio range is limited by the numerical relativity simulation ability \cite{PhysRevLett.106.041101}. For finite difference code, the adaptive mesh refinement is adopted to treat the multi-scale problem involved in the binary black hole system. But the grid numbers on each mesh level limited the strong parallel scaling ability. In numerical relativity about $100\times100\times100$ grid boxes are used. Not too many cores can be used in the simulation of binary black hole systems with finite difference code \cite{LOFFLER201679}. For pseudo-spectral code, the parallel scaling ability is much worse due to the global data change character. But thanks to the high convergence property, the pseudo-spectral code does not need too many cores for binary black hole system simulations \cite{PhysRevD.79.024003}. Unfortunately when mass ratio increases, it is quite hard to tune the pseudo-spectral code to make it work.

Finite element method can possibly combine the high convergence property of pseudo-spectral method and the high parallel scaling property of finite difference method. This is because the finite element discretization admit the local data property as finite difference. While in each element, high order polynomial function basis and/or spectral function basis can be used, which are similar to spectral method (spectral element method). So it is possible to use finite element method to treat unsmooth region with small element (h refinement) and to treat smooth region with large element but high order basis or spectral basis (p refinement). For finite difference method, data have to be transferred between different mesh levels. So the strong parallelization scalability for finite difference method is limited by the size of single mesh. In contrast, all elements in finite element method are treated uniformly. This possibly makes finite element method admit higher strong parallelization scalability than both finite difference method and spectral method.

There are two works aimed to develop the finite element method for numerical relativity already. In \cite{KIDDER201784} the authors used discontinuous Garlerkin finite element method to treat general relativistic hydrodynamics. The authors of \cite{0264-9381-34-1-015003} used local discontinuous Garlerkin finite element method at the level of the derivative operator to treat the BSSN formalism of the Einstein equations \cite{PhysRevD.78.124011}. In the current work, we will apply the local discontinuous Garlerkin finite element method at the level of the differential equations to the Einstein equations. In order to simply the problem, we only consider one dimensional space in the current work. Physically we consider spherical symmetric spacetime. In the next section, we will introduce the Einstein equations formalism used for our local discontinuous Garlerkin finite element method. After that we will present the numerical algorithm in Sec.~\ref{Sec::III}. Then the numerical test results are presented in Sec.~\ref{Sec::IV}. And finally some comments, discussions and conclusions are given in Sec.~\ref{Sec::V}. Besides some detail mathematical expressions are postponed in the Appendixes.

The Einstein summation convention, i.e. the repeated super and sub index mean a summation, is adopted throughout the paper. And the geometrized units with $G=c=1$ is used. We take the notation convention used in \cite{wald84,liang00}. Latin indices are spatial indices and run from 1 to 3, whereas Greek indices are space-time indices and run from 0 to 3.
\section{The evolution system with generalized harmonic gauge (GHG) formulation}
Due to the spherical symmetry, we can write out the metric in the following form \cite{Sorkin2010}
\begin{align}
g_{ab}=
\begin{pmatrix}
g_{00} & g_{01} &   0         & 0\\
g_{01} & g_{11} &   0         & 0\\
0      & 0      &   e^{2S}r^2 & 0\\
0      & 0      &   0         & e^{2S}r^2\sin^2\theta
\end{pmatrix}\label{metric_sphsymm}
\end{align}
with coordinate system $(t,r,\theta,\phi)$. From now on we use $A$, $B$, $\cdots$ to run from 0 to 1 (corresponding to $t$ and $r$), and use $I$, $J$, $\cdots$ to run from 2 to 3 (corresponding to $\theta$ and $\phi$). We define new variables
\begin{align}
\Pi_{AB}&\equiv-\frac{1}{\alpha}(\partial_t-\beta^r\partial_r)g_{AB},\\
\Pi_S&\equiv-\frac{1}{\alpha}(\partial_t-\beta^r\partial_r)S,\\
\Phi_{AB}&\equiv\partial_rg_{AB},\\
\Phi_S&\equiv\partial_rS,
\end{align}
where we have used notation lapse $\alpha$ and shift vector $\beta^i=(\beta^r,0,0)$.

Denote the state vector as $u^{\mu}=(g_{AB},\Pi_{AB},\Phi_{AB},S,\Pi_S,\Phi_S)^T$, we have the dynamical equations which is reduced from the Einstein equations \cite{PhysRevD.93.063006}
\begin{align}
\partial_tu^\mu+A^{\mu}{}_\nu\partial_r u^\nu&=S^\mu,\label{problem_eq}
\end{align}
with
\begin{align}
&{A^{\mu}}_\nu =\nonumber\\
&\begin{pmatrix}
-(1+\gamma_1)\beta^r         &  0                   &  0                  &0           &0          &0\\
-\gamma_1\gamma_2 \beta^r    &  -\beta^r            &  \alpha \gamma^{rr} &0           &0          &0\\
-\gamma_2 \alpha             &  \alpha              &  -\beta^r           &0           &0          &0\\
0 &0  &0 &-(1+\gamma_1)\beta^r         &  0                   &  0\\
0 &0  &0 &-\gamma_1\gamma_2 \beta^r    &  -\beta^r            &  \alpha \gamma^{rr}\\
0 &0  &0 &-\gamma_2 \alpha  &  \alpha   &  -\beta^r
\end{pmatrix}\\
&S^\mu=\nonumber\\
&\begin{pmatrix}
-\alpha\Pi_{AB}-\gamma_1\beta^r\Phi_{AB} \\
S^\mu_{\Pi_{AB}} \\
\alpha[\tfrac{1}{2}n^Cn^D\Phi_{CD}\Pi_{AB}+\gamma^{rr}n^C\Phi_{rC}\Phi_{AB}
-\gamma_2\Phi_{AB}]\\
-\alpha\Pi_S-\gamma_1\beta^r\Phi_S \\
S^\mu_{\Pi_{S}} \\
\alpha[\tfrac{1}{2}n^Cn^D\Phi_{CD}\Pi_{S}+\gamma^{rr}n^C\Phi_{rC}\Phi_{S}
-\gamma_2\Phi_{S}]
\end{pmatrix}\,,
\end{align}
with
\begin{widetext}
\begin{align}
S^\mu_{\Pi_{AB}}=&2\alpha g^{CD}\big(\gamma^{rr}
\Phi_{CA}\Phi_{DB}-\Pi_{CA}\Pi_{DB}-g^{EF}
\Gamma_{ACE}\Gamma_{BDF}\big)-\tfrac{1}{2}\alpha n^Cn^D\Pi_{CD}\Pi_{AB}
-\alpha n^C\gamma^{rr}\Pi_{Cr}\Phi_{AB}\nonumber\\
&-2\alpha\big[\partial_{(A}H_{B)}+g^{CD}\Gamma_{CAB}(\gamma_4 C_D-H_D)
-\tfrac{1}{2}\gamma_5\,g_{AB}g^{CD}\Gamma_CC_D\big]\nonumber\\
&+\alpha\gamma_0\big[2\delta^C{}_{(A}n_{B)}-g_{AB}n^C\big]C_C-\gamma_1\gamma_2\beta^r\Phi_{AB}\nonumber\\
&-\tfrac{4\alpha}{r^2}[\gamma^r{}_A(r\Phi_S+1)+n_A(r\Pi_S-n^r)][\gamma^r{}_B(r\Phi_S+1)+n_B(r\Pi_S-n^r)],\\
S^\mu_{\Pi_S}=&-\alpha\Phi_S\gamma^{rr}n^A\Pi_{Ar}
-\tfrac{\alpha}{2}\Pi_Sn^An^B\Pi_{AB}-\gamma_1\gamma_2\beta^r\Phi_S-\tfrac{1}{2}\alpha\gamma_0n^AC_A\nonumber\\
&-\tfrac{2\alpha}{r^2}g^{AB}[\gamma^r{}_A(r\Phi_S+1)+n_A(r\Pi_S-n^r)][\gamma^r{}_B(r\Phi_S+1)+n_B(r\Pi_S-n^r)]\nonumber\\
&-\tfrac{\alpha}{r}g^{AB}H_A[\gamma^r{}_B(r\Phi_S+1)+n_B(r\Pi_S-n^r)]\nonumber\\
&-2\alpha(\Pi_S^2-\gamma^{rr}\Phi_S^2)+\tfrac{4\alpha}{r}(n^r\Pi_S+\gamma^{rr}\Phi_S)+\tfrac{3}{r^2}(\alpha\gamma^{rr}+\beta^rn^r)
+\tfrac{\alpha}{r^2}e^{-2S},
\end{align}
\end{widetext}
where $H_A$ are the source functions for generalized harmonic formalism, $n^A$ is the unit vector normal to the spatial slices of constant coordinate time $t$, $\Gamma_A=g^{bc}\Gamma_{Abc}$ are the contracted Christoffel symbol and $C_A=H_A+\Gamma_A$ are the constraint functions. The terms multiplied with $\gamma_{0,1,2,4,5}$ are the additional constraint terms beyond original Einstein equations \cite{PhysRevD.93.063006}. The rule we added them is the consideration of the structure of the resulted partial differential equations. We would like to let the equations symmetrizable, most possibly linearly degenerate, and constraint damping. More complete equations related to the formulation for the Einstein equations are presented in the appendix \ref{App::I} and \ref{App::II}.

\section{Numerical algorithm for local discontinuous Garlerkin finite element method}\label{Sec::III}
Our problem equations (\ref{problem_eq}) take the form of Hamilton-Jacobi-like equations \cite{yan2011local}
\begin{align}
u_t+H(u_x)=S(u).
\end{align}

We discretize the computational domain as $I_j=(x_{j-\tfrac{1}{2}},x_{j+\tfrac{1}{2}}),j=1,...,N$. We denote $x_j=\tfrac{1}{2}(x_{j-\tfrac{1}{2}}+x_{j+\tfrac{1}{2}})$ as the center of the cell $I_j$ and $\Delta x_j=x_{j+\tfrac{1}{2}}-x_{j-\tfrac{1}{2}}$ as the size of the each cell. The numerical solution space is defined as the piecewise polynomial space where there is no continuity requirement at the cell interfaces $x_{j\pm\tfrac{1}{2}}$ which manifests the property of discontinuous Garlerkin method.

Regarding to the polynomial we use Legendre polynomials which admits the coefficient of the highest polynomial term 1. Then we can use the Legendre polynomials to decompose functions in the approximation space as
\begin{align}
&f(x_l)=\sum_{i=0}^{k_p}f_iP_i(x_l),
\end{align}
where $x_l\in(-1,1)$ is the local coordinate for a given cell. The Legendre polynomials we used are listed in the Appendix~\ref{App::III}. The local coordinate is related to the global coordinate through
\begin{align}
x_l=\frac{x-x_0}{\Delta x/2},
\end{align}
where $x$ is the global coordinate, $x_0$ is the global coordinate for the center point of the cell, and $\Delta x$ is the width of the cell. These $P_i$s are mutually orthogonal. Based on this property, we have
\begin{align}
&f_i=\frac{\langle f|P_i\rangle}{\langle P_i|P_i\rangle},\\
&\langle f|g\rangle\equiv\int_{-1}^1f(x_l)g(x_l)dx_l.\label{Eq::innerprod}
\end{align}

For integration we have
\begin{align}
&\int_{I_j} f dx=\tfrac{\Delta x}{2}\int_{-1}^1 \sum_{i=0}^{k_p}f_iP_idx_l=f_0\Delta x,\\
&\int_{I_j} f^2dx=\tfrac{\Delta x}{2}\int_{-1}^1 \sum_{i,j=0}^{k_p}f_iP_if_jP_jdx_l=\tfrac{\Delta x}{2}\sum_{i=0}^{k_p}f_i^2\langle P_i|P_i\rangle.
\end{align}

The local discontinuous Garlerkin finite element method we used includes two steps. Firstly we calculate the derivative $u_x$ through solving the equation for $\psi$
\begin{align}
u_x=\psi.
\end{align}
Numerically we have two approximate solutions $p_1$ and $p_2$ to the above equation as following
\begin{align}
&\int_{I_j}p_1vdx+\int_{I_j}u v_xdx-u^{j+1}_{j+\tfrac{1}{2}}v^j_{j+\tfrac{1}{2}}+
u^{j}_{j-\tfrac{1}{2}}v^j_{j-\tfrac{1}{2}}=0,\\
&\int_{I_j}p_2vdx+\int_{I_j}u v_xdx-u^{j}_{j+\tfrac{1}{2}}v^j_{j+\tfrac{1}{2}}+
u^{j-1}_{j-\tfrac{1}{2}}v^j_{j-\tfrac{1}{2}}=0.
\end{align}
$v$ is the test function used for finite element method. $u^{j}_{j-\tfrac{1}{2}}$ means the value at $x=x_{j-\tfrac{1}{2}}$ got by the polynomial expansion within the $j$-th cell. Then based on $p_{1,2}$ we construct the Lax-Friedrichs numerical Hamiltonian
\begin{align}
\hat{H}(p_1,p_2)=H(\frac{p_1+p_2}{2})-\frac{1}{2}\xi(p_1-p_2)\label{eq:xi}
\end{align}
where $\xi = \max\limits_{p\in D} |\frac{\partial H(p)}{\partial p}|$ with the relevant domain $D$. For a local Lax-Friedrichs scheme, $D$ is defined locally as $D = [\min(p_1, p_2), \max(p_1,p_2)]|_{I_j}$, and $D = [\min(p_1, p_2), \max(p_1,p_2)]|_{\Omega}$ with $\Omega$ the whole computational domain for a global Lax-Friedrichs scheme. In practice we simplify this operation and explicitly take $\xi=1$. Our test results imply this simplification works well for Einstein equations.

The second step of the local discontinuous Garlerkin finite element method is calculating $u_t$ through
\begin{align}
\int_{I_j}u_tvdx+\int_{I_j}\hat{H}(p_1,p_2)vdx=\int_{I_j}S(u)vdx.
\end{align}
Then based on the $u_t$ we use fourth order Runge-Kutta method to update $u$ \cite{PhysRevD.78.124011}. We have tested the total variational diminishing third order Runge-Kutta method \cite{j.1365-2966.2010.17859.x}. Marginal difference shows up compared to the fourth order Runge-Kutta method.

\begin{figure*}
\begin{tabular}{cc}
\includegraphics[width=0.5\textwidth]{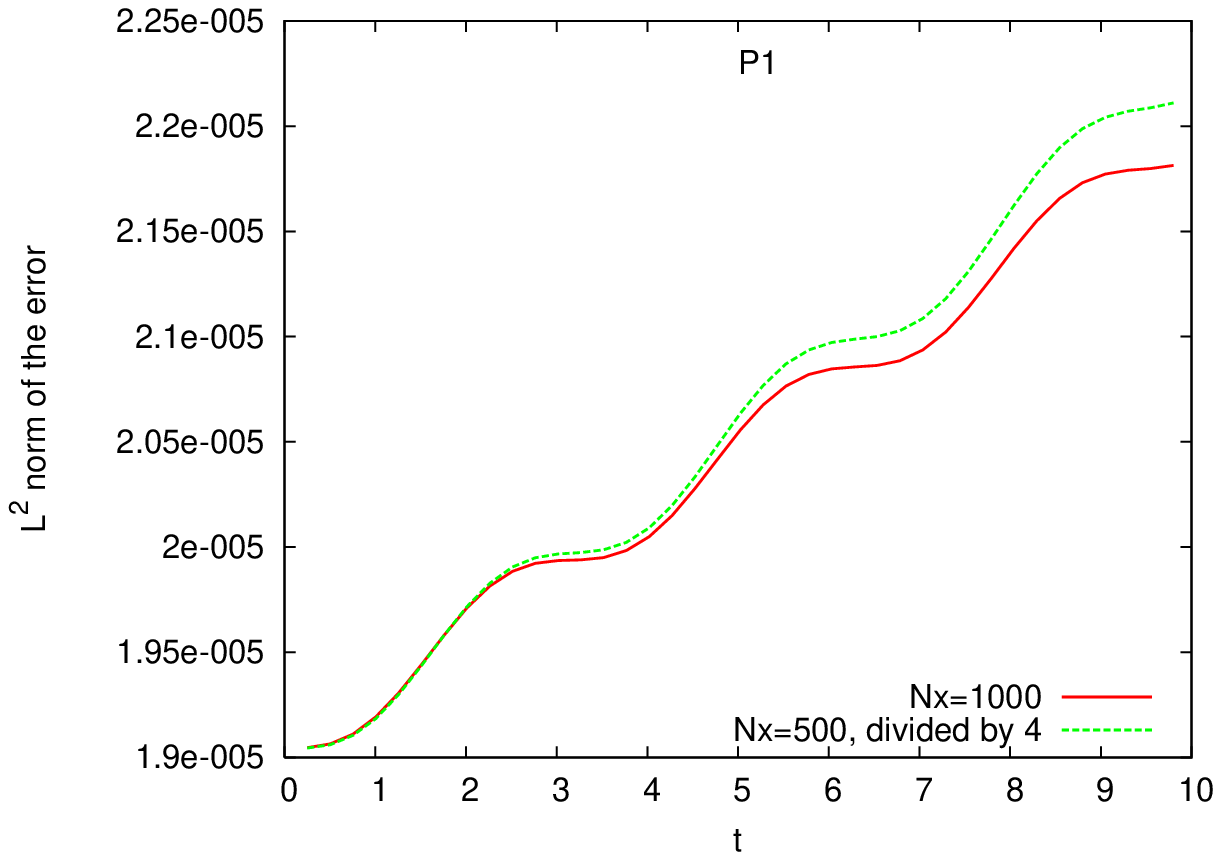}&
\includegraphics[width=0.5\textwidth]{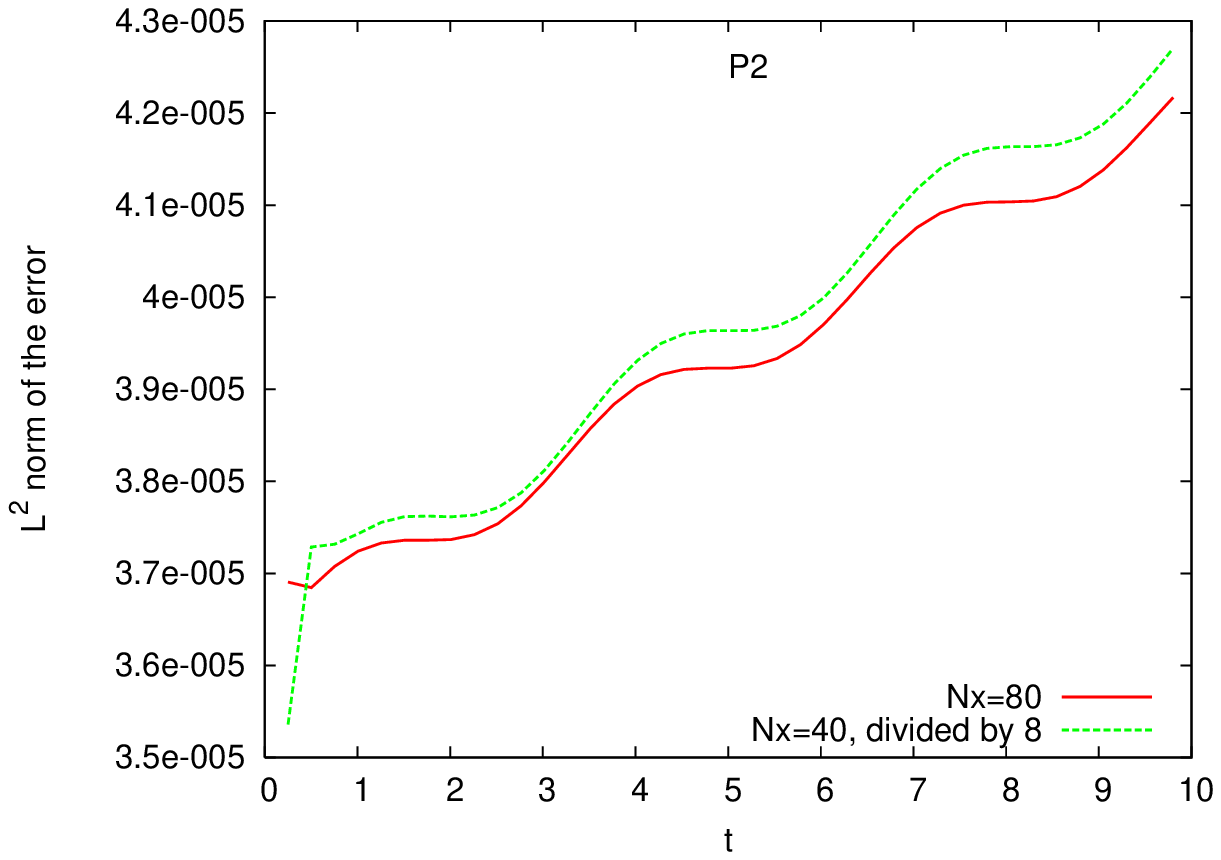}\\
\includegraphics[width=0.5\textwidth]{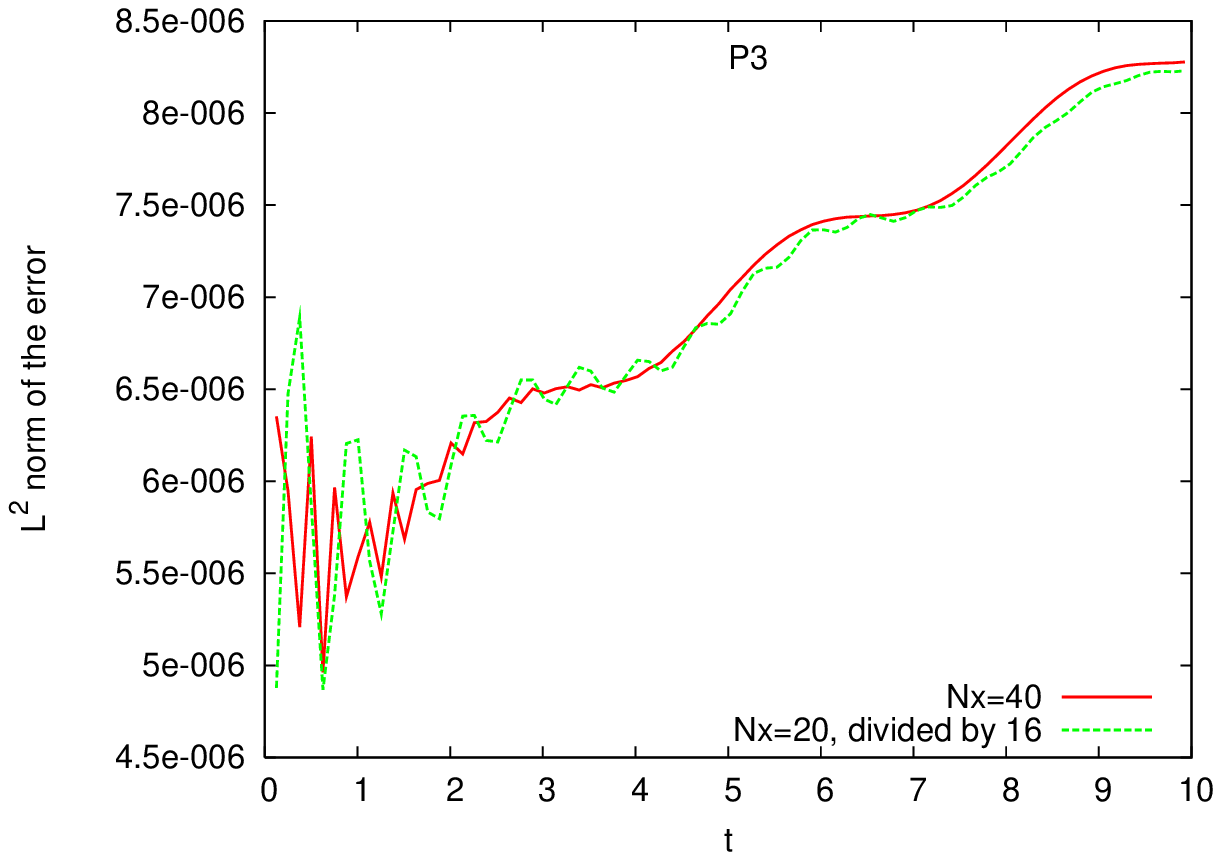}&
\includegraphics[width=0.5\textwidth]{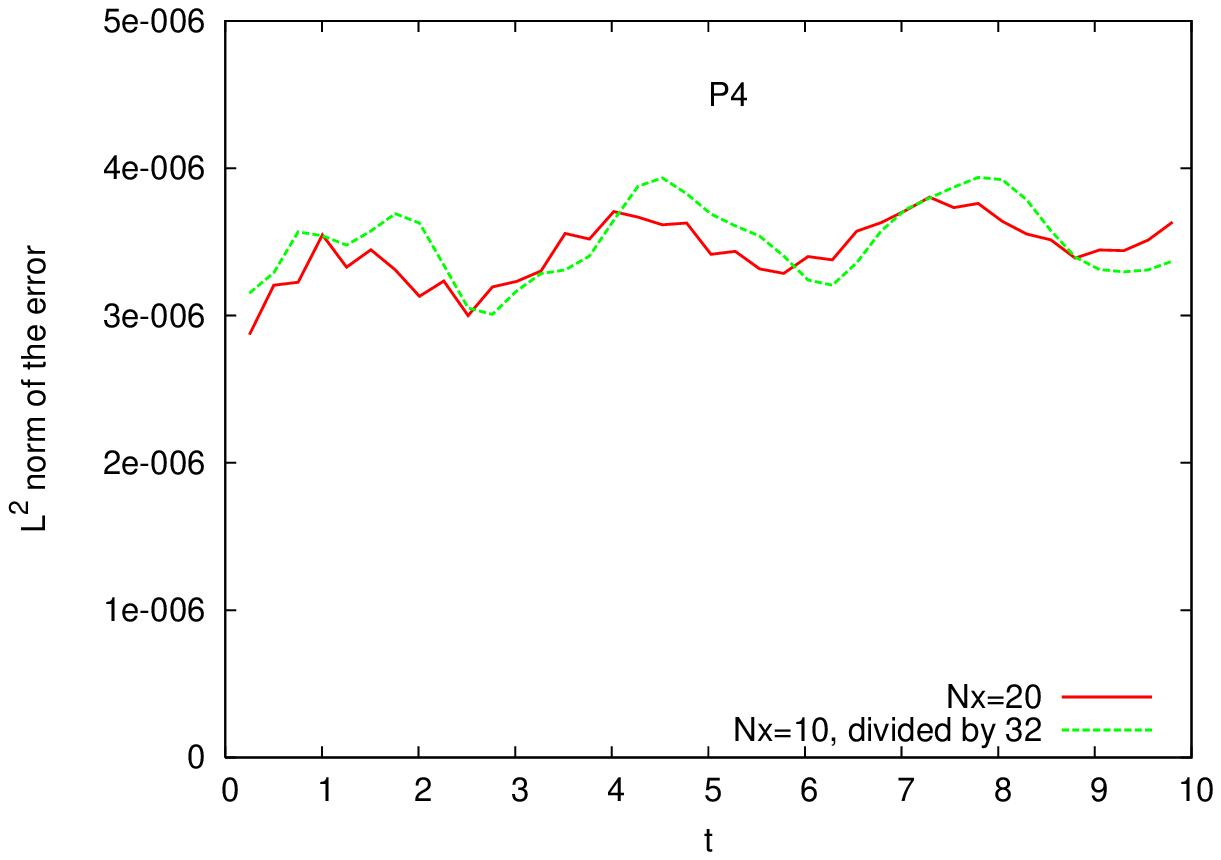}
\end{tabular}
\caption{Convergence behavior for the toy model problem (\ref{Eq::toymodel0}). $P_k$ marked in the plots means the highest polynomial order used in the numerical calculation. $N_x$ is the number of cells.}\label{fig1}
\end{figure*}

\begin{figure*}
\begin{tabular}{cc}
\includegraphics[width=0.48\textwidth]{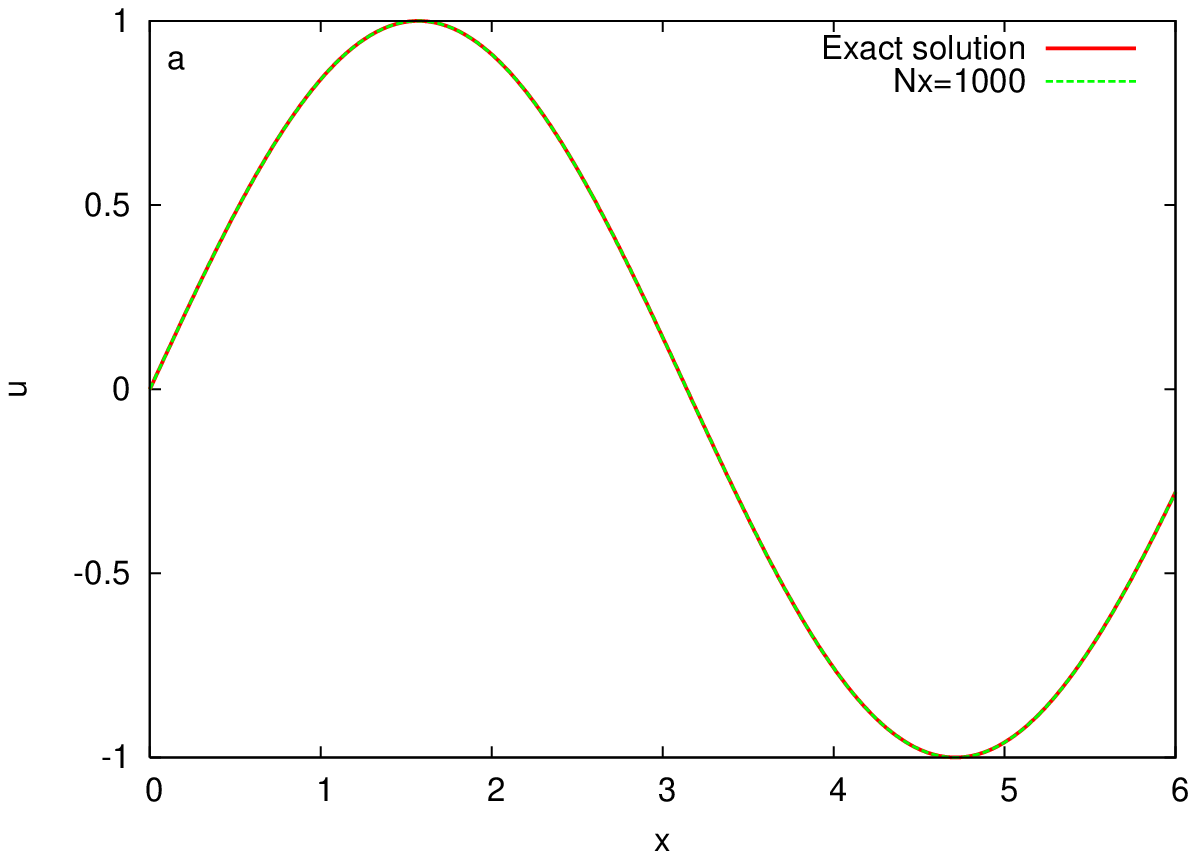}&
\includegraphics[width=0.48\textwidth]{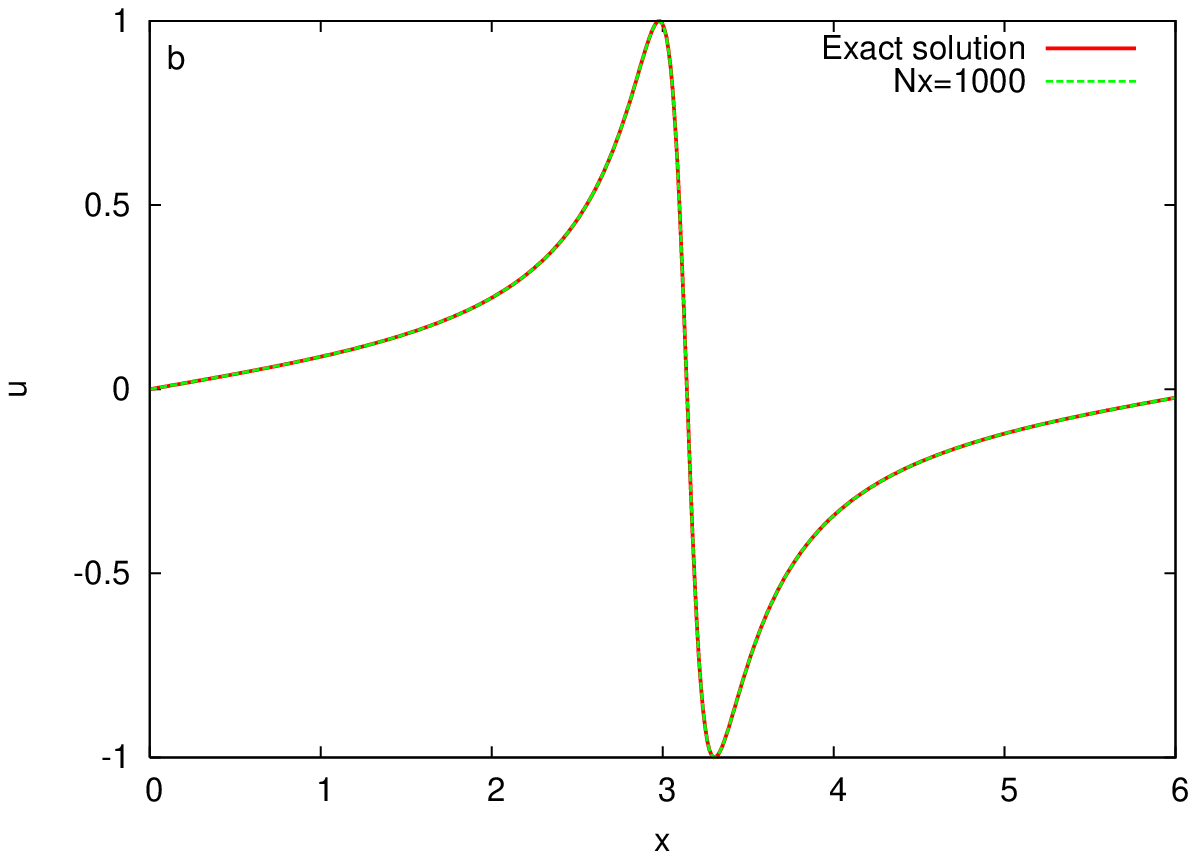}\\
\includegraphics[width=0.48\textwidth]{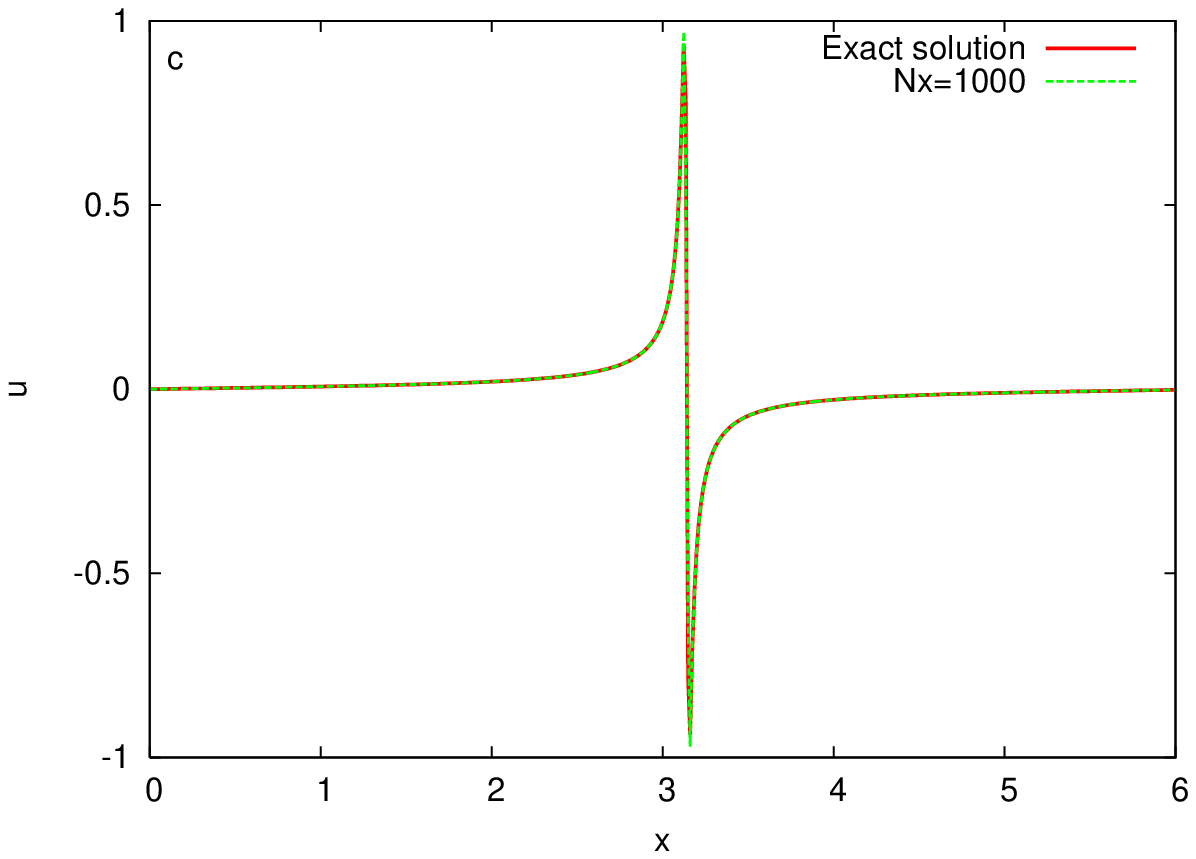}&
\includegraphics[width=0.48\textwidth]{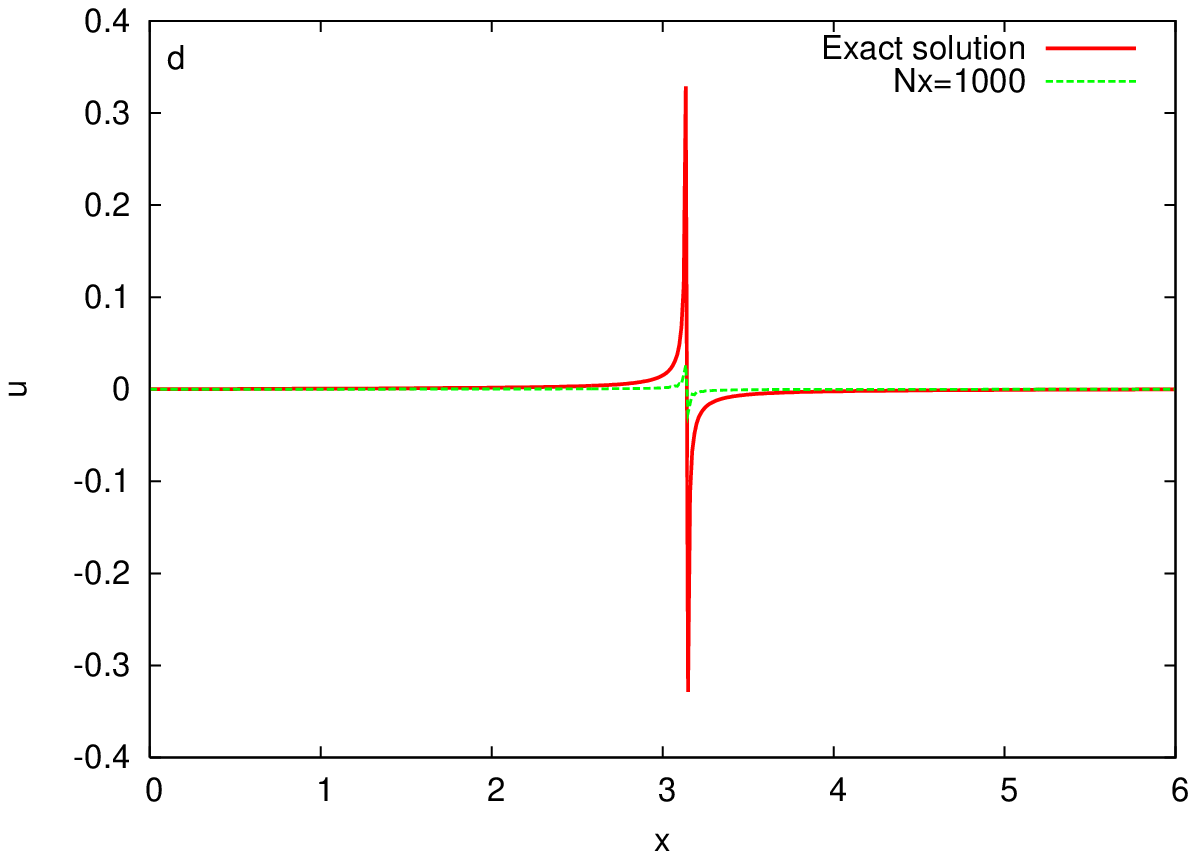}
\end{tabular}
\caption{Evolution results of the toy model (\ref{Eq::toymodel}). $N_x=1000$ cells and highest polynomial $P_1$ are used. Subplots a, b, c and d correspond to $t=0$, $2.51327$, $5.02655$ and $7.53982$ respectively.}\label{fig2}
\end{figure*}

\begin{figure*}
\begin{tabular}{cc}
\includegraphics[width=0.48\textwidth]{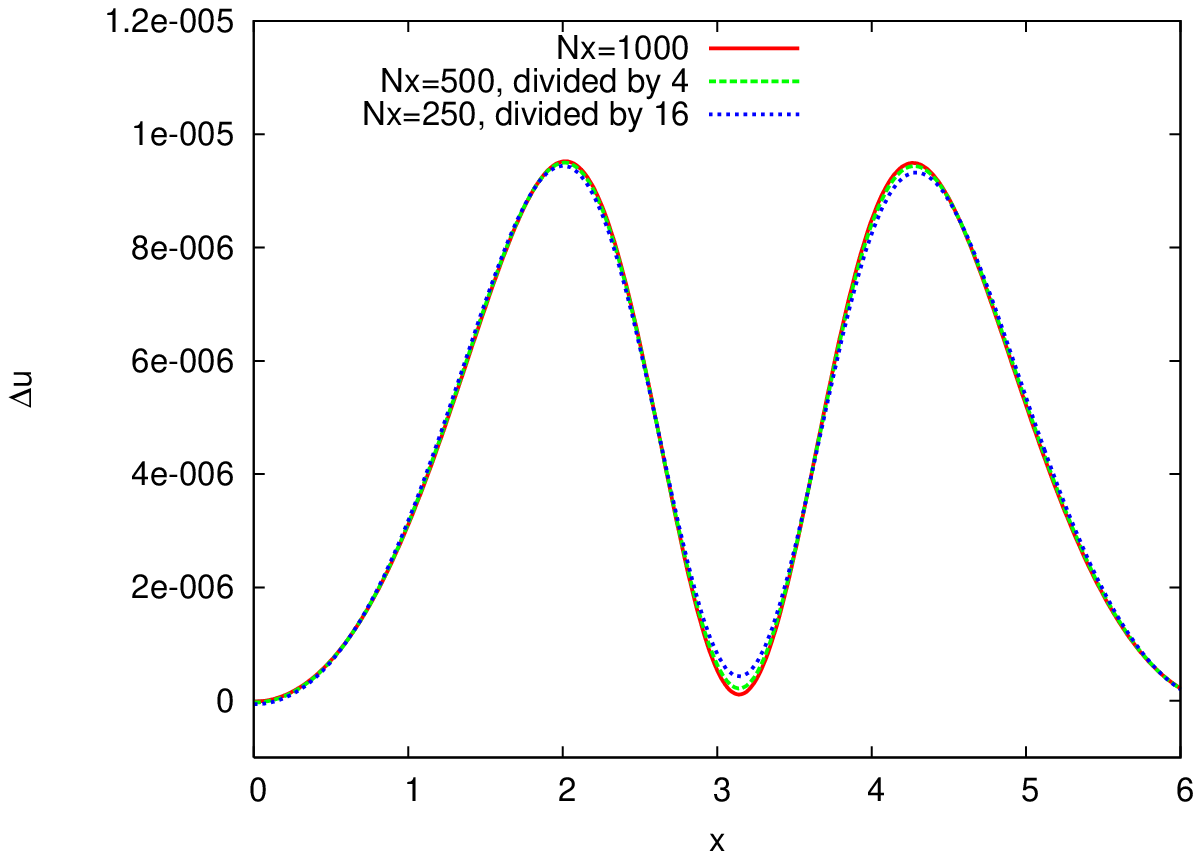}&
\includegraphics[width=0.48\textwidth]{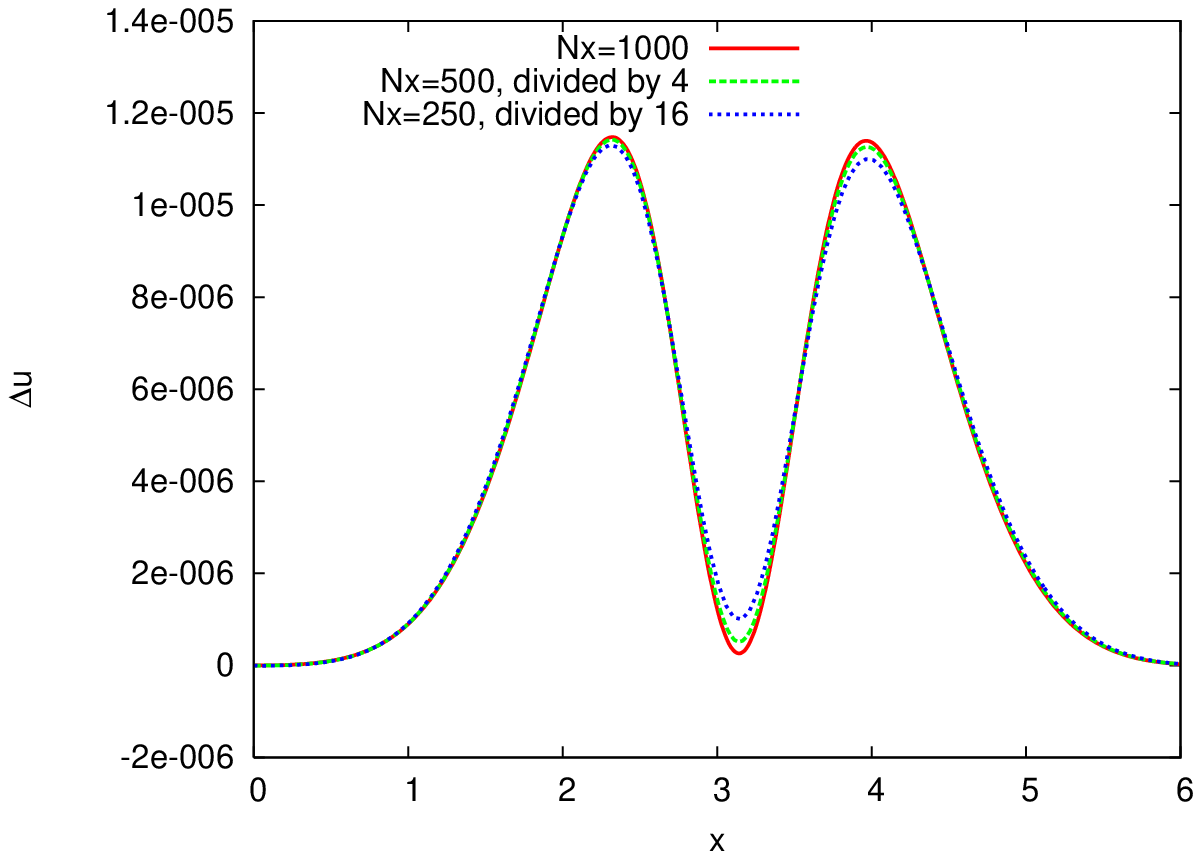}\\
\includegraphics[width=0.48\textwidth]{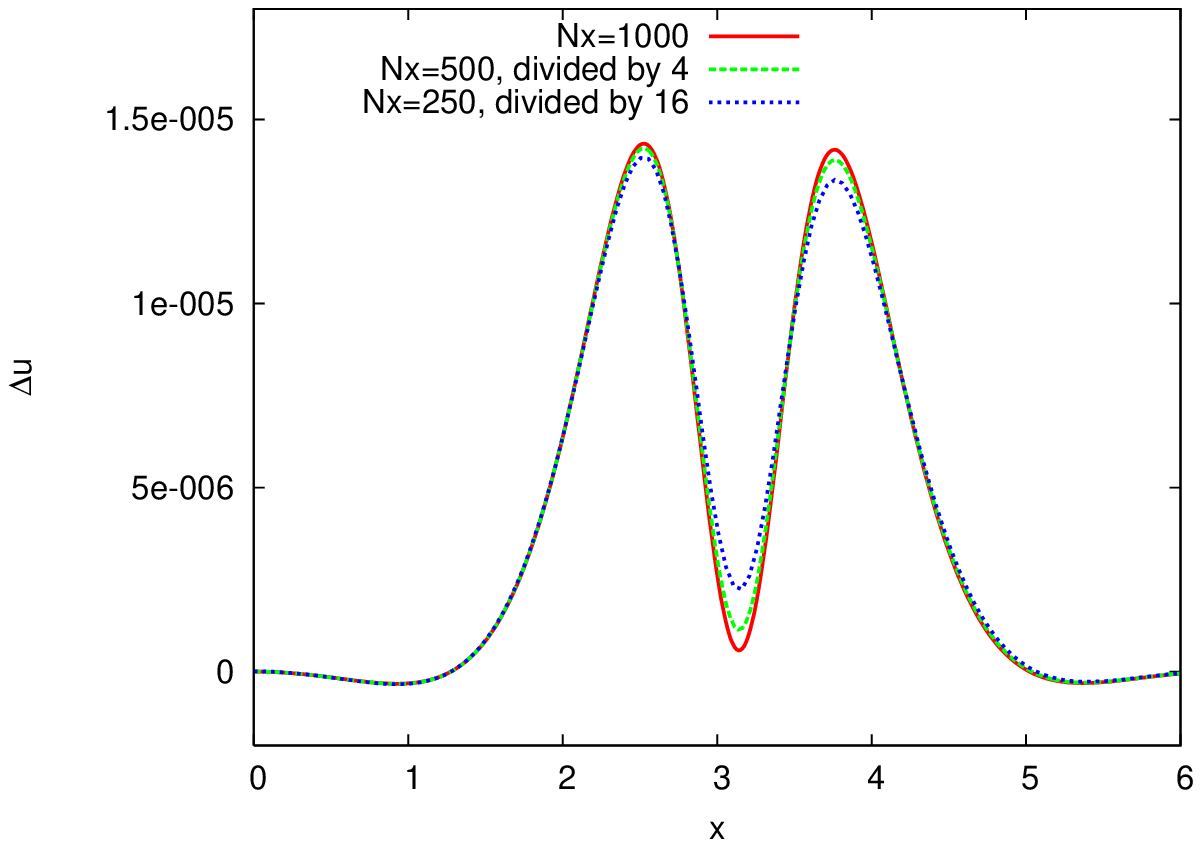}&
\includegraphics[width=0.48\textwidth]{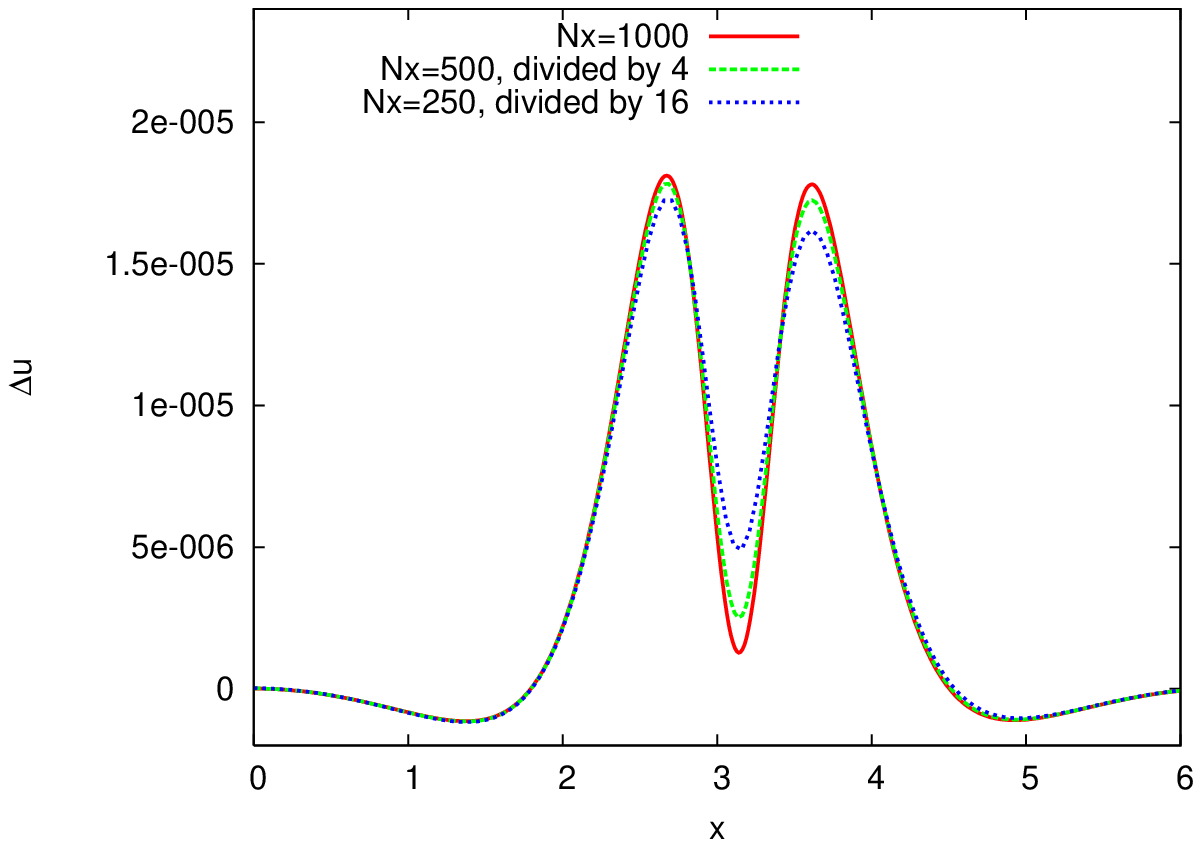}
\end{tabular}
\caption{Convergence behavior of the numerical solution of the toy model (\ref{Eq::toymodel}). $N_x=1000$, $N_x=500$ and $N_x=250$ cells are respectively used. 4 and 16 corresponds to the 2nd order convergence factor respectively for $N_x=500$ and $N_x=250$. Subplots a, b, c and d correspond to $t=0.251327$, $0.502655$, $0.753982$ and $1.00531$ respectively.}\label{fig3}
\end{figure*}

\begin{figure}
\begin{tabular}{c}
\includegraphics[width=0.5\textwidth]{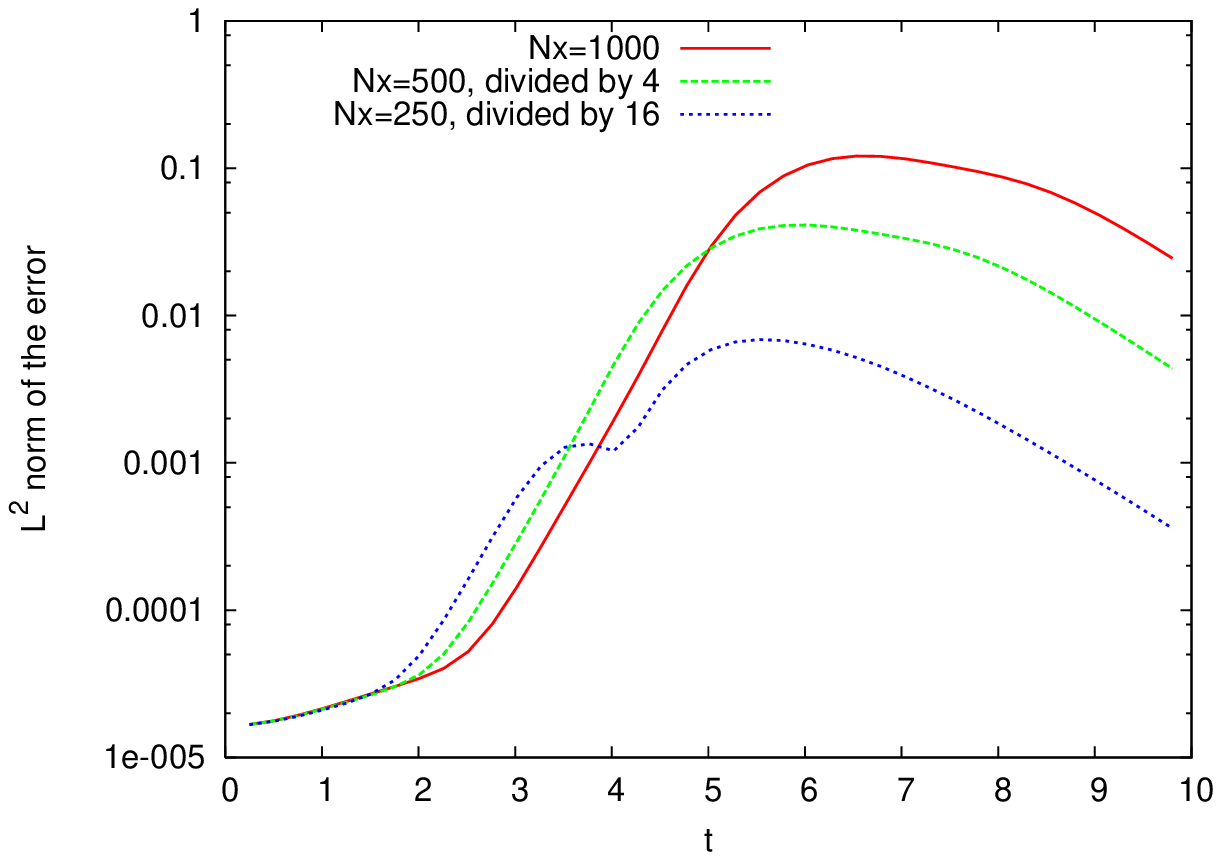}
\end{tabular}
\caption{The convergence behavior of the $L^2$ norm of the numerical error corresponding to the Fig.~\ref{fig2}. Before about $t=2$ quite good second order convergence can be seen.}\label{fig4}
\end{figure}

During the evolution, high frequency noise may develop and make the calculation unstable. We can use a limiter to alleviate the high frequency numerical error. Our limiter only treat the Garlerkin coefficient of $P_1$. We denote such a coefficient $u_1$. If $u_1>50\Delta x^2$, we update $u_1$ with following steps
\begin{align}
&\Delta u^j=u_0^j-u_0^{j-1},\\
&a_1=\Delta u^j,\\
&a_2=\Delta u^{j+1},\\
&a_3=\min(\max(0,a_1),\max(\min(0,a_1),u_1^j)),\\
&u_1^j=\min(\max(0,a_2),\max(\min(0,a_2),a_3)).
\end{align}
We have used $u_0^j$ and $u_1^j$ to denote the coefficients of $P_0$ and $P_1$ respectively for $j$-th cell.

At the boundary, we add one more buffer cell for boundary condition treatment. If periodic boundary condition is adopted, the data in the buffer cell takes the same values as the end cell on the other side. If Dirichlet boundary condition is adopted, the $P_0$ coefficient is set according to the Dirichlet boundary condition, while other polynomial coefficients are set 0. This buffer cell treatment follows the idea we used in the constraint preserving boundary condition implementation work \cite{PhysRevD.88.084057}. So it can be extended straightforwardly for Sommerfeld boundary condition \cite{PhysRevD.78.124011} and constraint preserving boundary condition \cite{0264-9381-23-16-S09,PhysRevD.93.063006}.

Note that our limiter algorithms admits first order convergence. Besides the limiter our local discontinuous Garlerkin finite element method admits ($k+1$)-th order convergence where $k$ is the highest order polynomial $P_k$ used for the function decomposition (proposition 2.1 of \cite{yan2011local}). But this ideal convergence can be achieved only for smooth problem. Unsmooth function may decrease the related convergence order. And more, the detail convergence order may affected by the flux factor $\xi$ defined in (\ref{eq:xi}). This phenomena has been found in previous studies such as in \cite{Klingenberg2017An}. For global Lax-Friedrichs scheme, if the theory expected convergence order is even, the convergence order will not be affected. But if the expected order is odd, the convergence order may decrease half. For local Lax-Friedrichs scheme, if the theory expected convergence order is odd, the convergence order will not be affected. But if the expected order is even, the convergence order may decrease one.
\section{Numerical test results}\label{Sec::IV}
\subsection{toy model problems}
In order to test our numerical algorithm and the major part of our code, we investigate two toy model problems. The first one is the simple wave equation
\begin{align}
u_t+u_x=0, -2\pi\leq x \leq 2\pi \label{Eq::toymodel0}
\end{align}
with initial data $u(x)=\sin(x)$ and periodic boundary condition. The analytic solution to this toy model problem is $u(t,x)=\sin(x-t)$. This toy problem is smooth. As expected our numerical solutions show very good $(k+1)$-th order convergence as presented in the Fig.~\ref{fig1}. Here $k$ is the highest order polynomial $P_k$ used for the function decomposition. In this plot, we have chosen $N_x$ the number of cells properly. If the $N_x$ is too large, the round off error will dominate the numerical solution, so the convergence order will be not clear. If the $N_x$ is too small, the numerical solution has not entered the convergence region. We use the Courant-Friedrichs-Lewy (CFL) rule to determine the time step $\Delta t=\alpha_{C}\min_j\Delta x_j$. In general we find the CFL factor $\alpha_{C}$ should decrease along with the increase of the polynomial order. For example, we use $\alpha_{C}=0.2$ for $P_1$ and $P_2$. But $\alpha_{C}=0.1$ is needed for $P_3$ and $P_4$ to make the numerical calculation stable.

Our second toy model problem is
\begin{align}
u_t+\sin(x)u_x=0, -2\pi\leq x \leq 2\pi \label{Eq::toymodel}
\end{align}
with initial data $u(x)=\sin(x)$ and periodic boundary condition. This toy model problem admits analytic solution $u(t,x)=\sin(2\tan^{-1}(e^{-t}\tan(\frac{x}{2})))$.

The evolution results with $N_x=1000$ uniformly populated cells are plotted in Fig.~\ref{fig2}. Since the behavior for $x<0$ part is symmetric to the part $x>0$, we only plot $x>0$ part. During the numerical evolution, the solution becomes more and more sharp. The sharper and sharper solution needs higher and higher resolution to do numerical calculation. Otherwise the numerical error will increase and convergence will be lost at some time point. This is what we exactly see in our numerical solutions. As shown in the plots of Fig.~\ref{fig2}. Before $t=5$, roughly we can not distinguish the exact solution and the numerical solution (shown in the subplot a-c of Fig.~\ref{fig2}). But later the difference becomes larger and larger as shown in the subplot d of Fig.~\ref{fig2}.

In order to check the convergence behavior, we have also done evolutions with $N_x=500$ and $N_x=250$. The difference between our numerical solution and the exact solution is plotted in Fig.~\ref{fig3} for different time. In this figure we can see very good second order convergence behavior. Since we have used highest polynomial order $P_1$, second order convergence is expected. Our numerical result is completely consistent to the theoretical expectation. In Fig.~\ref{fig4} we plot the $L^2$ norm of the difference between our numerical solution and the exact solution respect to time. And the convergence behavior is checked respect to $N_x=1000$, $N_x=500$ and $N_x=250$. For $N_x=250$ the second order convergence can be kept till $t=1.6$. For $N_x=500$ the second order convergence behavior can be kept till $t=1.8$ which is a little bit longer time. This is consistent to our expectation. Since the solution becomes sharper and sharper along time, higher and higher solution is needed to keep warranted convergence. We have tested other polynomial orders. When the polynomial order becomes higher, the Courant factor for the time evolution must be set smaller correspondingly to make sure the evolution stable. This is the same as we have seen in our first toy model.

Our setting $\xi=1$ corresponds to global Lax-Friedrichs scheme for our second toy model problem. We have tested other order polynomials. For $P_2$, $P_3$ and $P_4$ we get 2.5, 4 and 4.5 order convergence respectively.

\subsection{Schwarzschild black hole in Kerr-Schild coordinate}
\begin{figure*}
\begin{tabular}{cccc}
\includegraphics[width=0.25\textwidth]{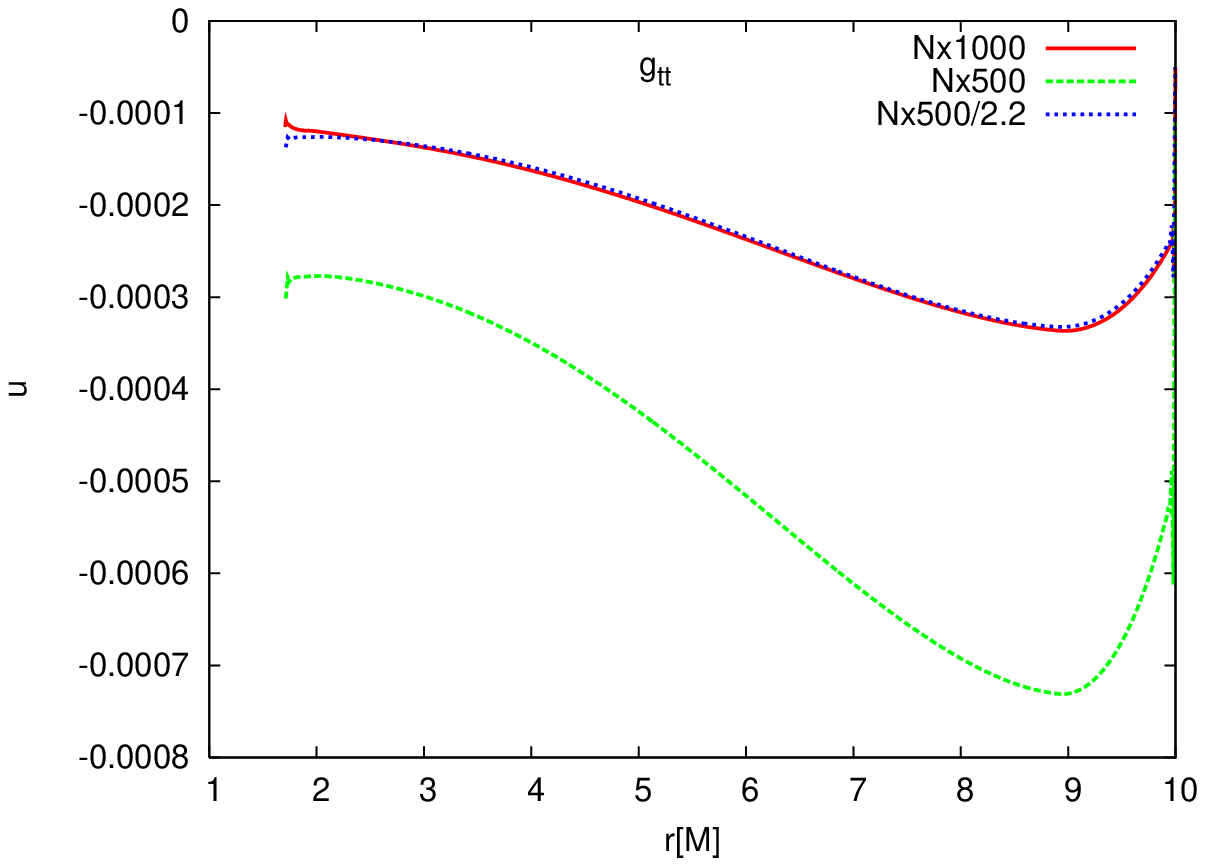}&
\includegraphics[width=0.25\textwidth]{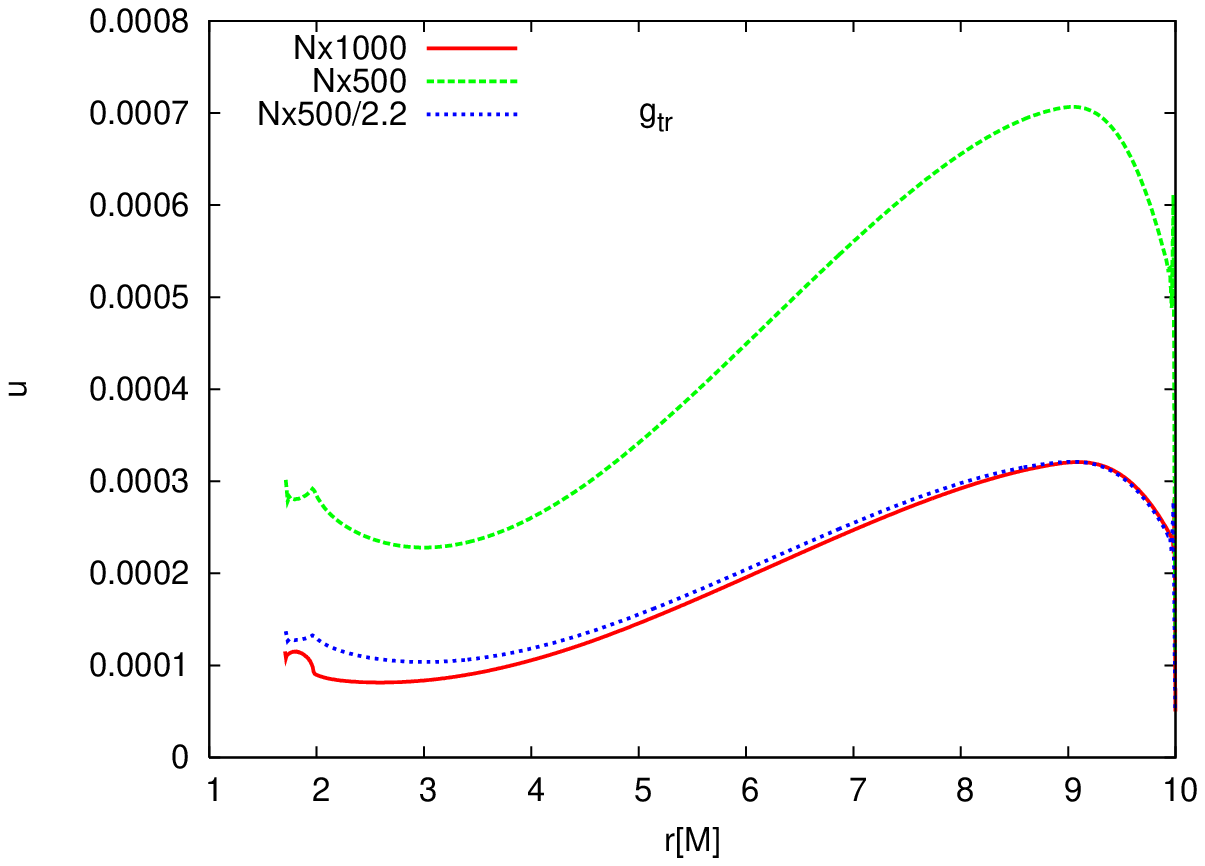}&
\includegraphics[width=0.25\textwidth]{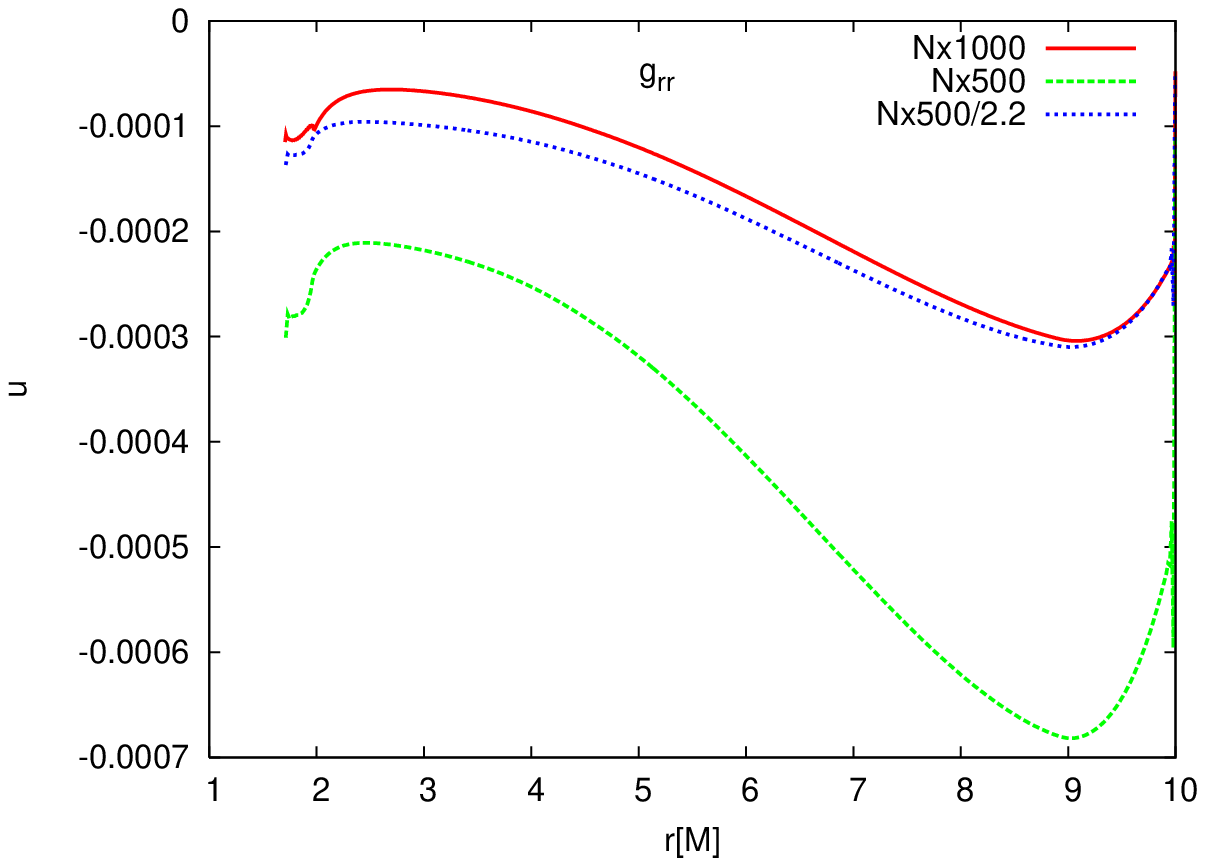}&
\includegraphics[width=0.25\textwidth]{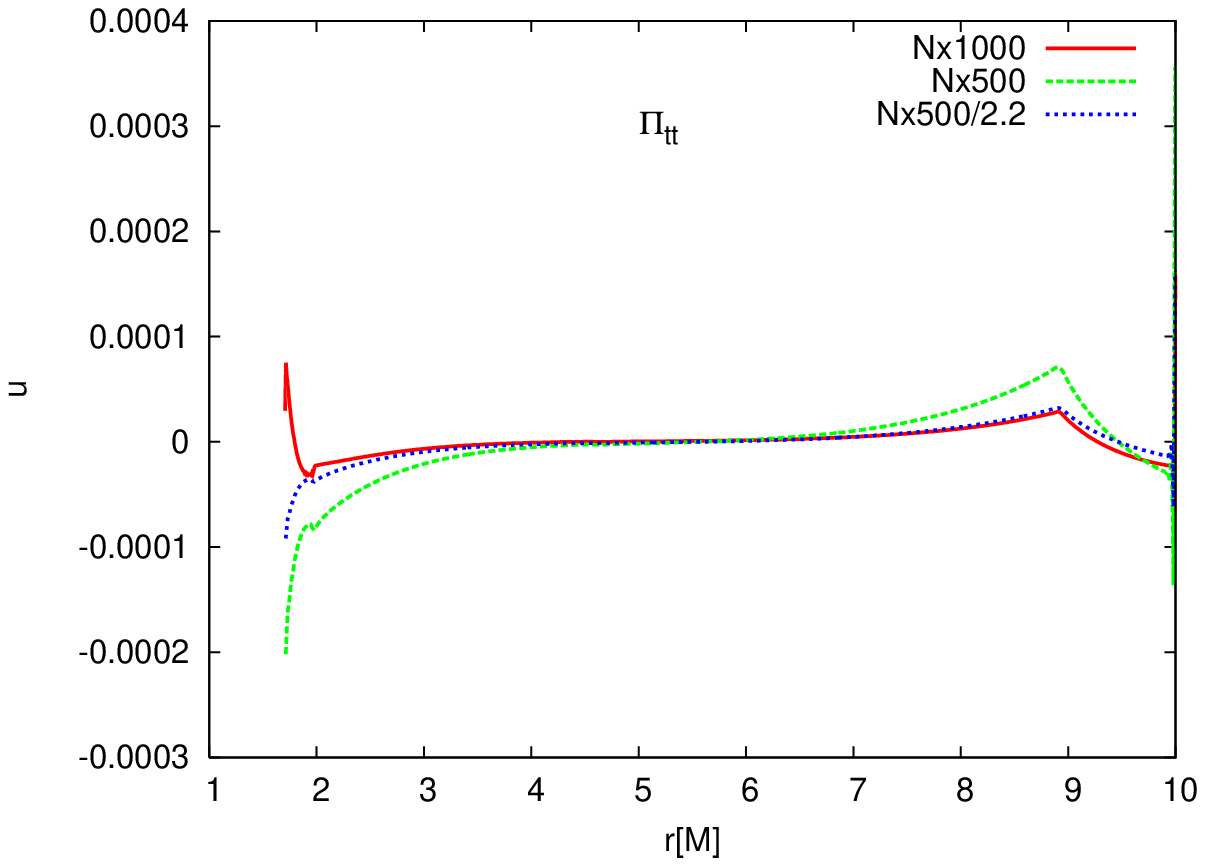}\\
\includegraphics[width=0.25\textwidth]{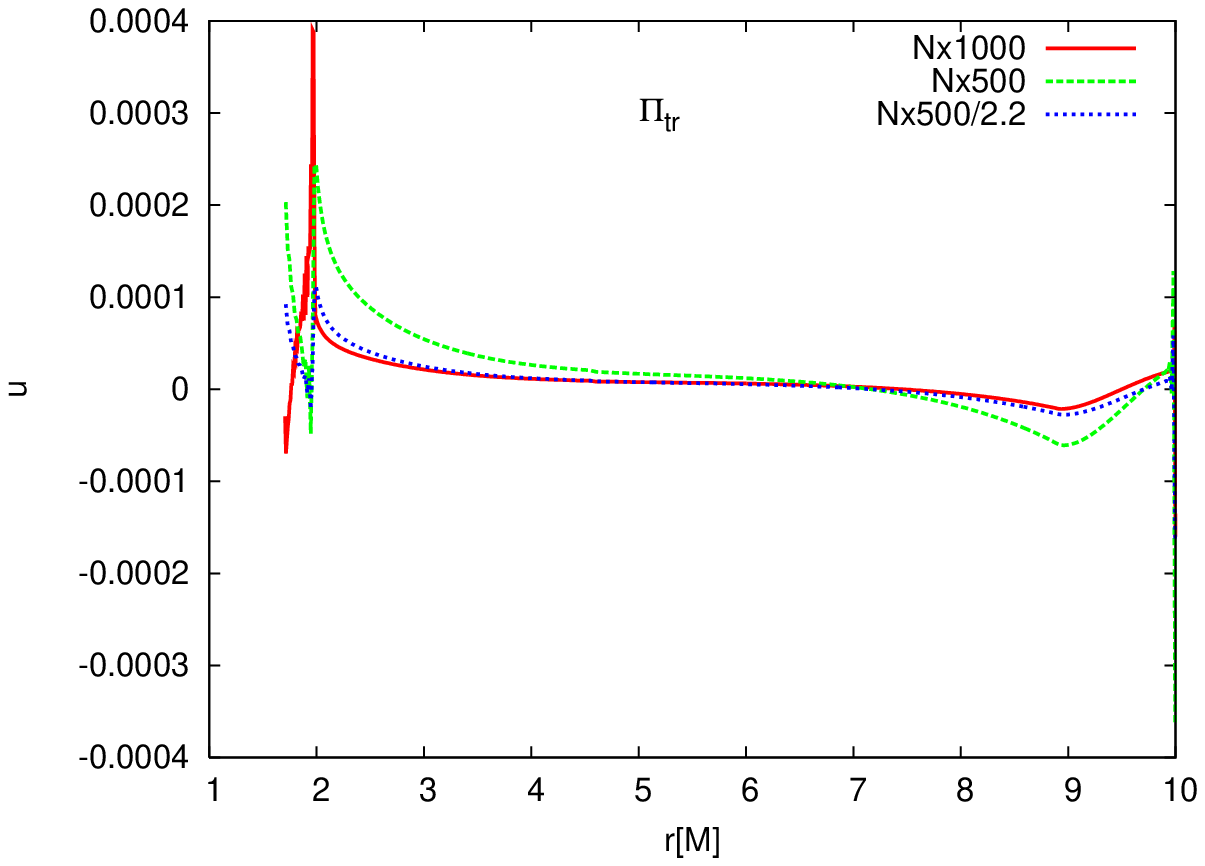}&
\includegraphics[width=0.25\textwidth]{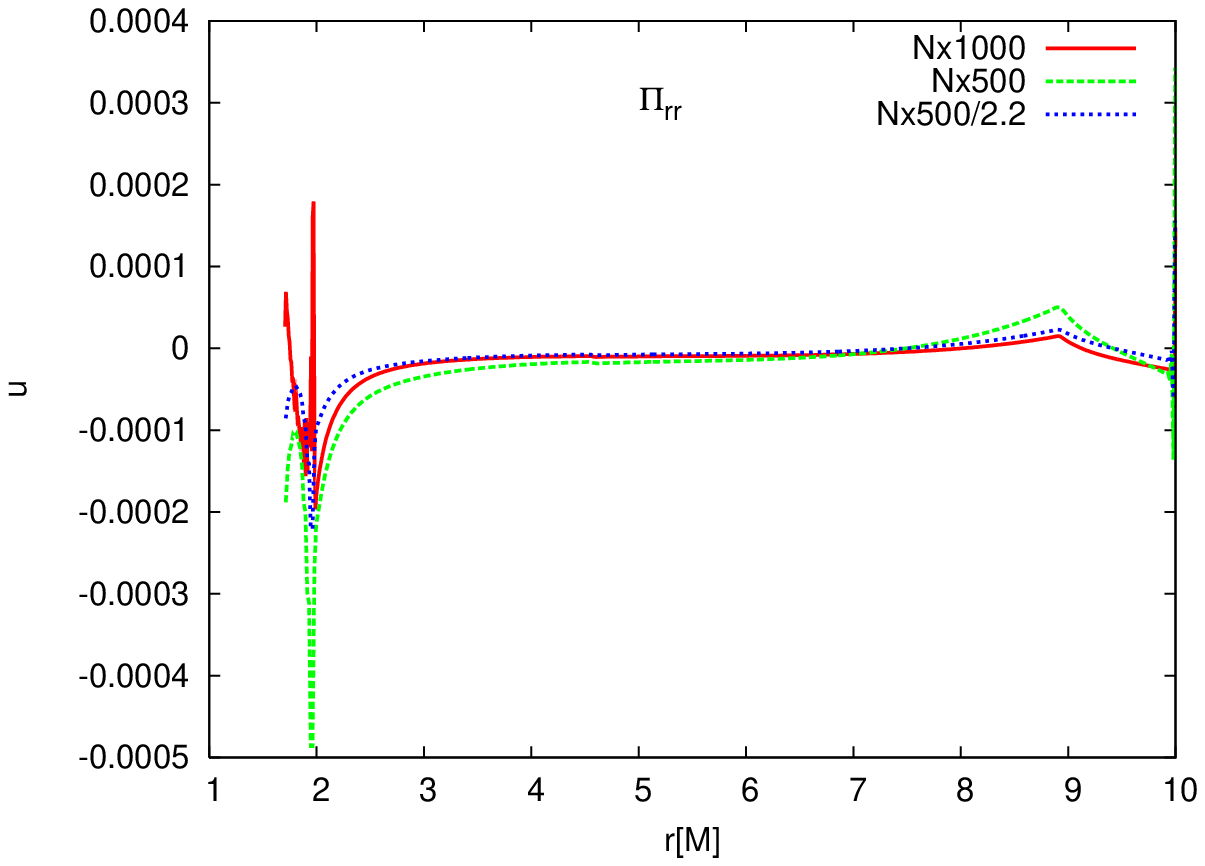}&
\includegraphics[width=0.25\textwidth]{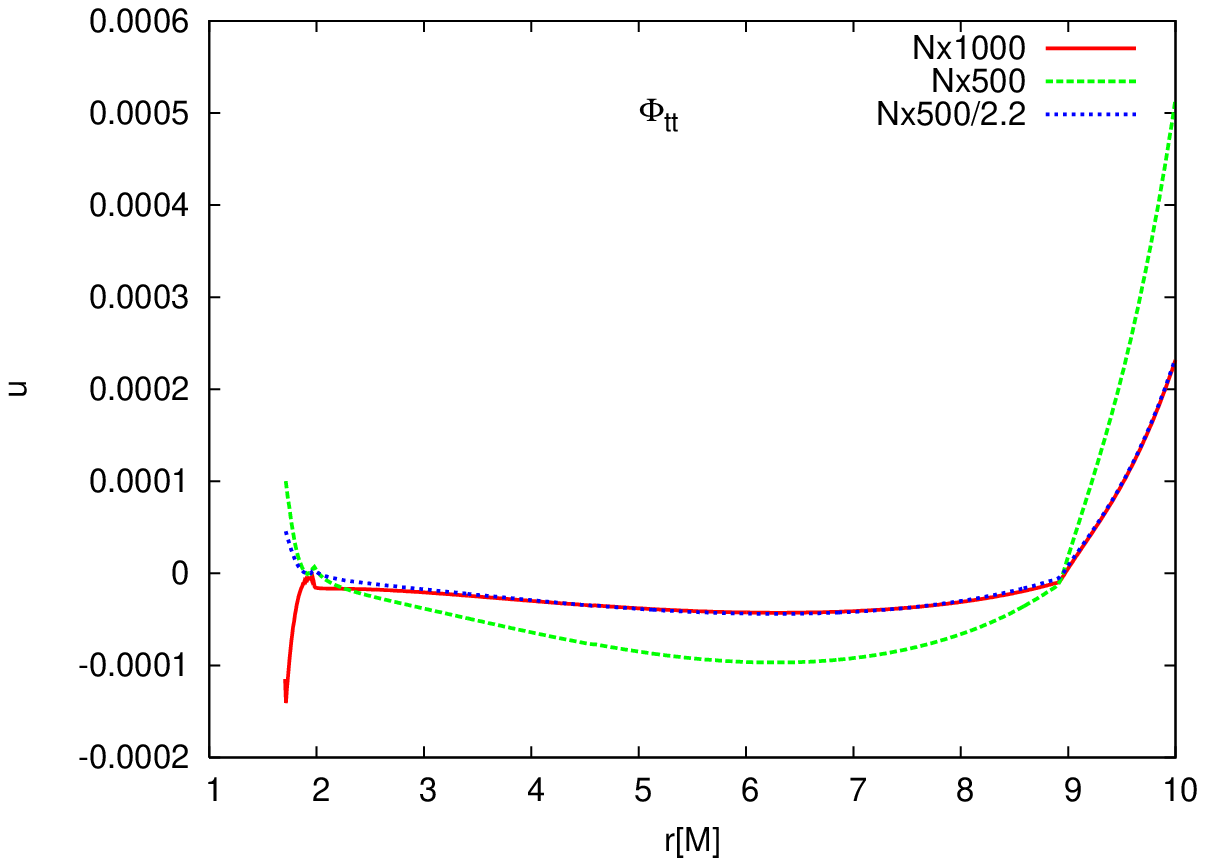}&
\includegraphics[width=0.25\textwidth]{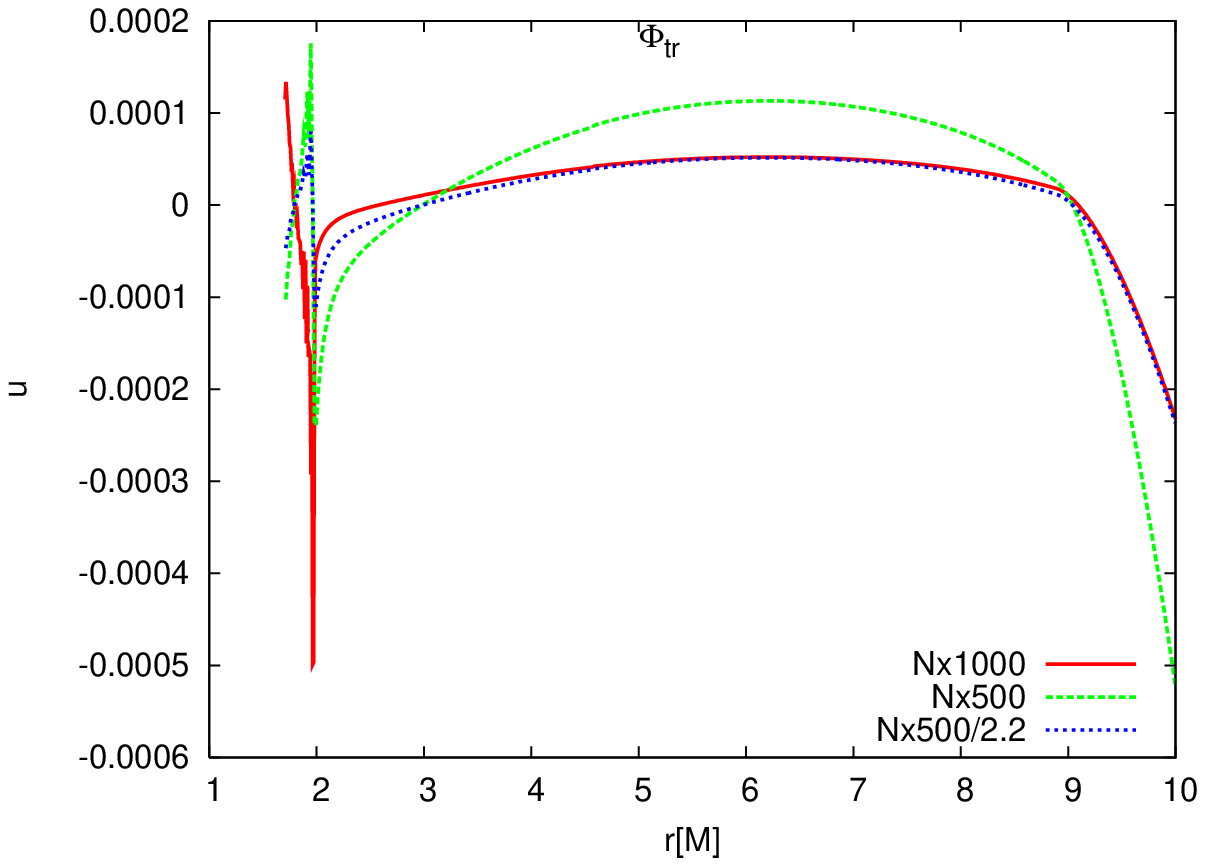}\\
\includegraphics[width=0.25\textwidth]{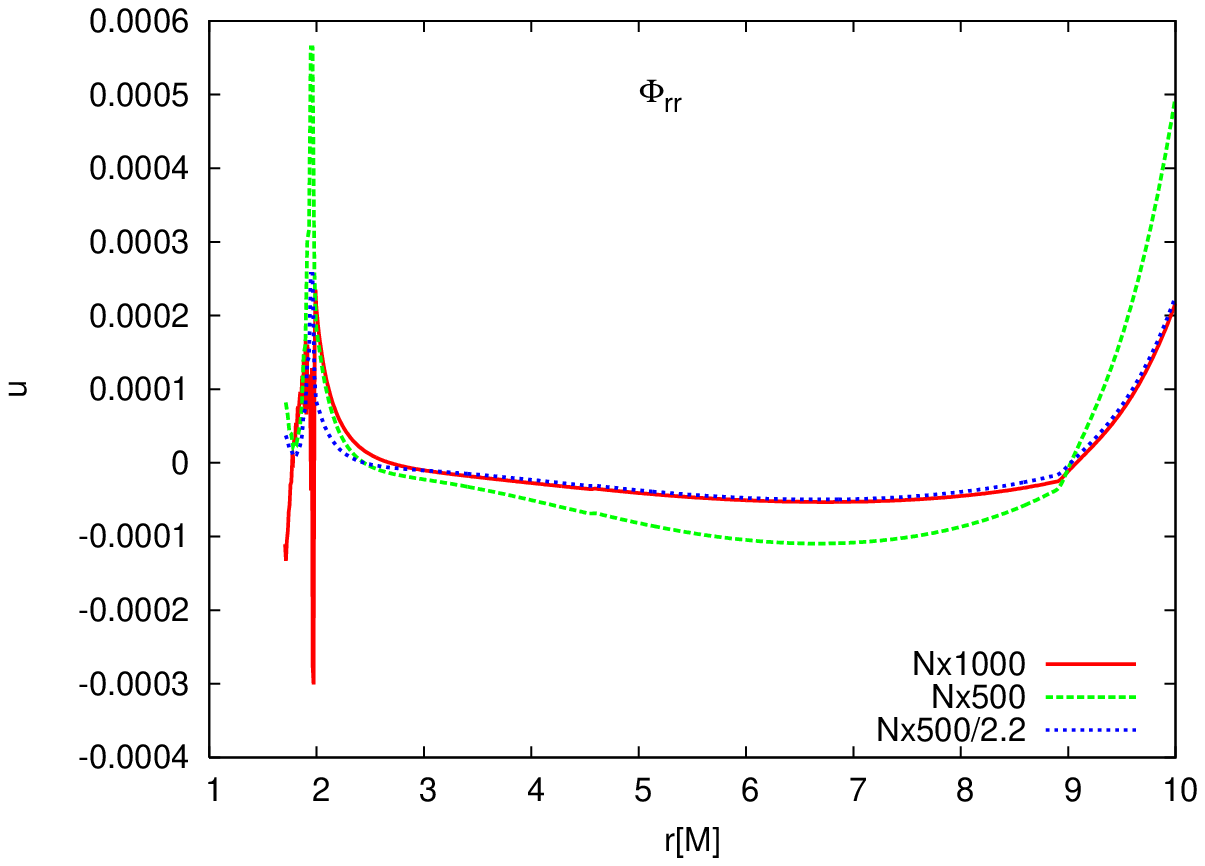}&
\includegraphics[width=0.25\textwidth]{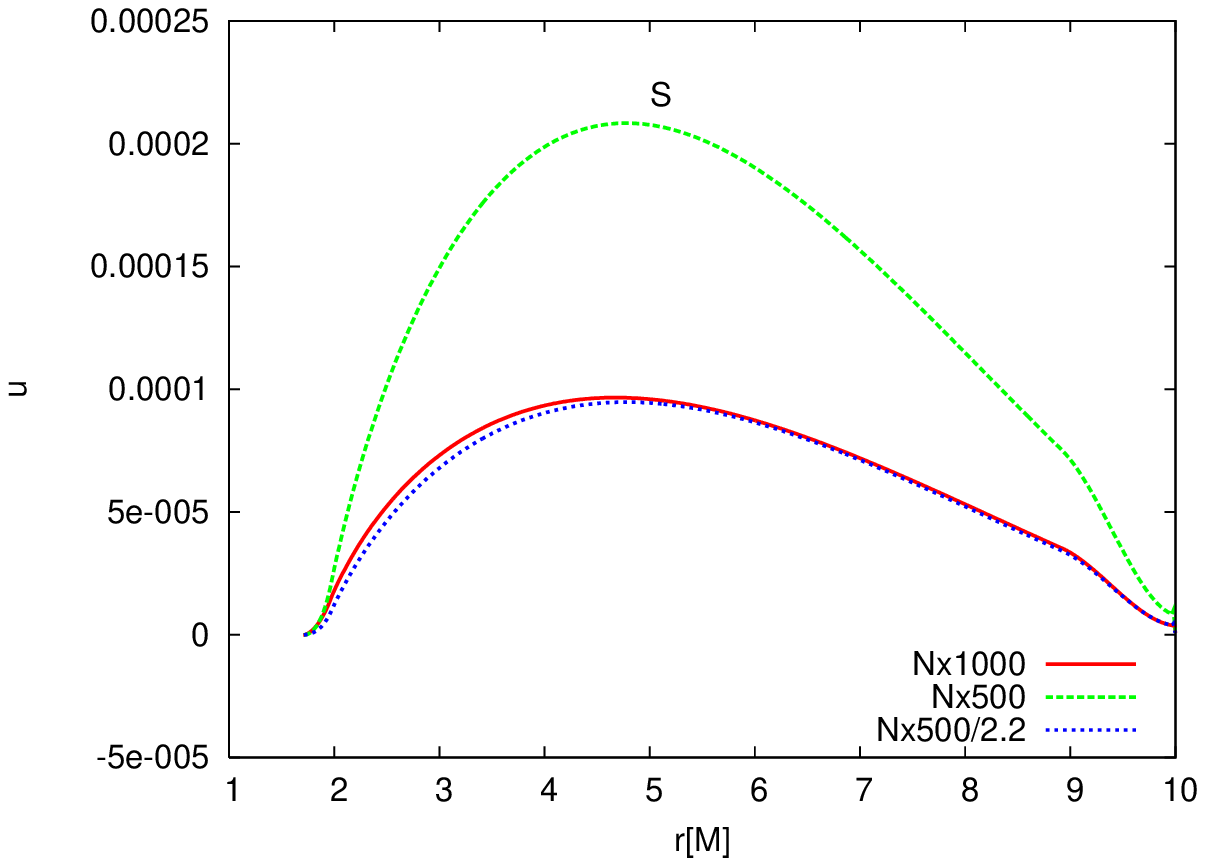}&
\includegraphics[width=0.25\textwidth]{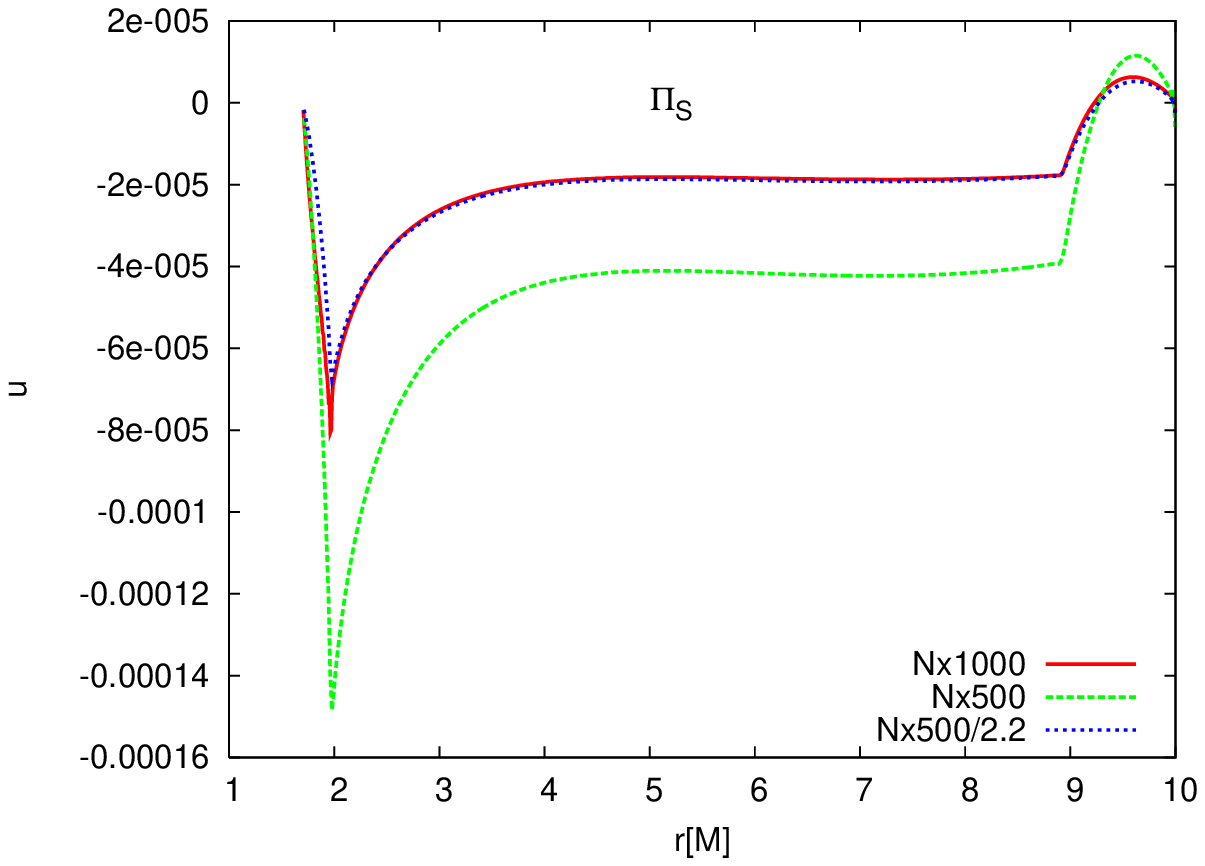}&
\includegraphics[width=0.25\textwidth]{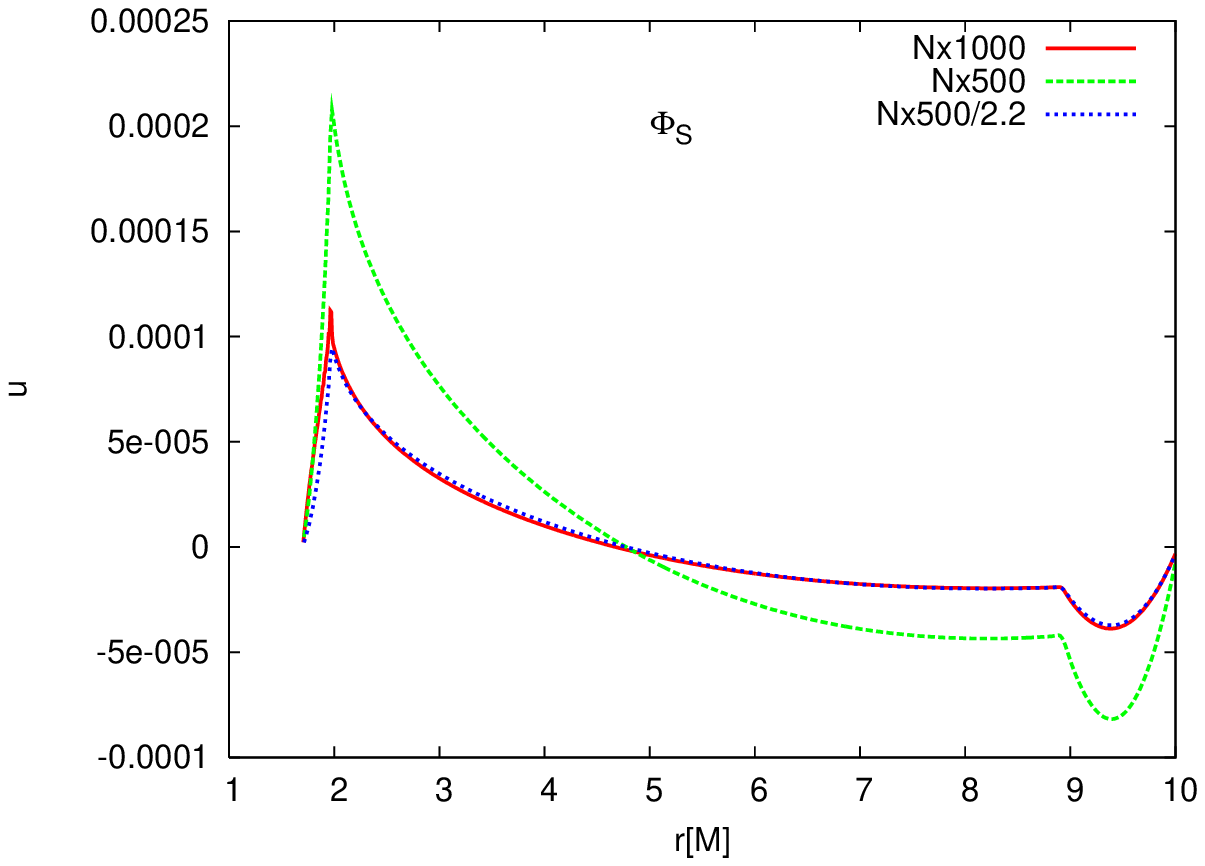}
\end{tabular}
\caption{The numerical error for different variables at $t=9.96M$. Results for $N_x=1000$ and $N_x=500$ are compared. Here highest polynomial $P_1$ is used. First order convergence behavior which corresponds to the factor 2.2 is apparent. Kerr-Schild coordinate is used.}\label{fig5}
\end{figure*}

The Schwarzschild metric in Kerr-Schild coordinate takes the form (c.f., Eq.~(26) of \cite{PhysRevD.66.084026})
\begin{align}
ds^2=&-(1-\frac{2M}{r})dt^2-\frac{4M}{r}dtdr\nonumber\\
&+(1+\frac{2M}{r})dr^2+r^2d\theta^2+r^2\sin^2\theta d\phi^2.\label{ksmetric}
\end{align}
Correspondingly the source function is
\begin{align}
&H_t=-\frac{2M}{r^2},\\
&H_r=2\frac{M+r}{r^2},\\
&H_\theta=\cot\theta,\\
&H_\phi=0.
\end{align}
We plug these source functions into our dynamical system (\ref{problem_eq}). We chose our computational domain $1.7M<r<10M$. The inner boundary is lightly inside the event horizon which is similar to the excision treatment for black hole \cite{PhysRevLett.95.121101}. At the boundaries, we apply the Dirichlet boundary condition based on the given metric form (\ref{ksmetric}).

\begin{figure}
\begin{tabular}{cc}
\includegraphics[width=0.25\textwidth]{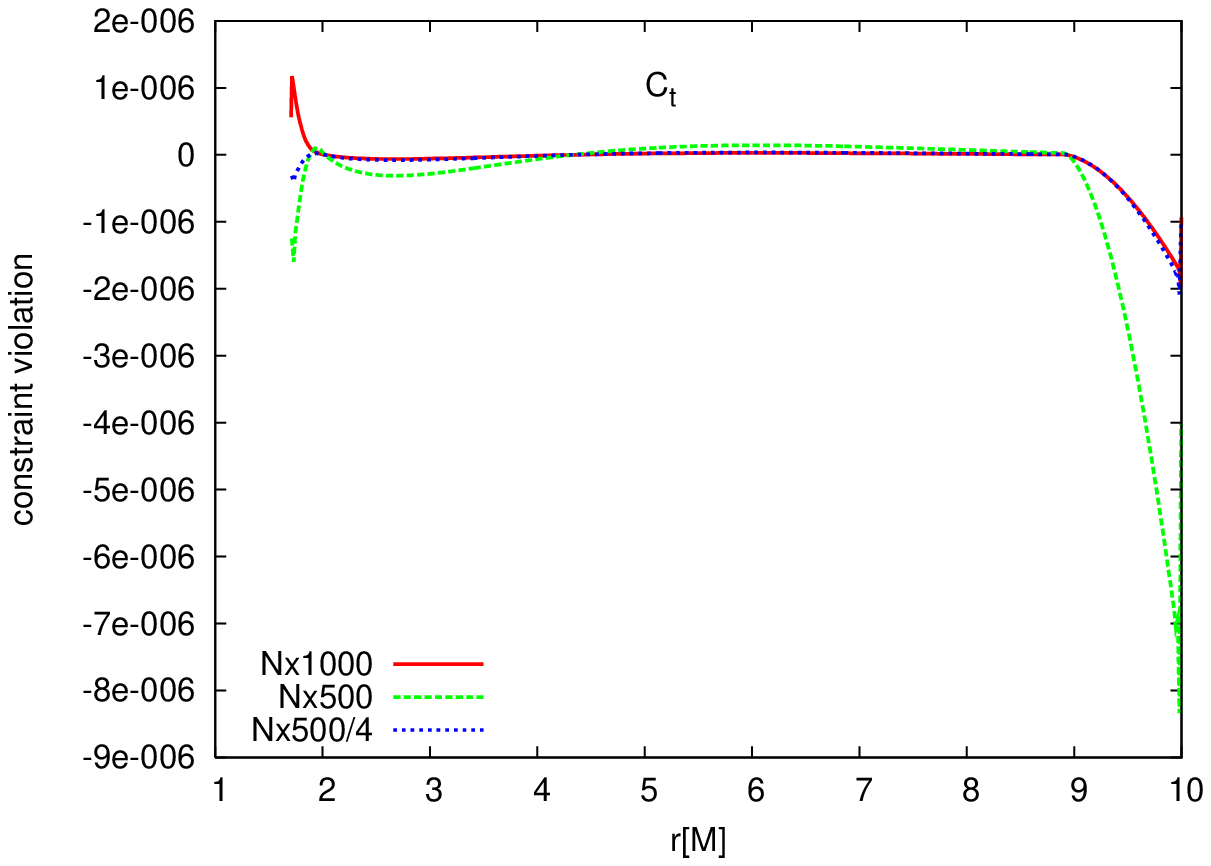}&
\includegraphics[width=0.25\textwidth]{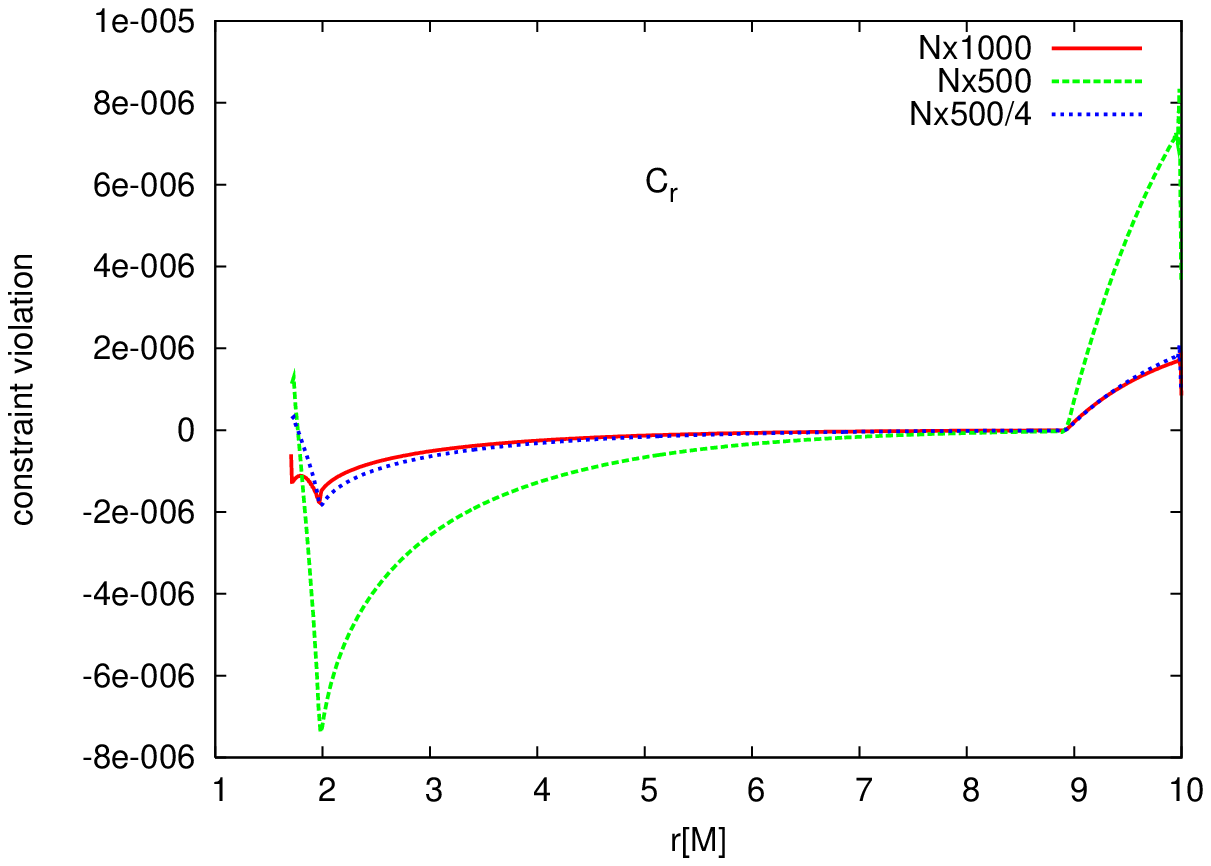}
\end{tabular}
\caption{The constraint violation $C_{t,r}$ corresponding to the Fig.~\ref{fig5}. Kerr-Schild coordinate is used.}\label{fig6}
\end{figure}
In the Fig.~\ref{fig5} we show the resulted numerical error for different dynamical variables. In this figure, we have used highest polynomial $P_1$ to do the evolution. Two resolutions have been tested. Compare the results for the two different resolutions, we can see clearly first order convergence behavior. One more interesting phenomena can also be seen. There are two unsmooth regions showup for $\Pi$ and $\Phi$ variables. One such position locates at the event horizon $r=2M$. The other one locates near out boundary. During the evolution, the second unsmooth point will move. Differently the horizon one does not move. Thanks to our discontinuous Garlerkin method, the unsmooth behavior can be numerically kept.

The constraint related to our dynamical system (\ref{problem_eq}) can be divided into three classes \cite{0264-9381-23-16-S09,PhysRevD.93.063006}. One is the intrinsic constraint of generalized harmonic formulation $C_A$. The second class is the one reduced from the dynamical variables $\Phi$ definition. The third class is the higher order constraint reduced from the evolution. In the current work we focus on the first class constraint violation. The corresponding constraint violation is plotted in the Fig.~\ref{fig6}. The constraint violation shows second order convergence.

\begin{figure*}
\begin{tabular}{cccc}
\includegraphics[width=0.25\textwidth]{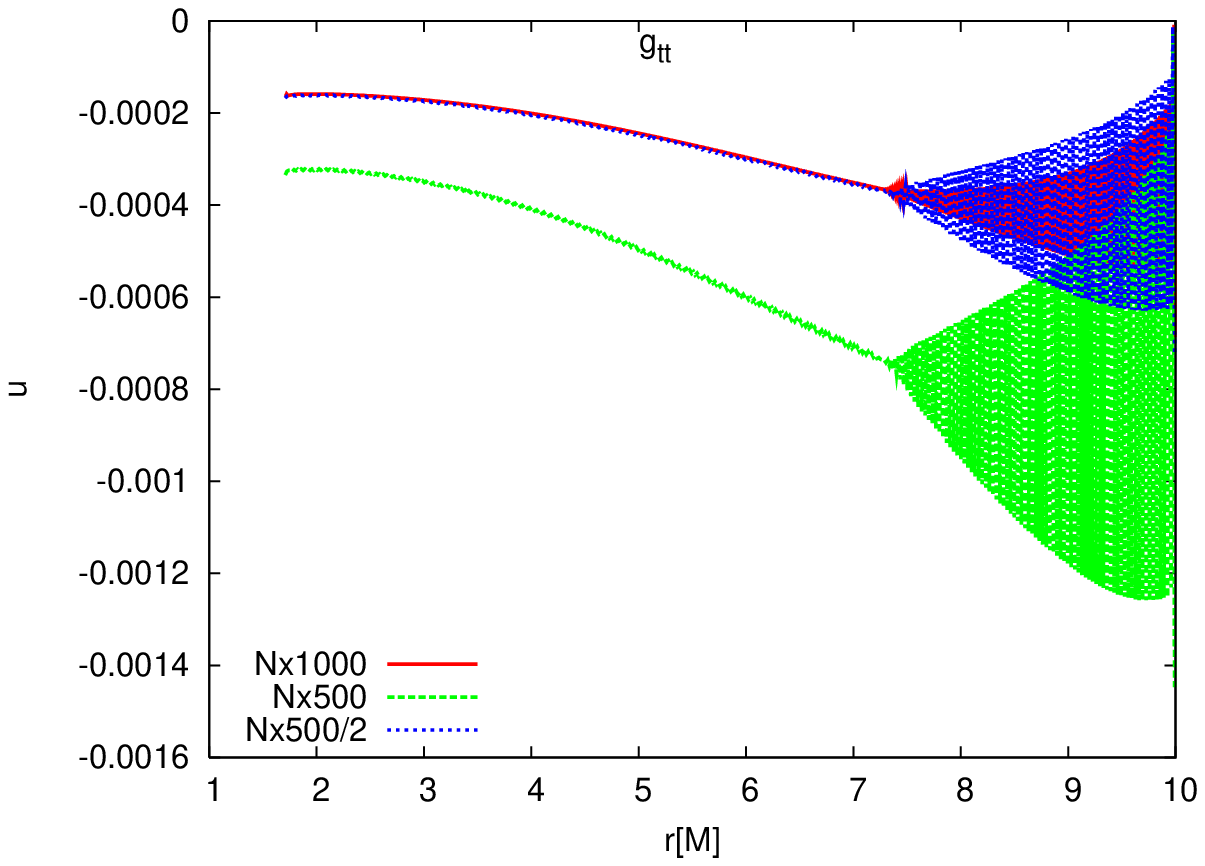}&
\includegraphics[width=0.25\textwidth]{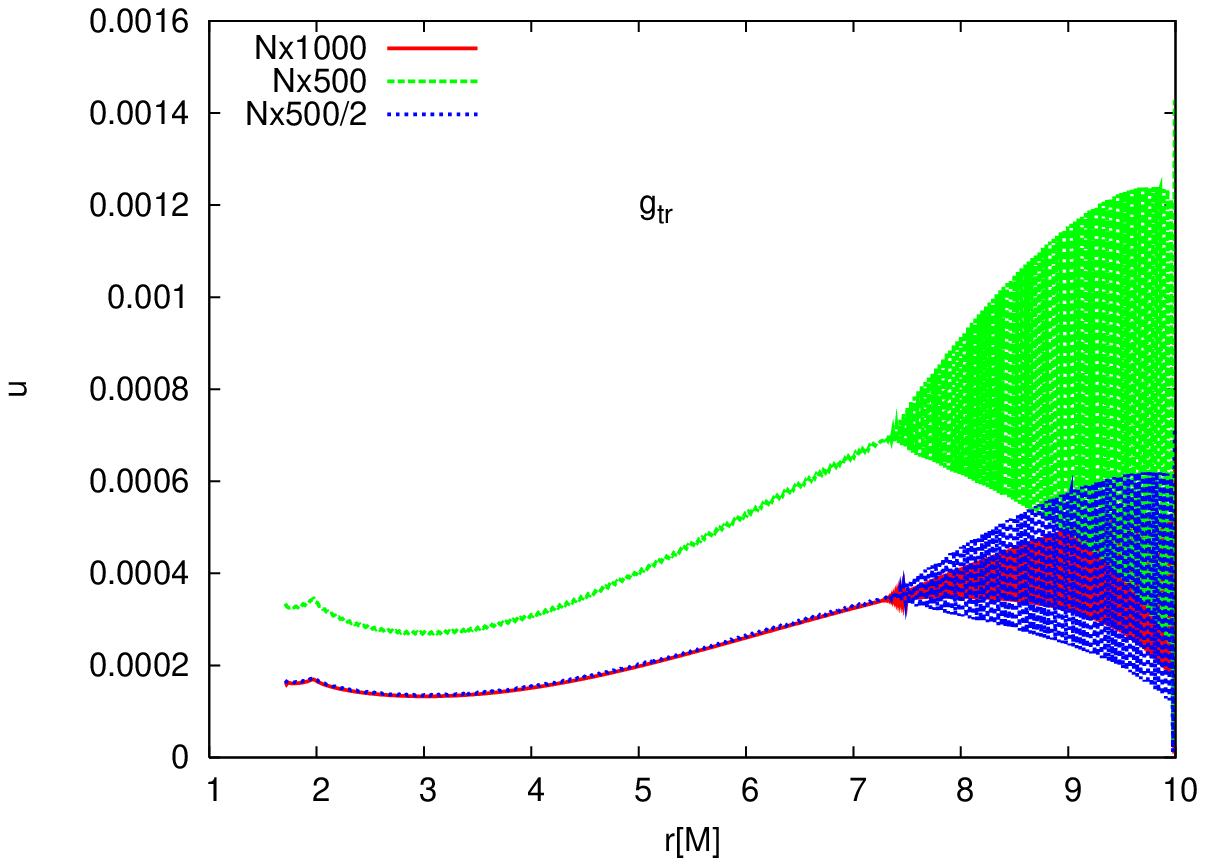}&
\includegraphics[width=0.25\textwidth]{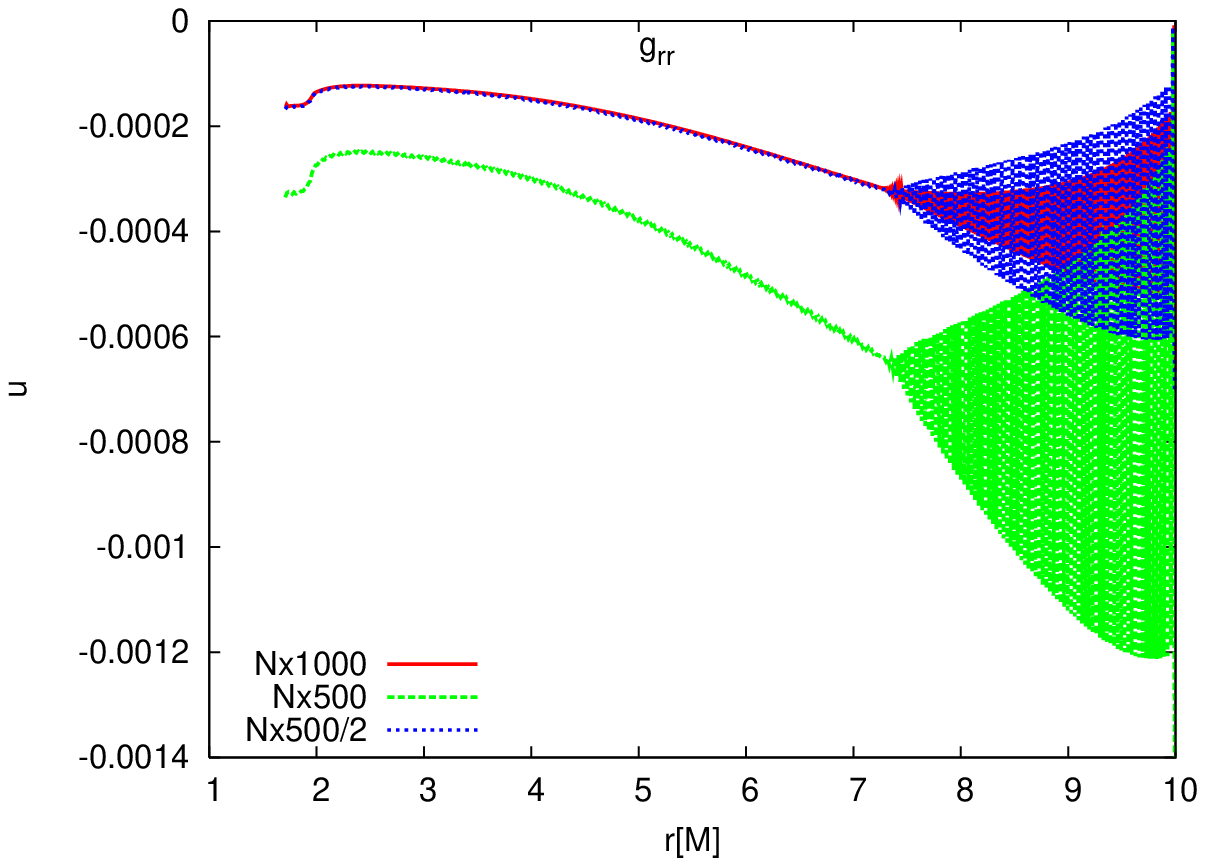}&
\includegraphics[width=0.25\textwidth]{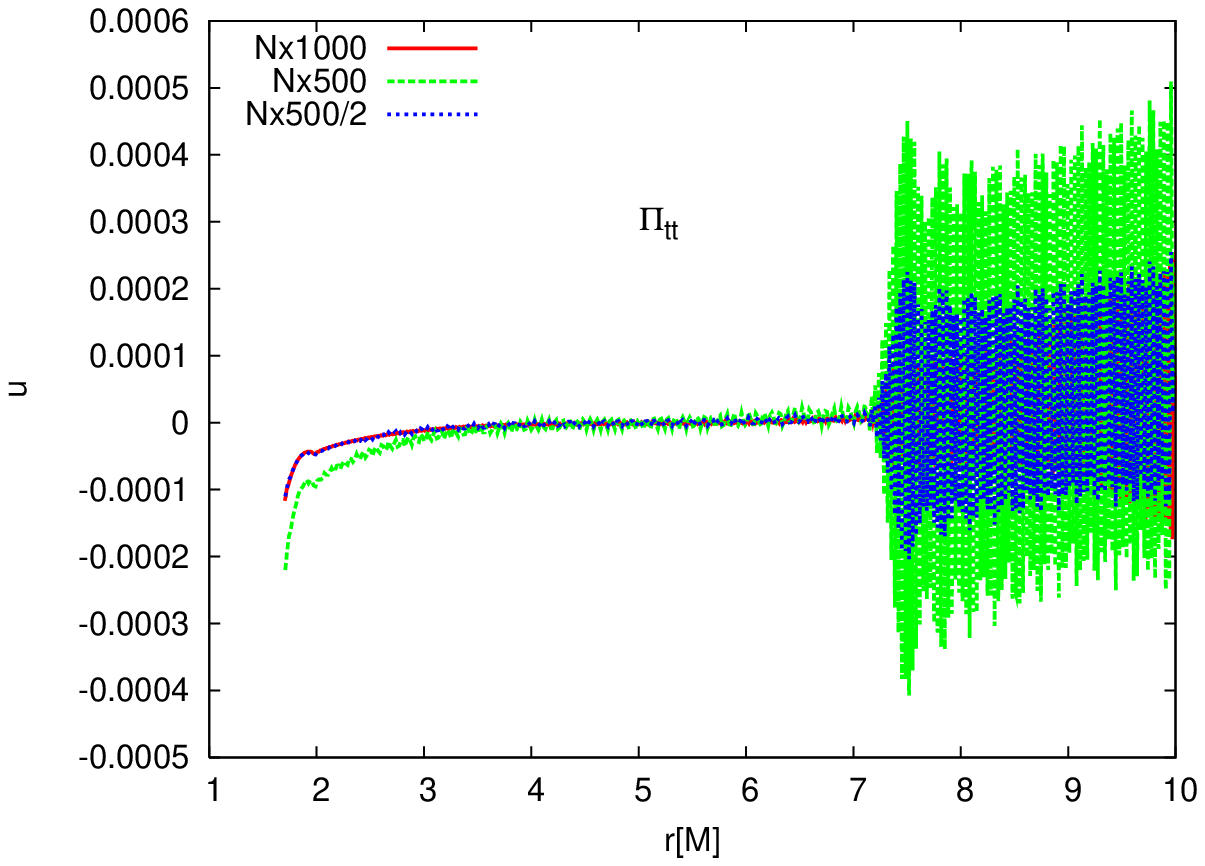}\\
\includegraphics[width=0.25\textwidth]{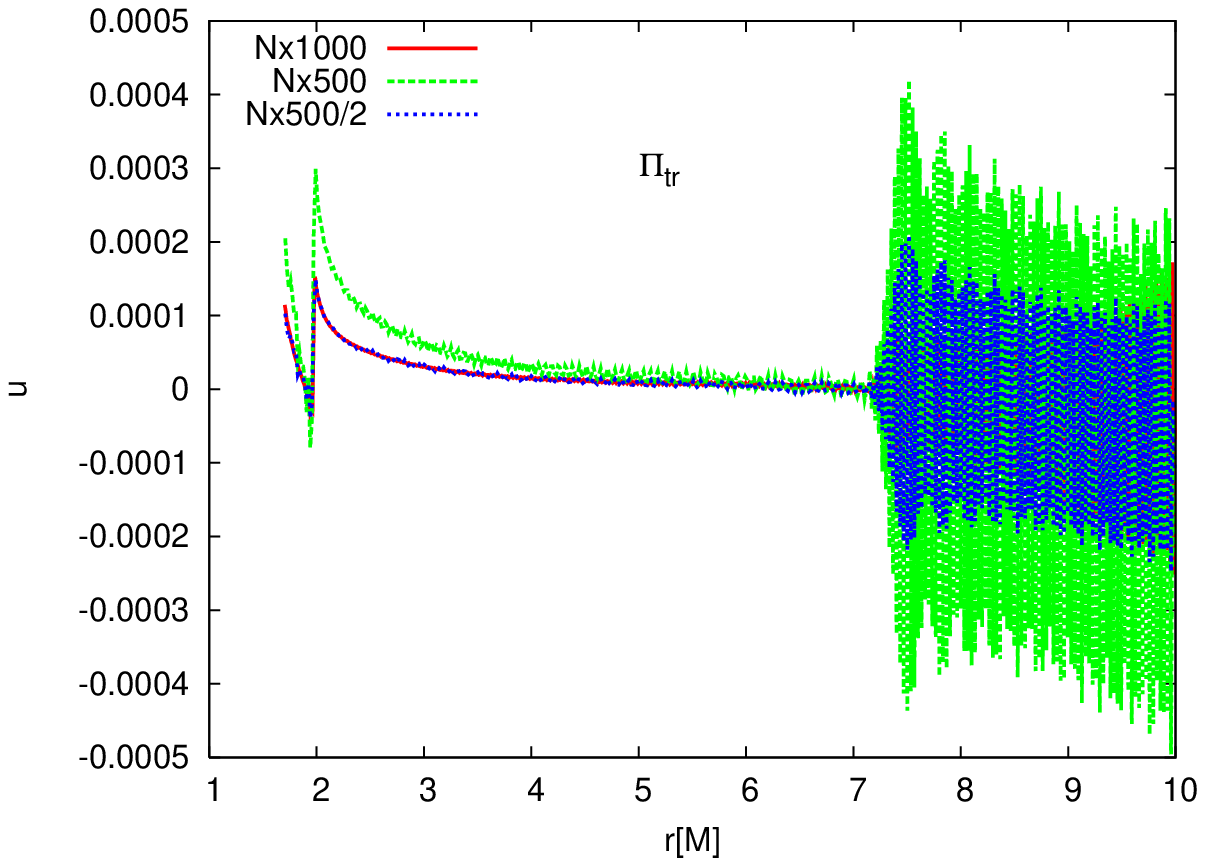}&
\includegraphics[width=0.25\textwidth]{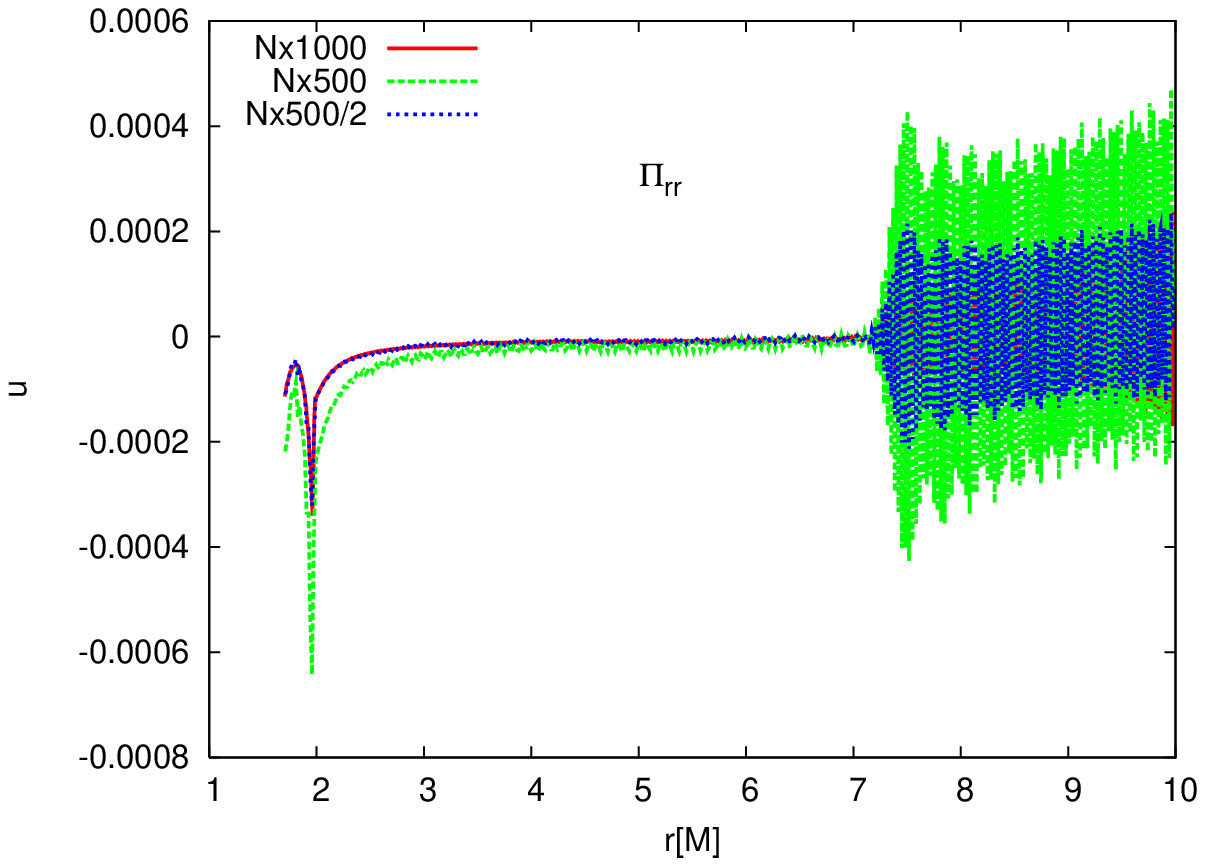}&
\includegraphics[width=0.25\textwidth]{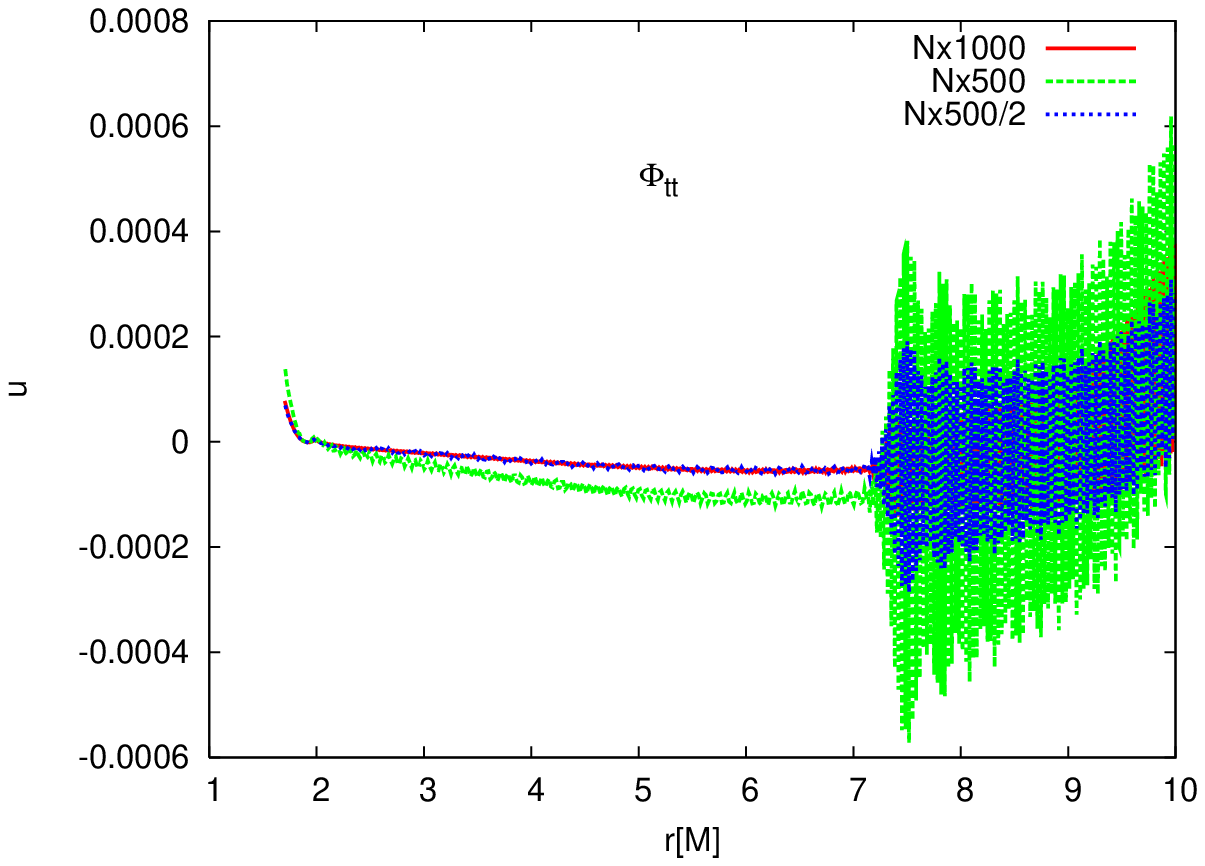}&
\includegraphics[width=0.25\textwidth]{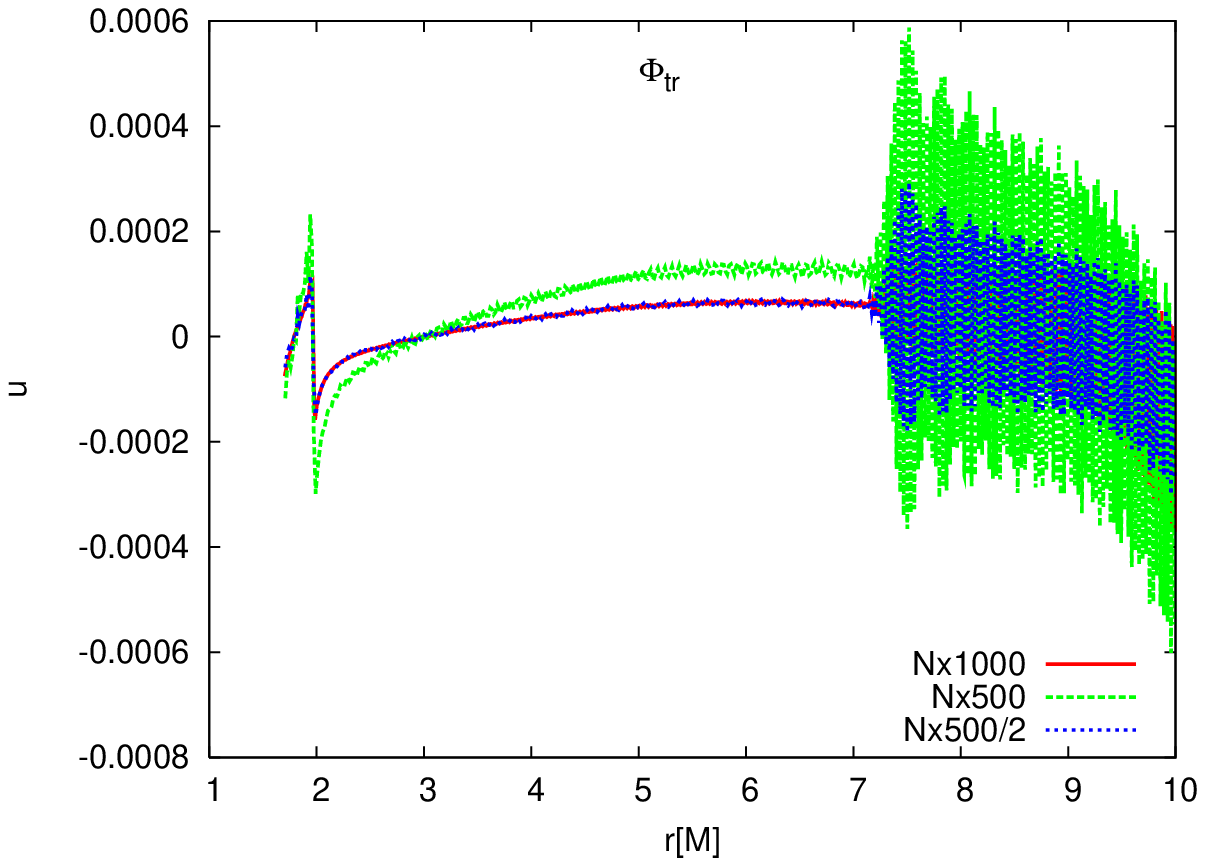}\\
\includegraphics[width=0.25\textwidth]{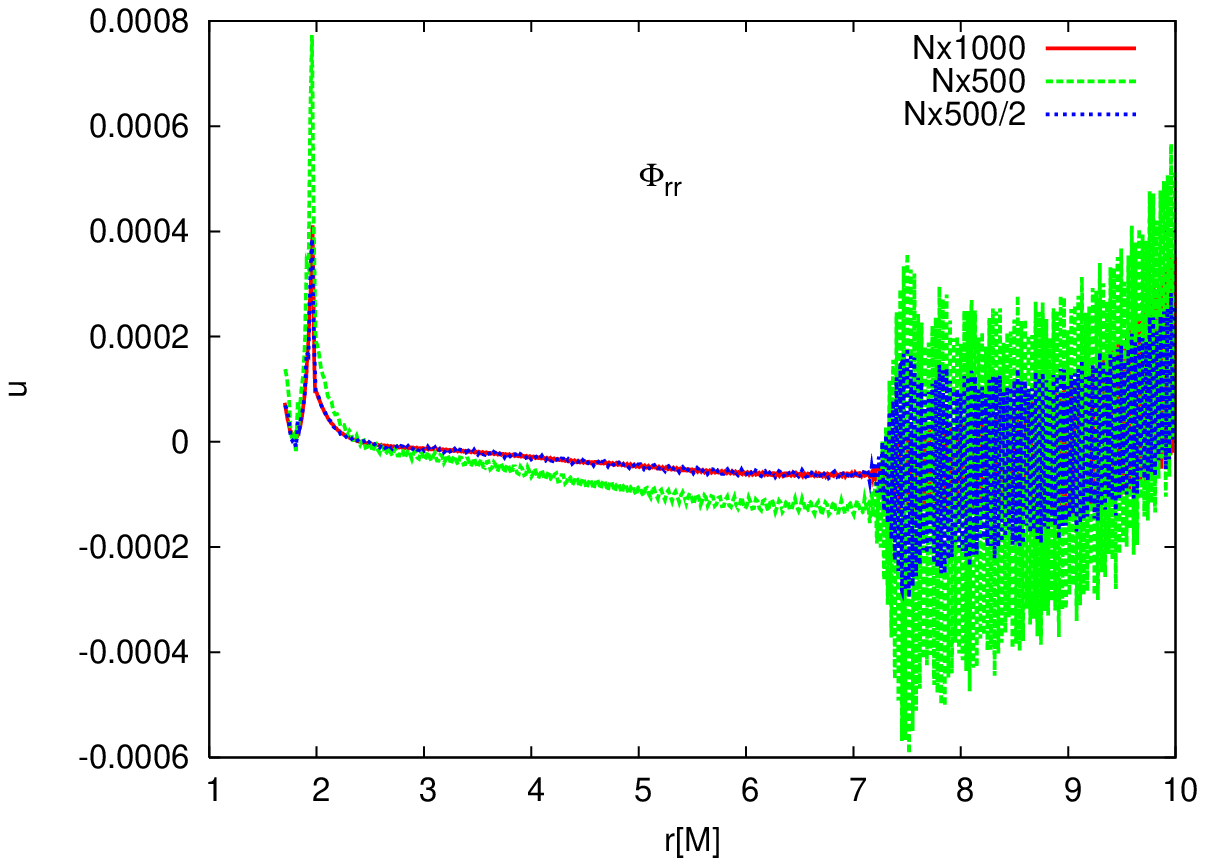}&
\includegraphics[width=0.25\textwidth]{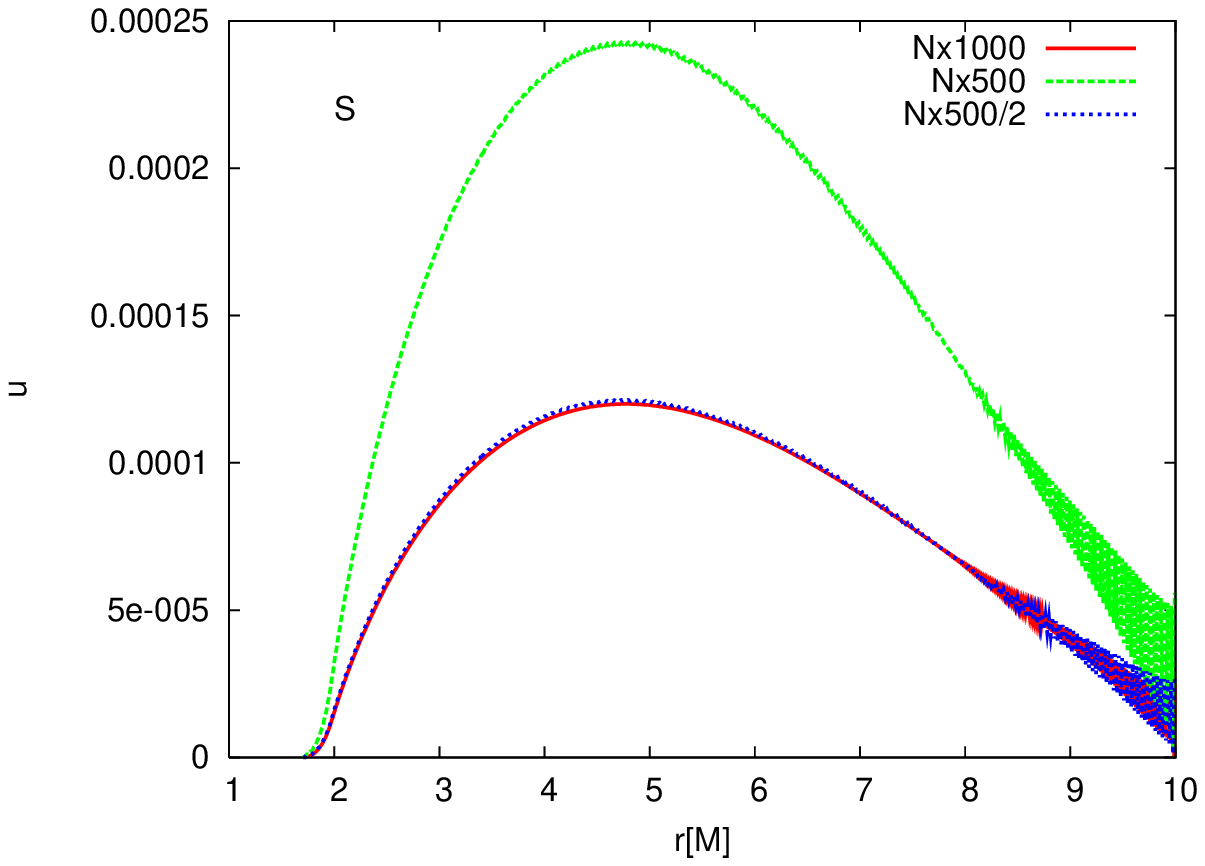}&
\includegraphics[width=0.25\textwidth]{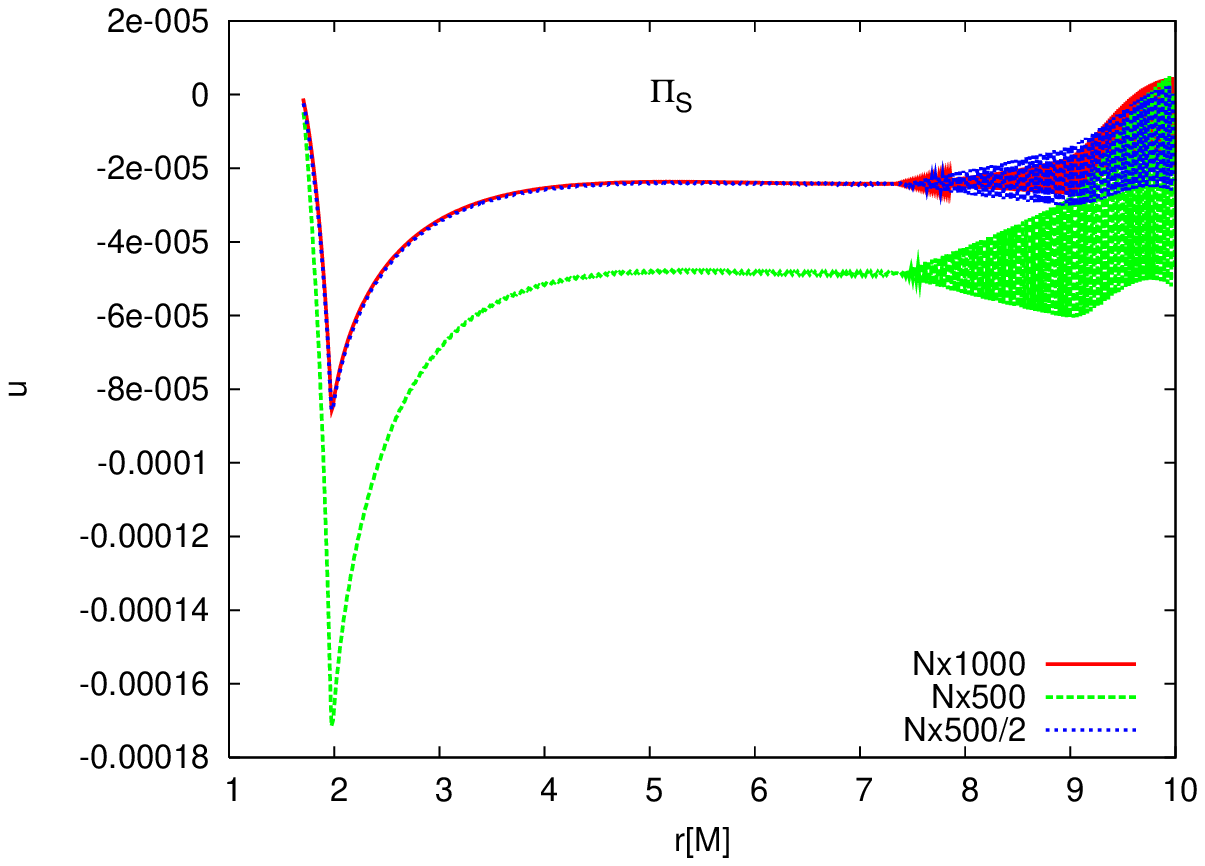}&
\includegraphics[width=0.25\textwidth]{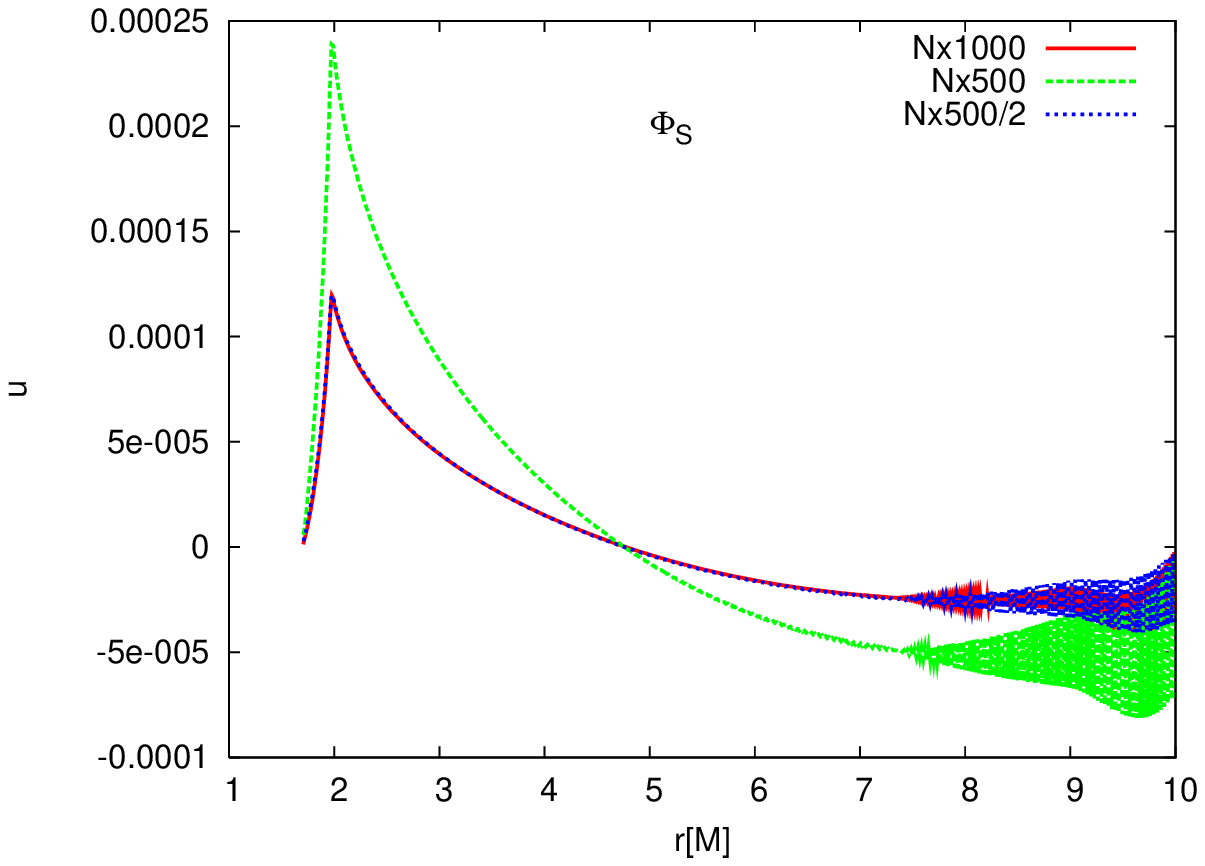}
\end{tabular}
\caption{Finite difference method comparison to the results shown in the Fig.~\ref{fig5}. The numerical error for different variables at $t=9.96M$. Results for $N_x=1000$ and $N_x=500$ are compared. Fourth order finite difference stencil is used. First order convergence behavior which corresponds to the factor 2 is apparent. Kerr-Schild coordinate is used.}\label{fig7}
\end{figure*}
\begin{figure}
\begin{tabular}{cc}
\includegraphics[width=0.25\textwidth]{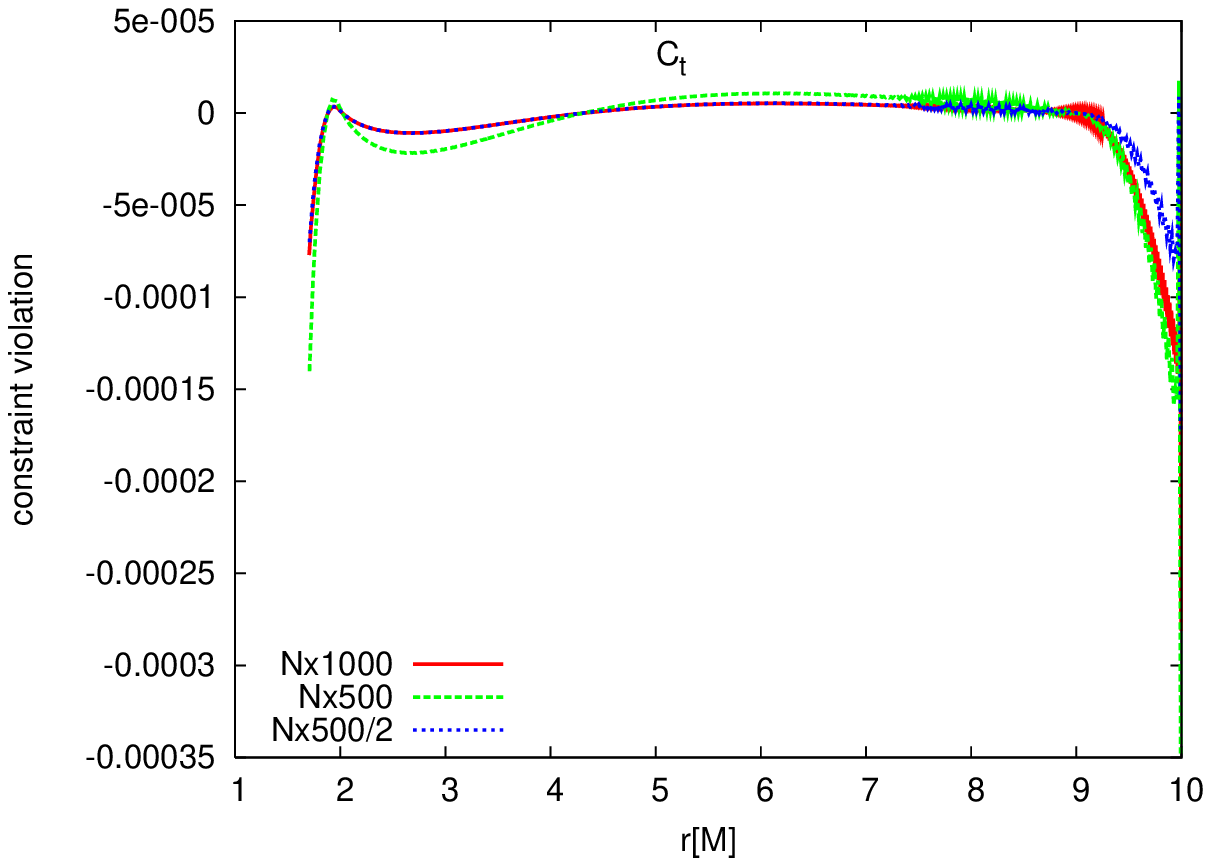}&
\includegraphics[width=0.25\textwidth]{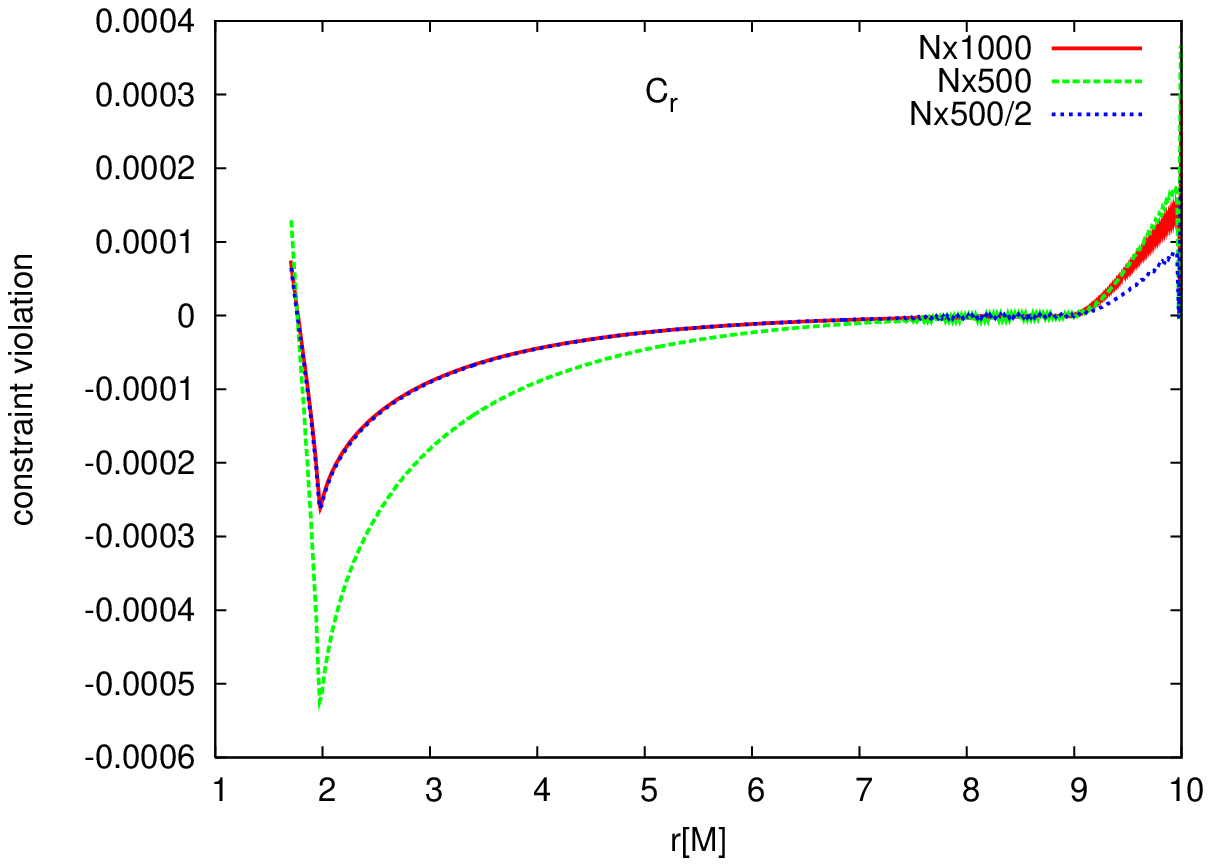}
\end{tabular}
\caption{The constraint violation $C_{t,r}$ corresponding to the Fig.~\ref{fig7} for finite difference evolution. Kerr-Schild coordinate is used.}\label{fig8}
\end{figure}
\begin{figure}
\begin{tabular}{c}
\includegraphics[width=0.5\textwidth]{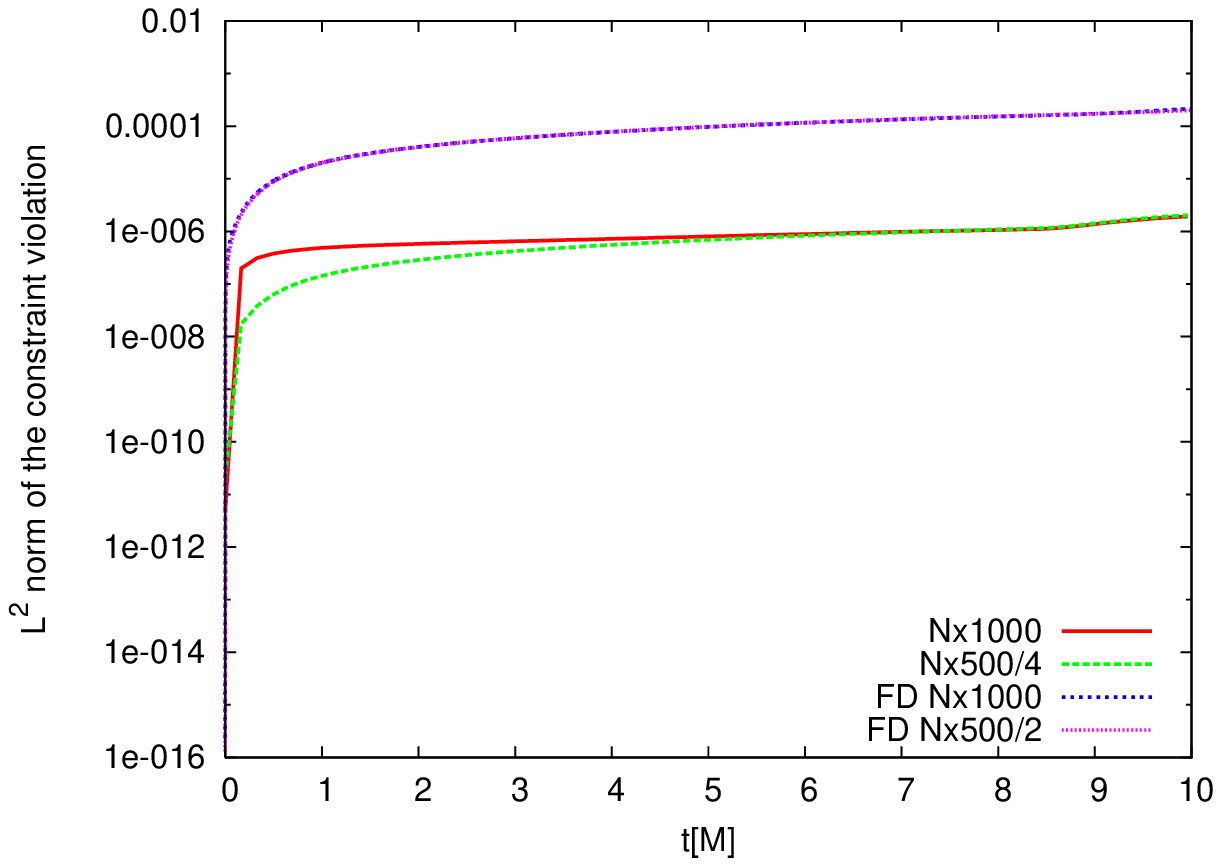}
\end{tabular}
\caption{$L^2$ norm of the constraint violation $\sqrt{\int (C^2_t+C_r^2)dx}$. Kerr-Schild coordinate is used. FD means `Finite Difference'.}\label{fig9}
\end{figure}

For comparison usage we have also used finite difference method to solve the dynamical system (\ref{problem_eq}). We use the fourth order finite difference stencil as we done in \cite{PhysRevD.78.124011}. The evolution results are plotted in the Fig.~\ref{fig7} which corresponds to the Fig.~\ref{fig5}. Firstly we can see first order convergence has been got instead of fourth order as the finite difference order. We suspect this is due to the unsmooth property of the numerical solution. We have seen that both finite difference method and the finite element method show first order convergence. One more point needed to be pointed out is that there are some high frequency numerical error near the out boundary. We suspect this is due to the Dirichlet boundary condition which is not perfectly consistent to the numerical solution. But interestingly, this phenomena does not appear in the finite element evolution case shown in the Fig.~\ref{fig5}. Instead a kink wave is generated from the boundary. Corresponding to the Fig.~\ref{fig6} we plot the constraint violation of the finite difference evolution in the Fig.~\ref{fig8}. Compared to the Fig.~\ref{fig7}, we can see the finite element method not only gives roughly one order smaller constraint violation, but also the convergence order is higher than finite difference. The convergence order of constraint violation for finite element method is 2, while the finite difference method is 1. We compare the constraint violation convergence behavior in the Fig.~\ref{fig9}. Here the $L^2$ norm is used. Once again we see that the convergence order of constraint violation for finite element method is 2, while the finite difference method is 1. This is consistent to the results of Fig.~\ref{fig6} and Fig.~\ref{fig8}.

\begin{figure}
\begin{tabular}{c}
\includegraphics[width=0.5\textwidth]{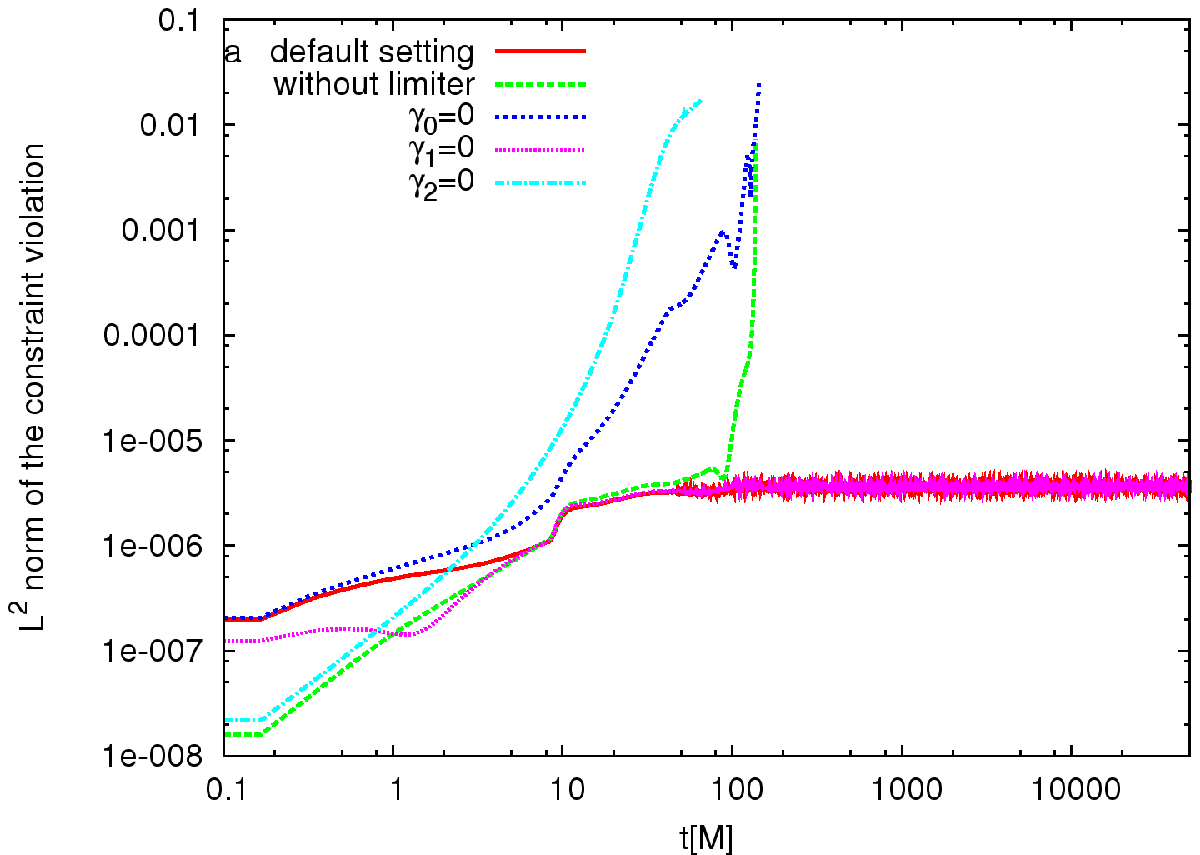}\\
\includegraphics[width=0.5\textwidth]{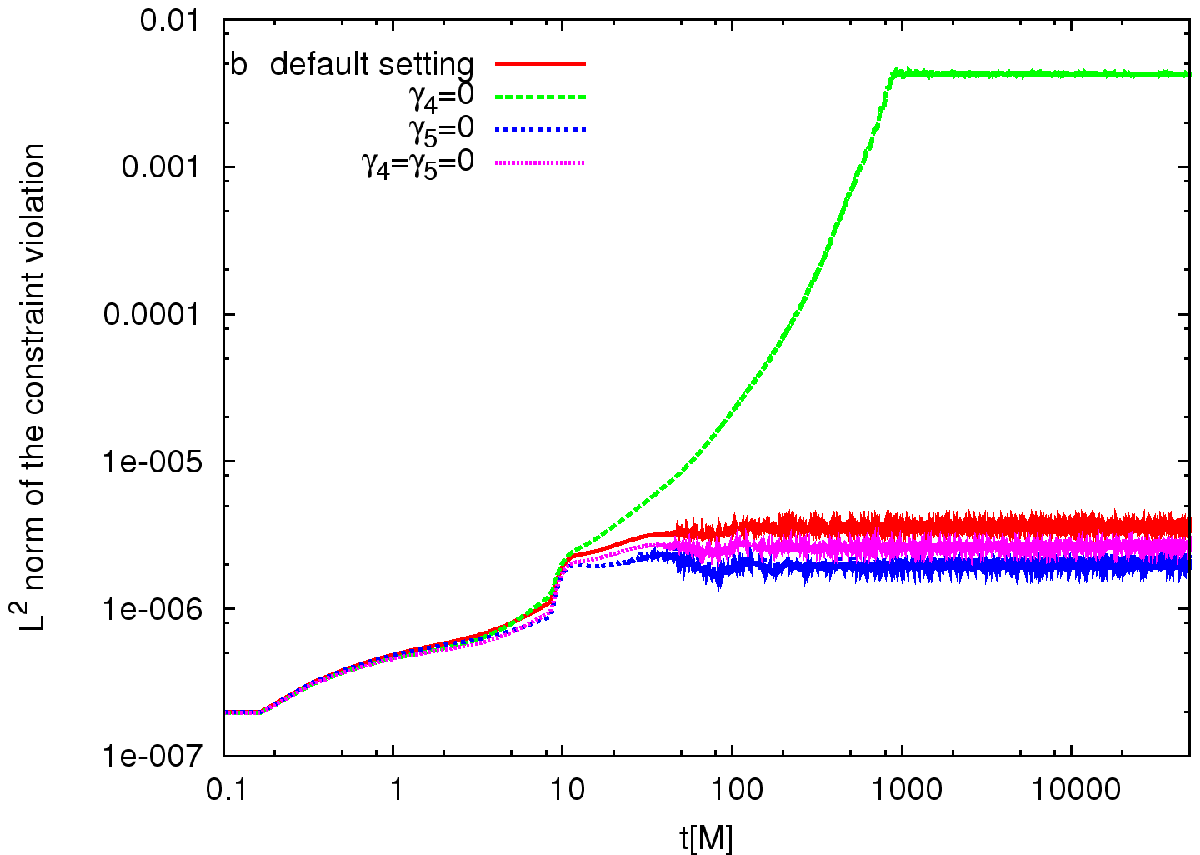}
\end{tabular}
\caption{Long term evolution behavior for the local discontinuous Garlerkin finite element method. $L^2$ norm of the constraint violation $\sqrt{\int (C^2_t+C_r^2)dx}$ is plotted. Kerr-Schild coordinate is used here. The default setting is $\gamma_0=-\gamma_1=\gamma_2=1$, $\gamma_4=\gamma_5=\frac{1}{2}$ together with limiter which is marked with ``default setting". Other markers indicate the specific setting respect to the default one.}\label{fig10}
\end{figure}

\begin{figure*}
\begin{tabular}{cccc}
\includegraphics[width=0.25\textwidth]{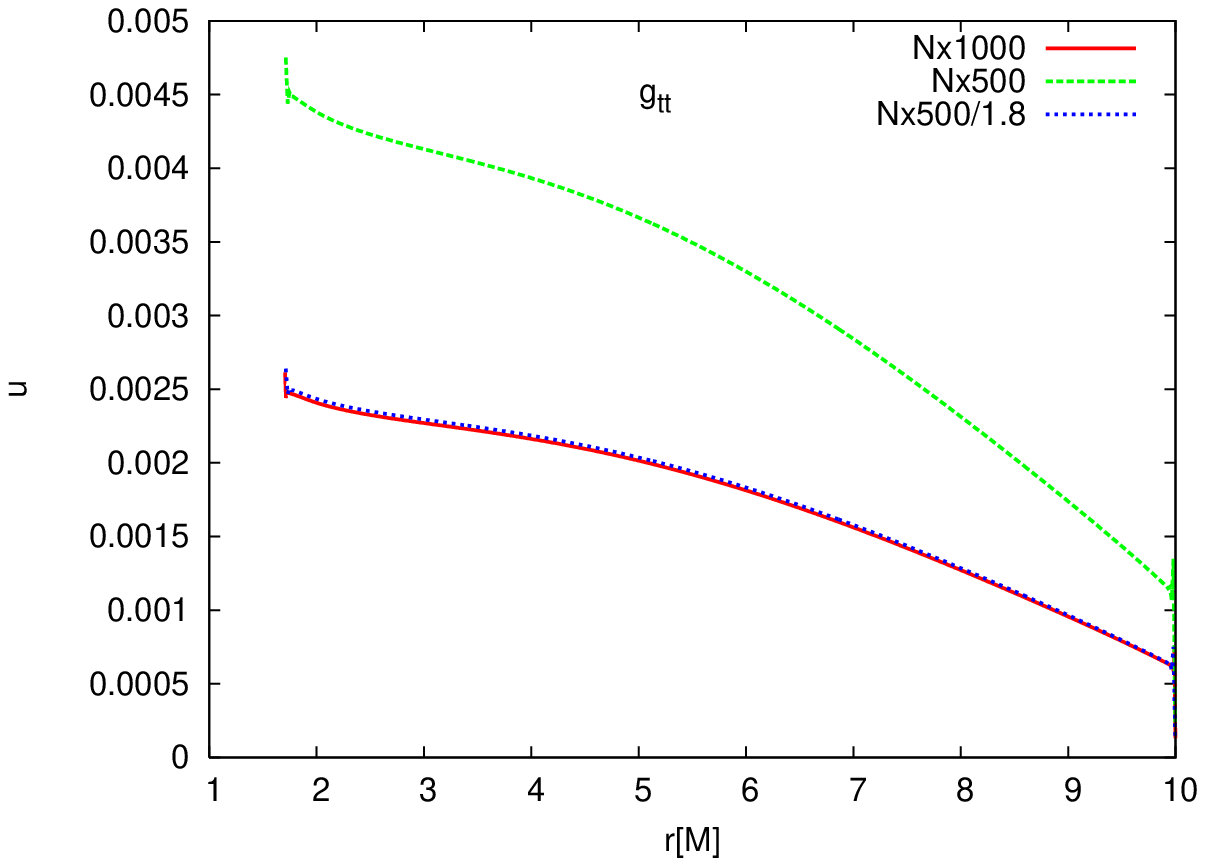}&
\includegraphics[width=0.25\textwidth]{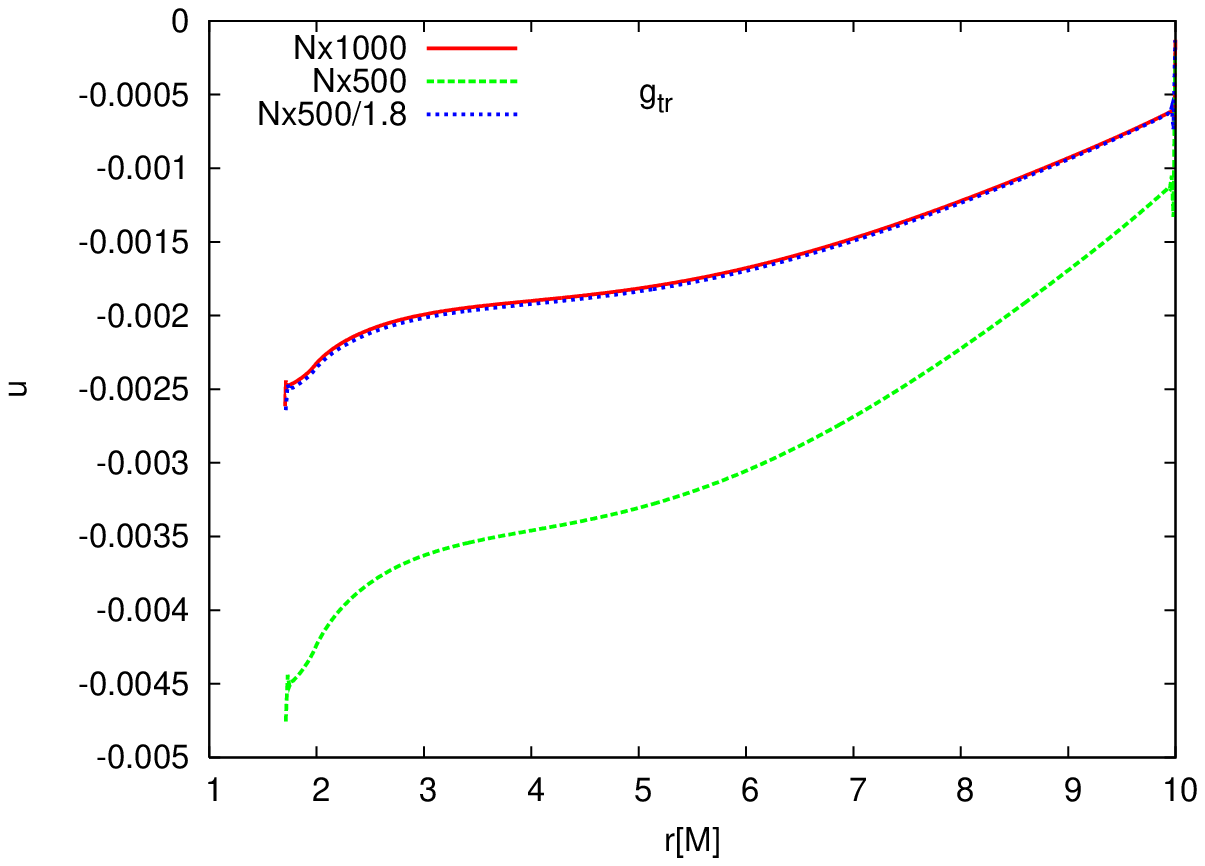}&
\includegraphics[width=0.25\textwidth]{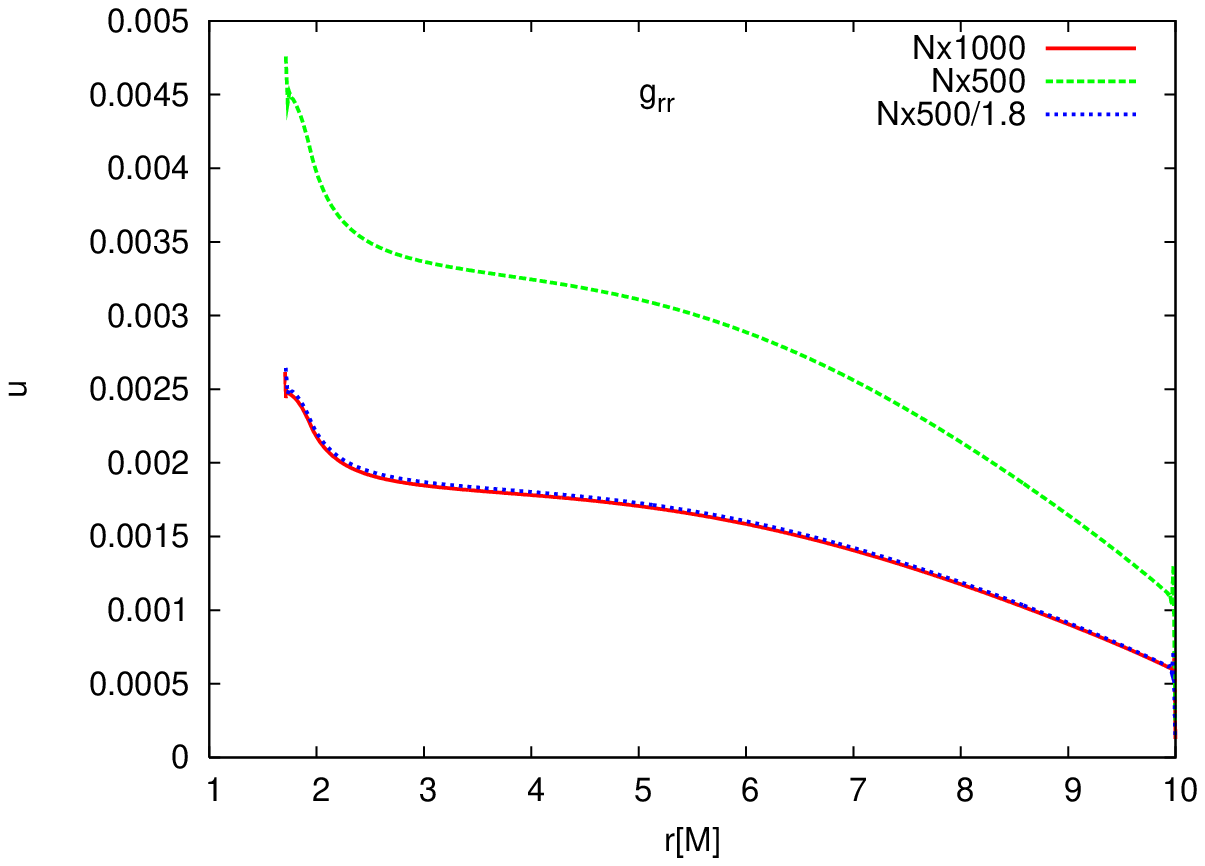}&
\includegraphics[width=0.25\textwidth]{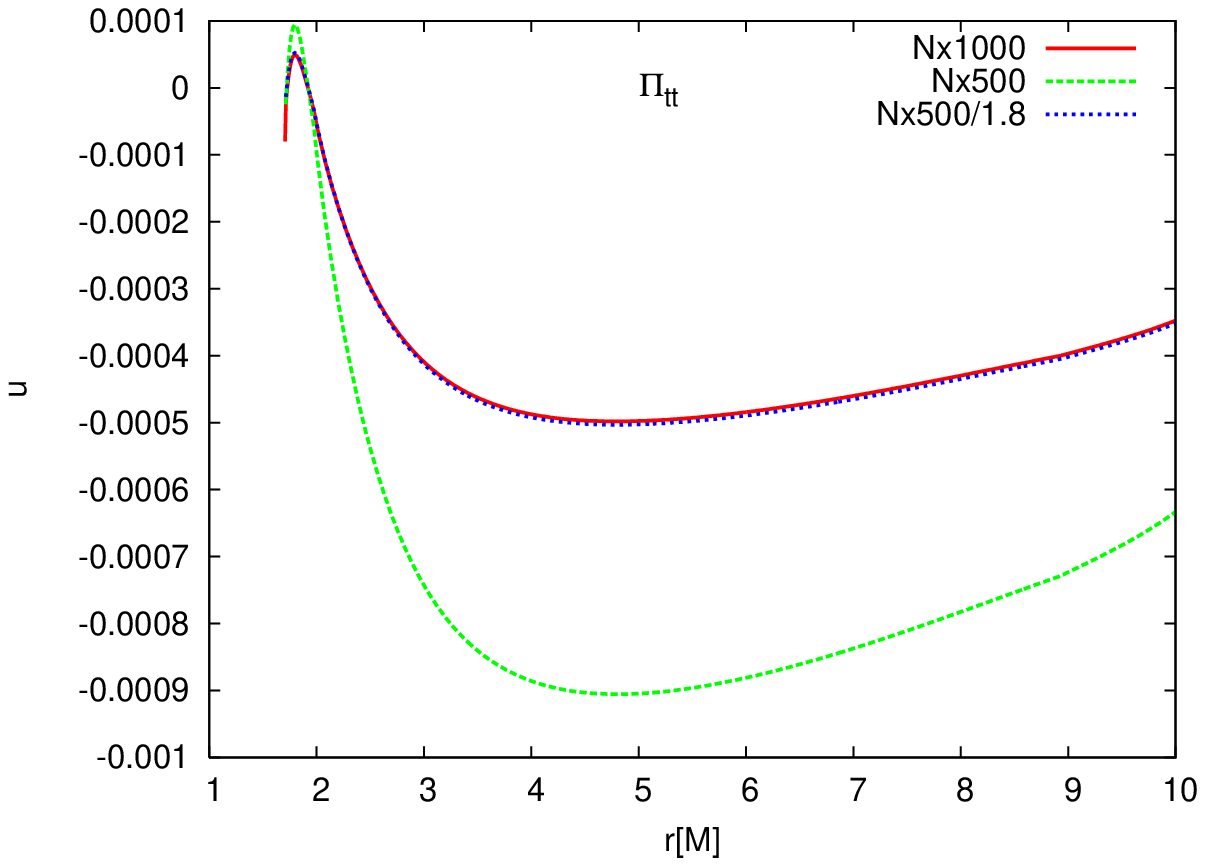}\\
\includegraphics[width=0.25\textwidth]{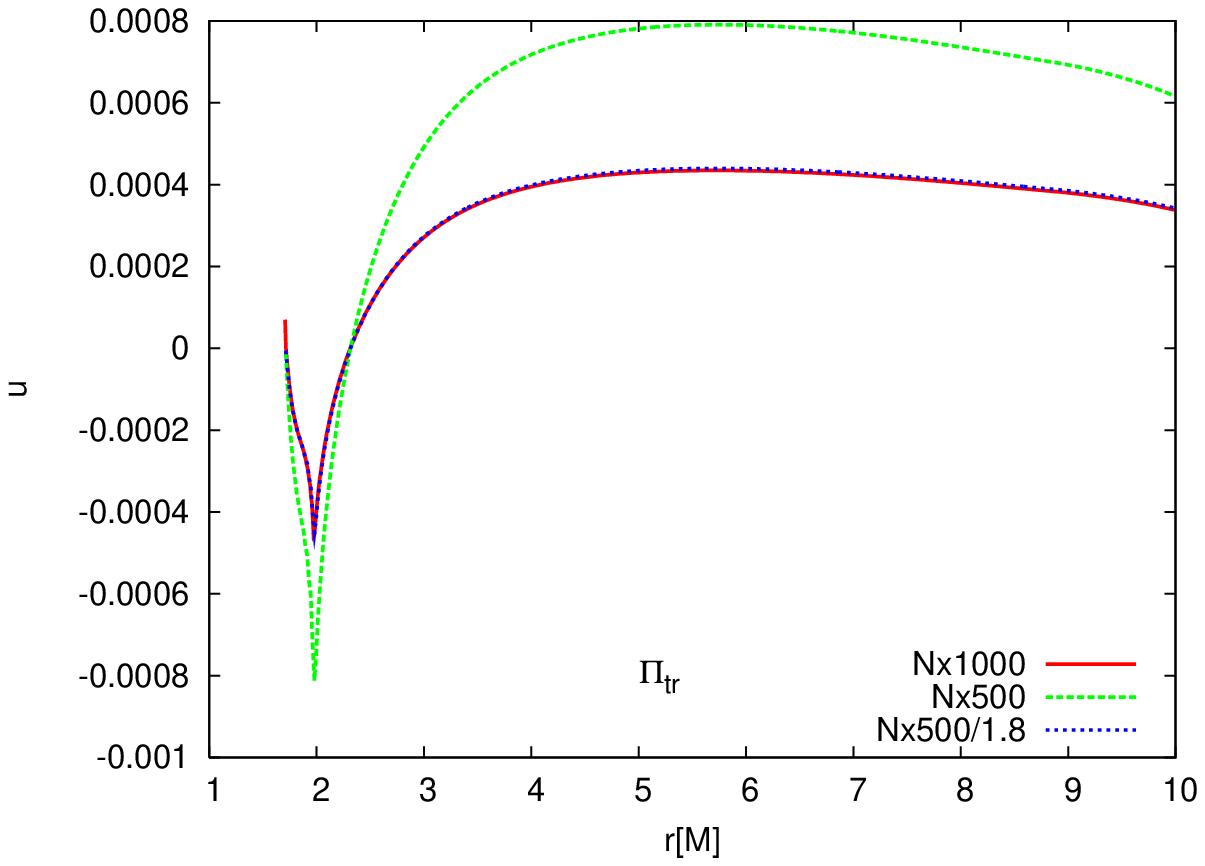}&
\includegraphics[width=0.25\textwidth]{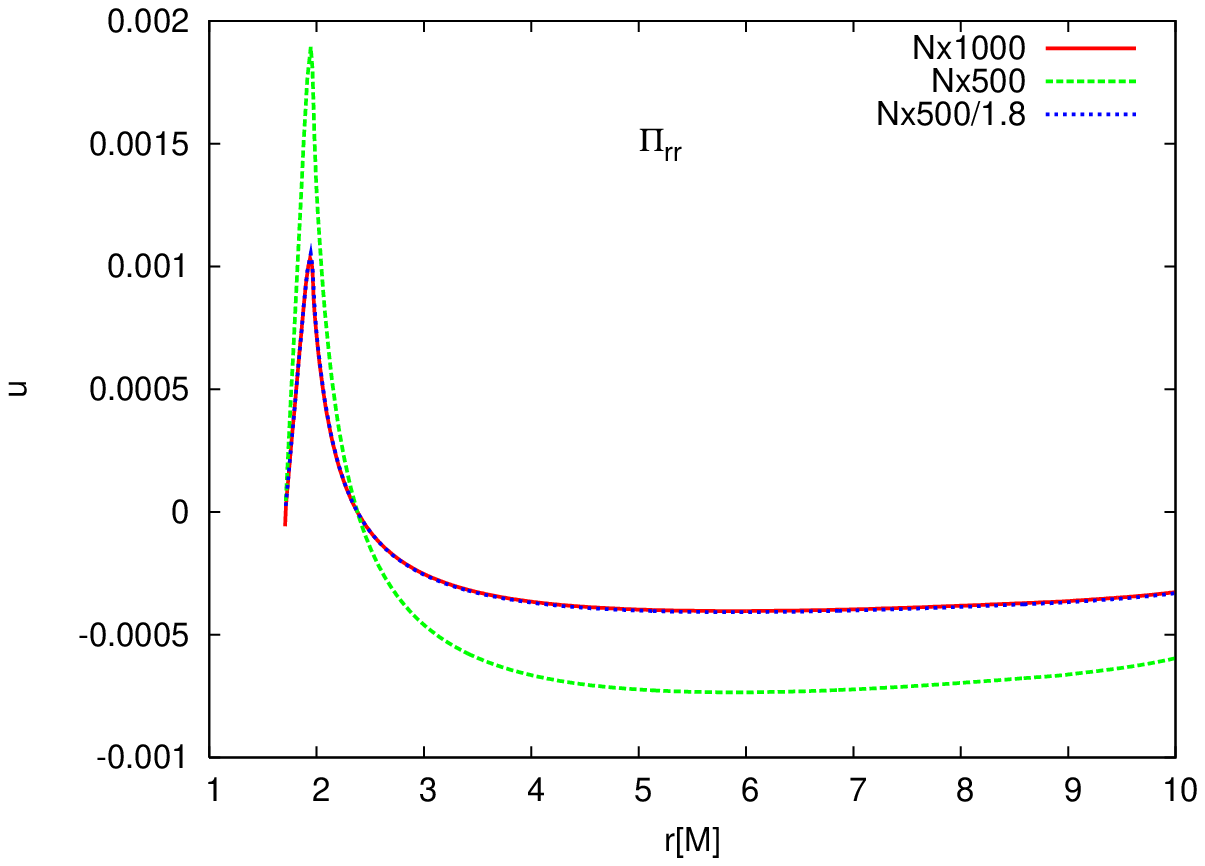}&
\includegraphics[width=0.25\textwidth]{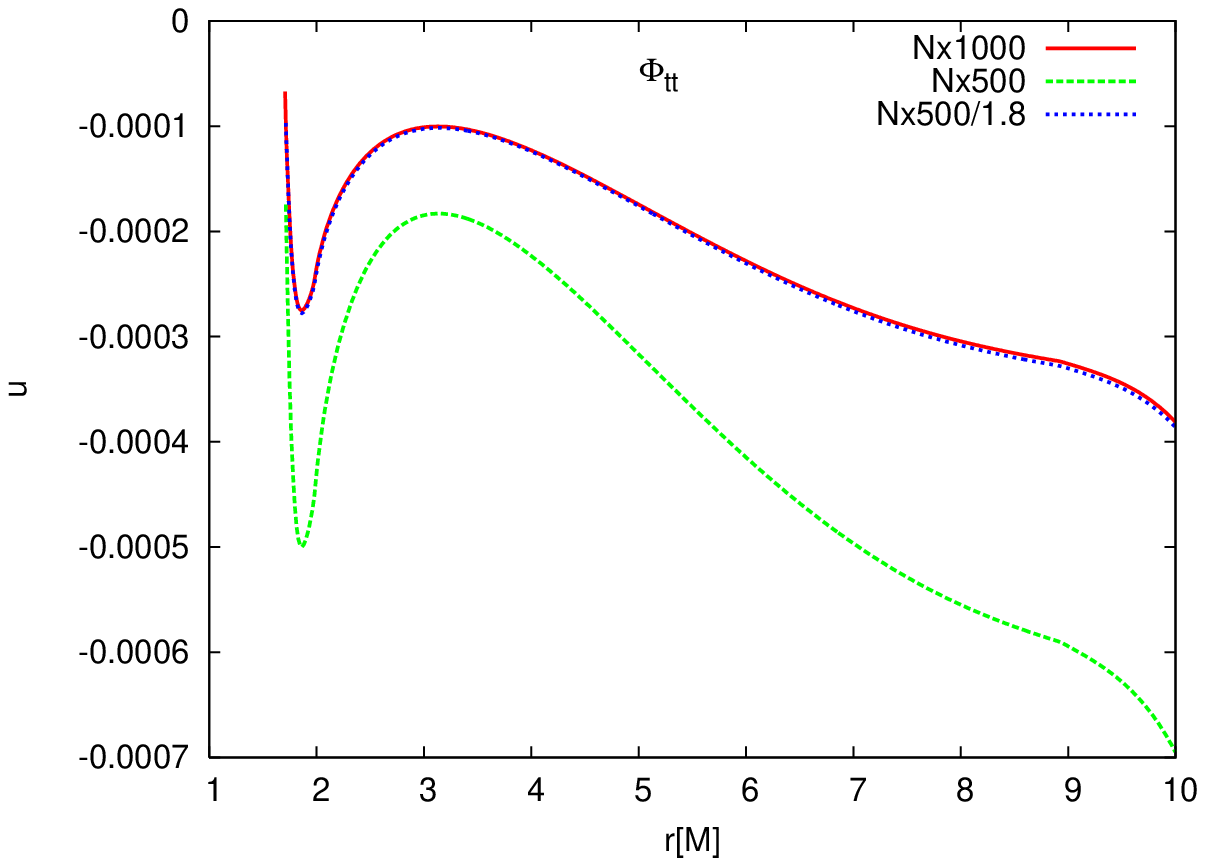}&
\includegraphics[width=0.25\textwidth]{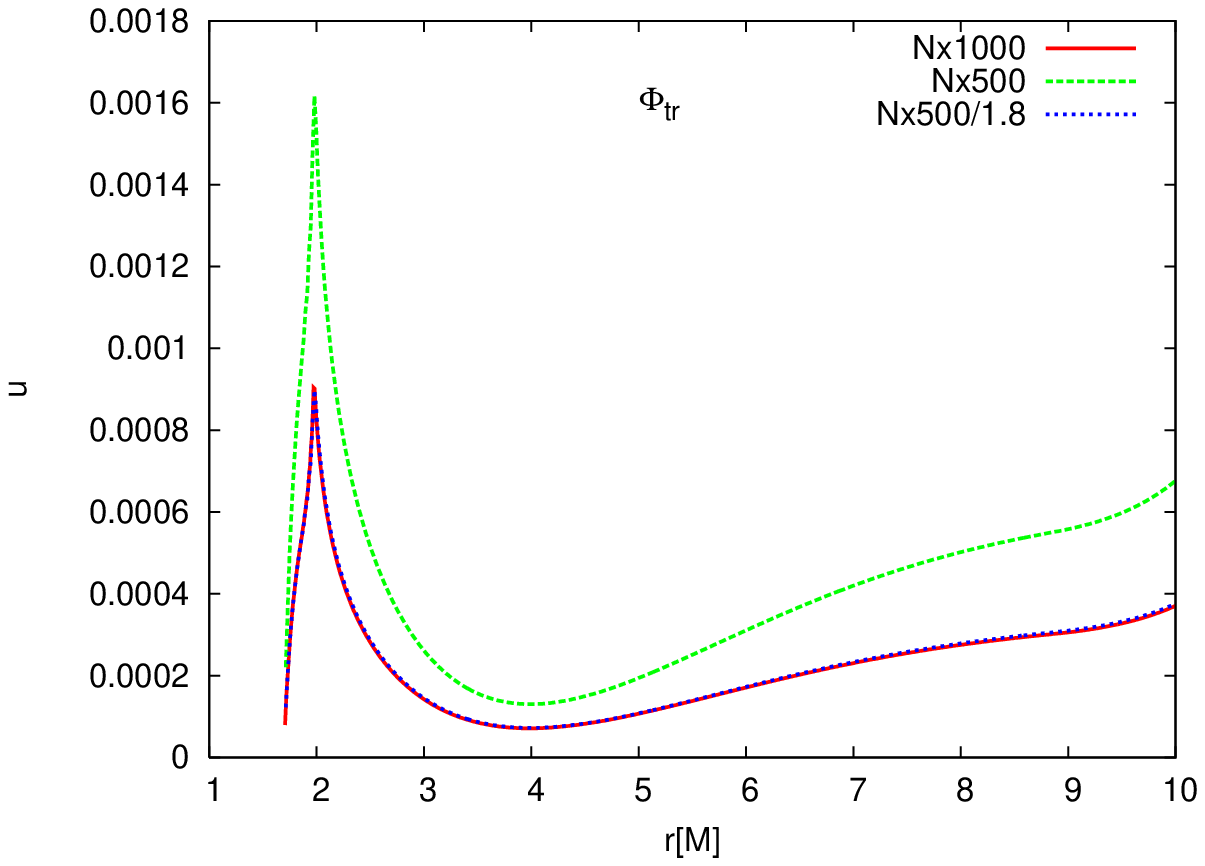}\\
\includegraphics[width=0.25\textwidth]{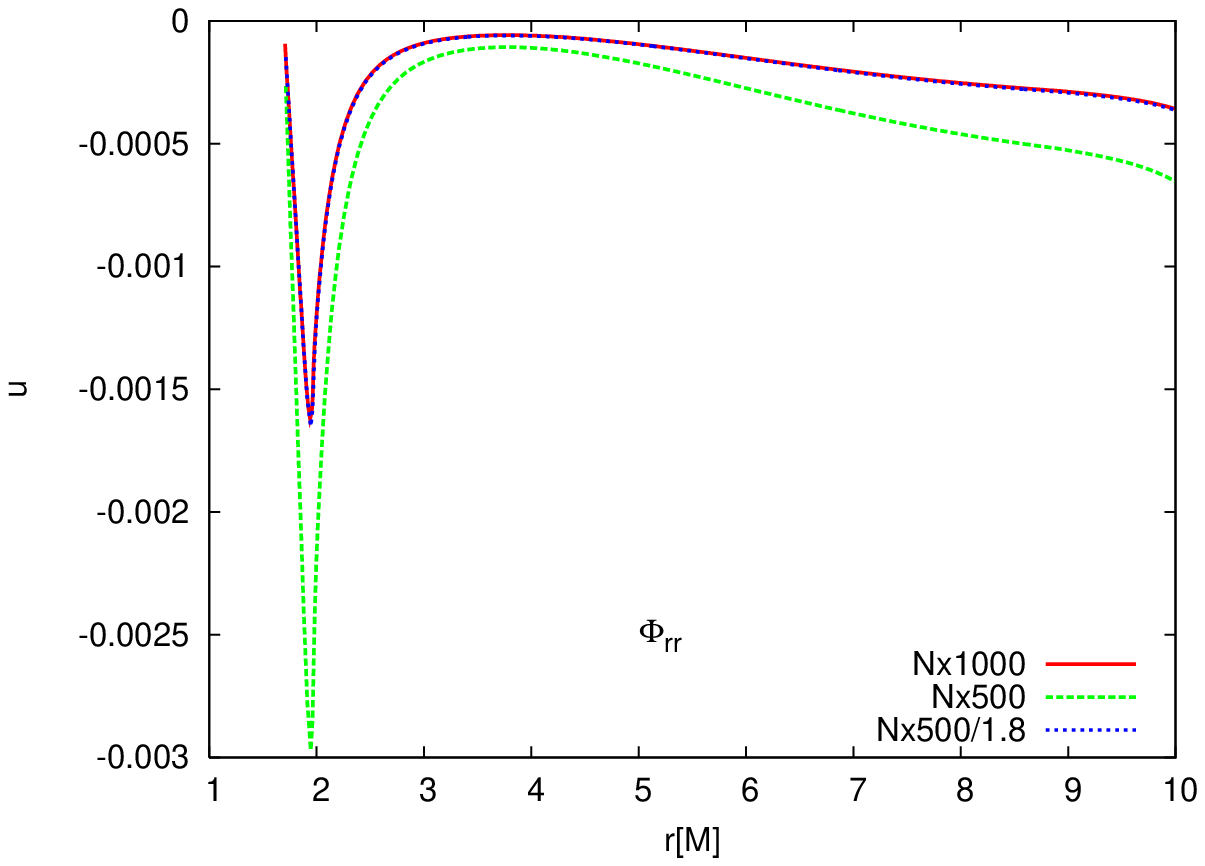}&
\includegraphics[width=0.25\textwidth]{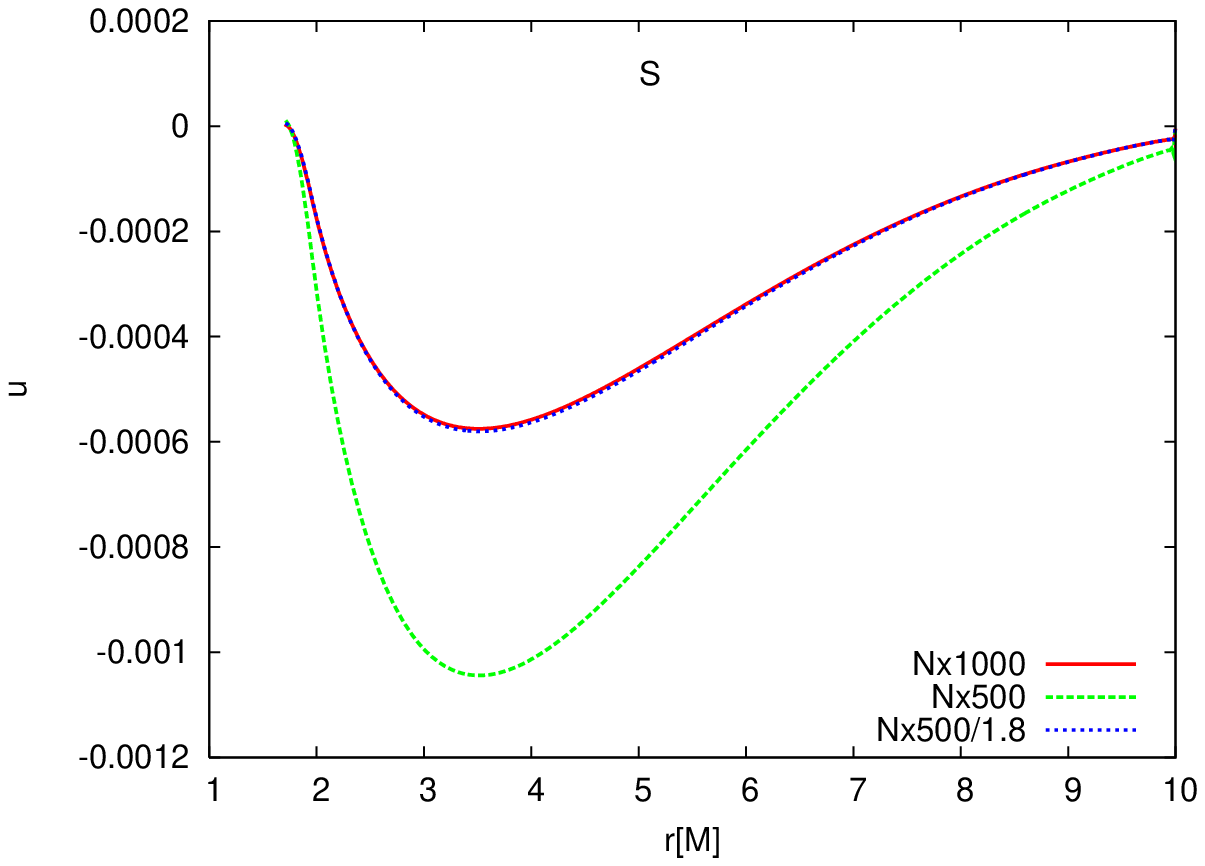}&
\includegraphics[width=0.25\textwidth]{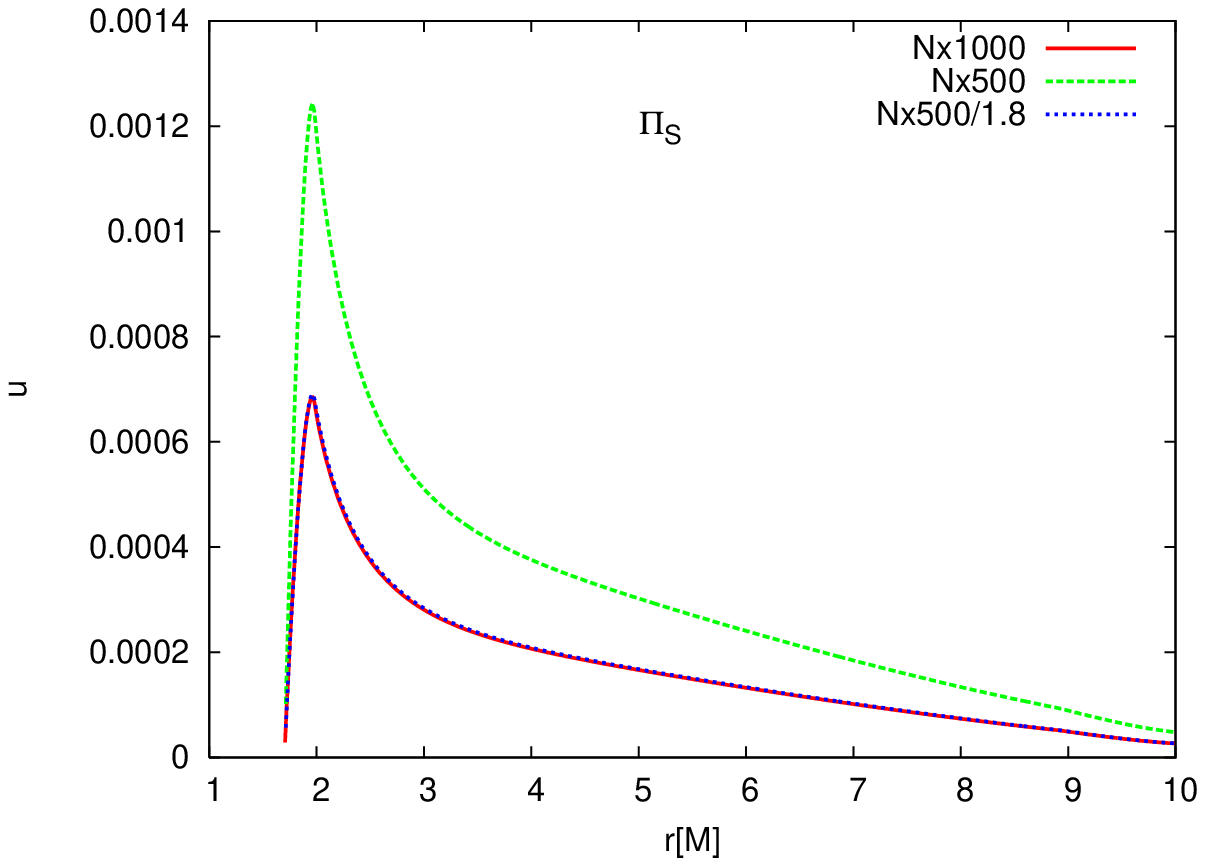}&
\includegraphics[width=0.25\textwidth]{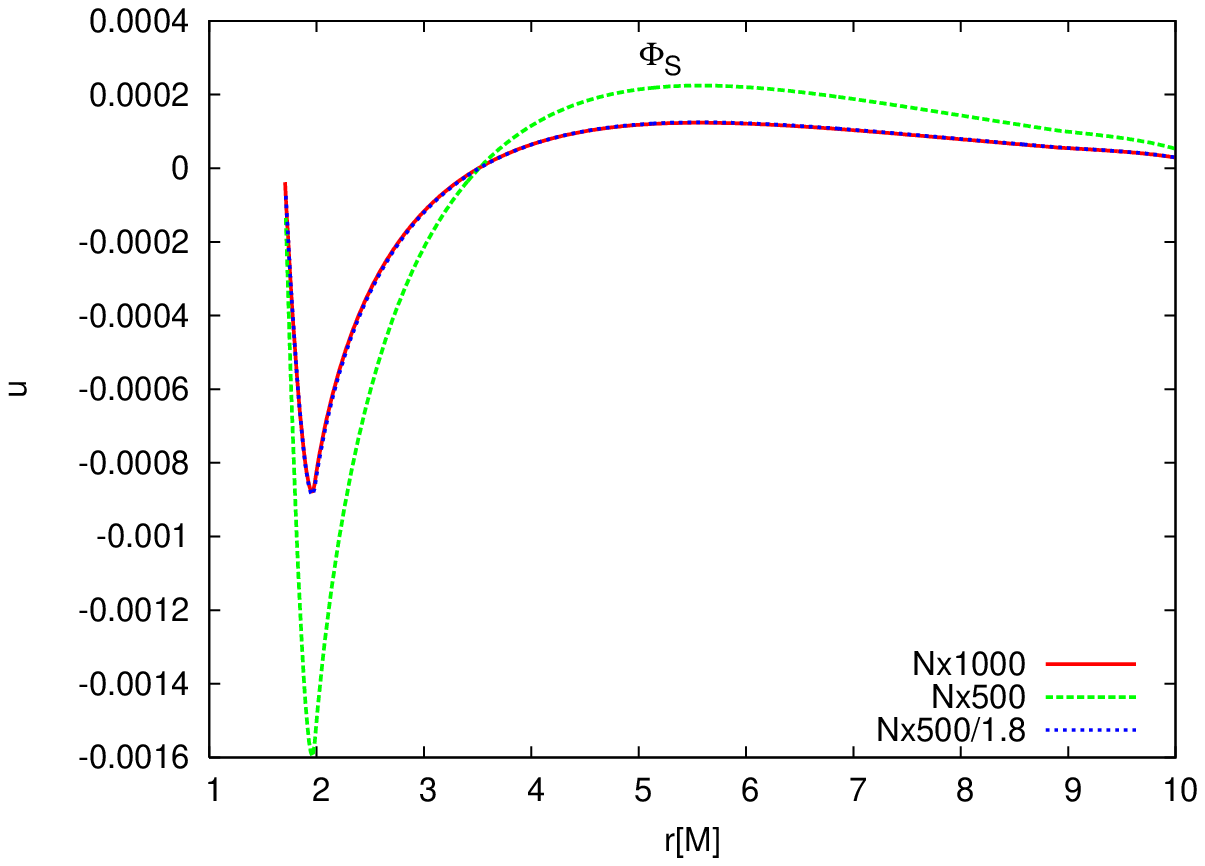}
\end{tabular}
\caption{As the Fig.~\ref{fig5} but for setting $\gamma_1=-1$ and $\gamma_2=0$. First order convergence and smooth variable configurations can be seen. Kerr-Schild coordinate is used.}\label{fig11}
\end{figure*}

For the finite difference method, we have used Kreiss-Oliger dissipation technique to weaken the high frequency noise. Otherwise such noise shown in the right region of the Fig.~\ref{fig7} will grow and make the evolution unstable. Similarly the limiter for the finite element method is also essential. Without the limiter high frequency noise will show up around the event horizon $r=2M$. In Fig.~\ref{fig10} we compare the long term evolution behavior for different setting. The $L^2$ norm of the constraint violation is shown in the plot. The setting $\gamma_0=-\gamma_1=\gamma_2=1$, $\gamma_4=\gamma_5=\frac{1}{2}$ together with limiter can make the evolution stable. And the constraint violation can be kept around $3\times10^{-6}$. The limiter is important for stability. Otherwise the code may crash at about $100M$. None zero setting of $\gamma_{0,2}$ is also important for stability. But the setting of $\gamma_{1,4,5}$ is less essential for stability.

As analyzed in the Appendix~\ref{App::I}, the setting $\gamma_1=-1$ and $\gamma_2=0$ will make our dynamical system (\ref{problem_eq}) linearly degenerate. For a linearly degenerated system, shocks will not form. Our numerical tests do indicate this behavior. Corresponding to the Fig.~\ref{fig5}, we show the results for setting $\gamma_1=-1$ and $\gamma_2=0$ in the Fig.~\ref{fig11}. Comparing to Fig.~\ref{fig5} and Fig.~\ref{fig11} (especially $\Phi_{tr}$ near $r=2M$), we can see the resulted variable configures for setting $\gamma_1=-1$ and $\gamma_2=0$ are really smooth. But due to the constraint grows resulted from $\gamma_2=0$, this setting is not stable as shown in the subplot a of Fig.~\ref{fig10}. Simply speaking, linearly degenerate property is not necessary to numerical stability (c.f. the $\gamma_1=0$ case in the subplot a of Fig.~\ref{fig10}), in contrast constraint damping is important to numerical stability (c.f. the $\gamma_2=0$ case in the subplot a of Fig.~\ref{fig10}).

As mentioned above, the setting of $\gamma_{4,5}$ is less crucial for stability. In order to simplify the constraint evolution equation (c.f. the Equation (16) of \cite{PhysRevD.93.063006}), Hilditch and his coworkers set $\gamma_4=\gamma_5=\frac{1}{2}$. If we ignore the nonlinear terms and consider only the linear terms related to $\gamma_{4,5}$ for the Equation (16) of \cite{PhysRevD.93.063006}, we can get
\begin{align}
C_\mu(t)\sim e^{(1-2\gamma_4) t}C_\mu(0),e^{(\gamma_4-\gamma_5) t}C_\mu(0).
\end{align}
In order to let $\gamma_{4,5}$ damp the constraint, we need
\begin{align}
\gamma_4<\frac{1}{2},\gamma_5>\gamma_4.
\end{align}
$\gamma_5=1$, $\gamma_4=0$ belongs to this case, while other tested cases shown in the subplot b of Fig.~\ref{fig10} do not satisfy this criteria. But the numerical results indicate that the settings $\gamma_4=\gamma_5=\frac{1}{2}$, $\gamma_5=0$ and $\gamma_4=\gamma_5=0$ are several orders better than the setting $\gamma_5=1$, $\gamma_4=0$. In particular, the setting $\gamma_4=\gamma_5=0$ is even two times better than the default setting $\gamma_4=\gamma_5=\frac{1}{2}$ which is borrowed from \cite{PhysRevD.93.063006}. We suspect this is due to the nonlinear effect of the formulation equations. More detail analysis is needed to understand this interesting behavior. That is out of the scope of current work.
\subsection{Schwarzschild black hole in isotropic coordinate}
\begin{figure*}
\begin{tabular}{cccc}
\includegraphics[width=0.25\textwidth]{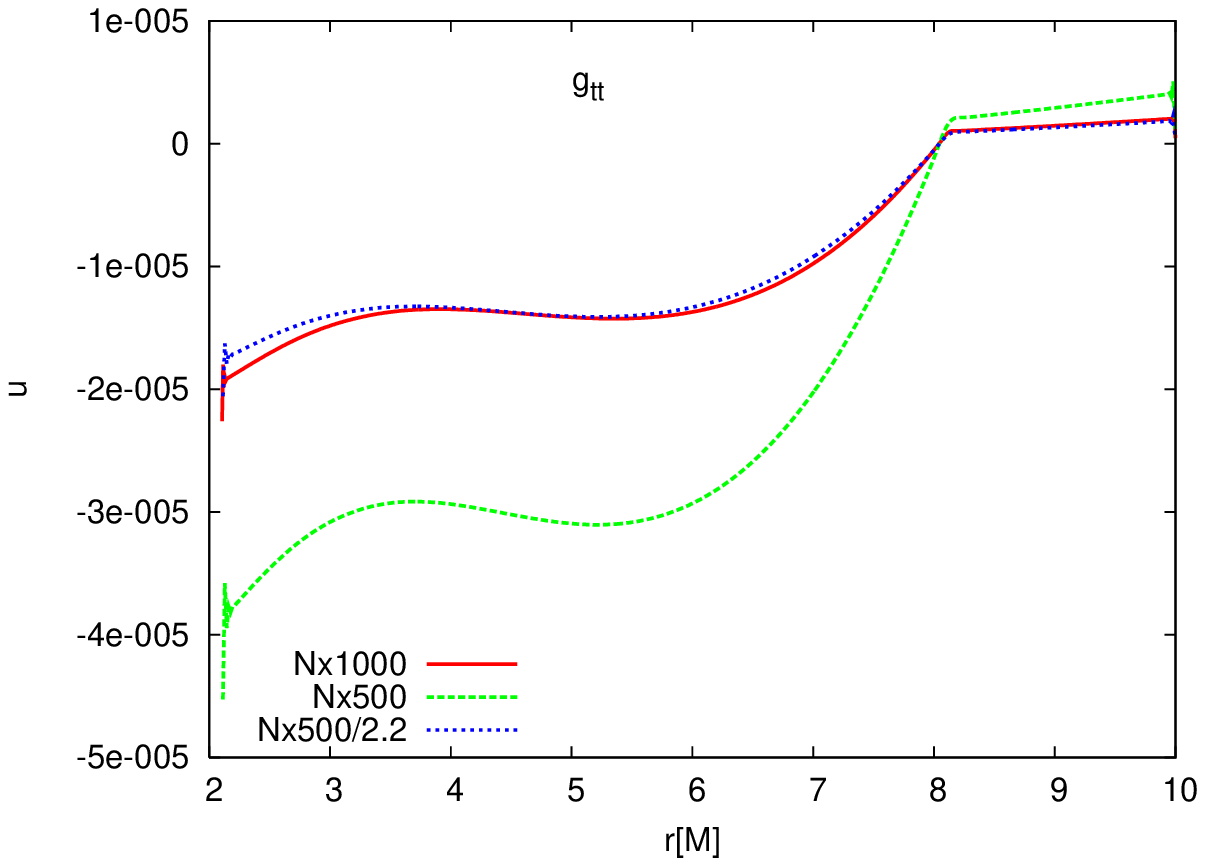}&
\includegraphics[width=0.25\textwidth]{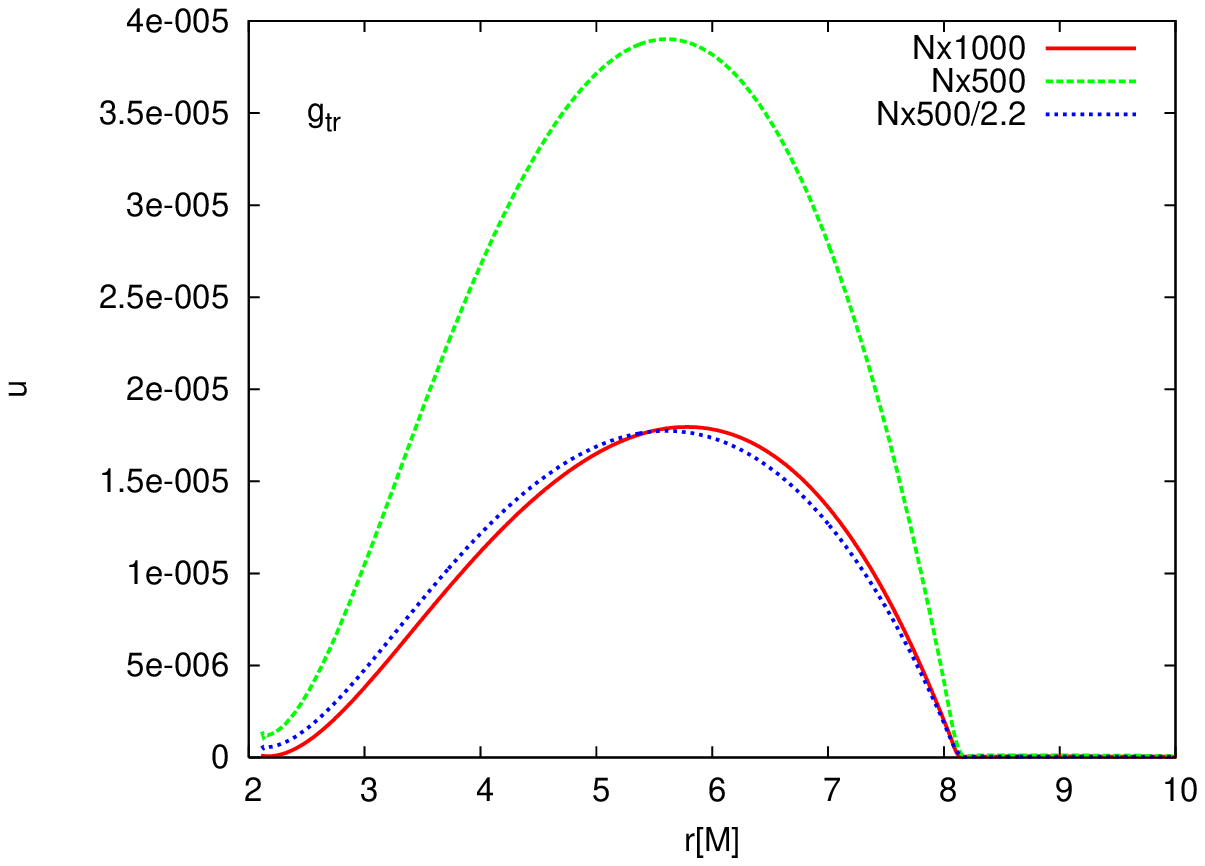}&
\includegraphics[width=0.25\textwidth]{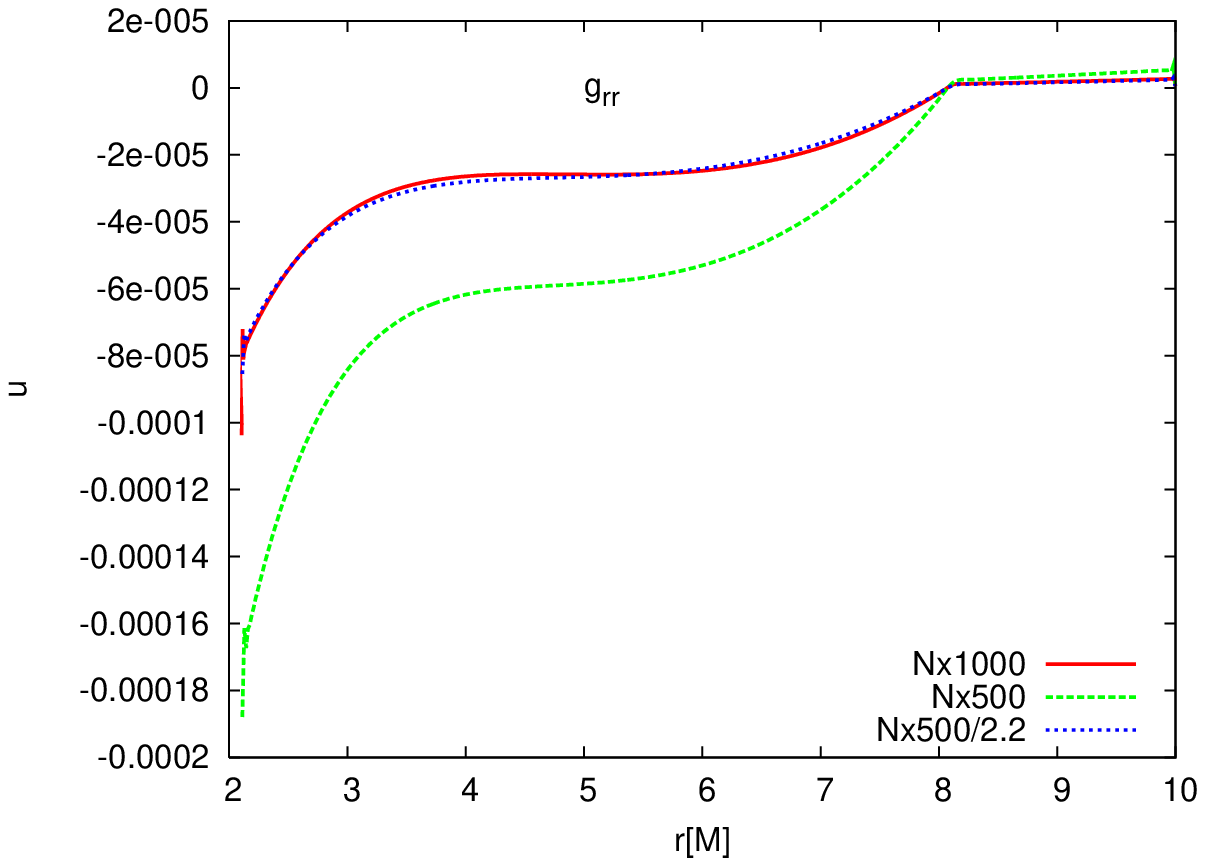}&
\includegraphics[width=0.25\textwidth]{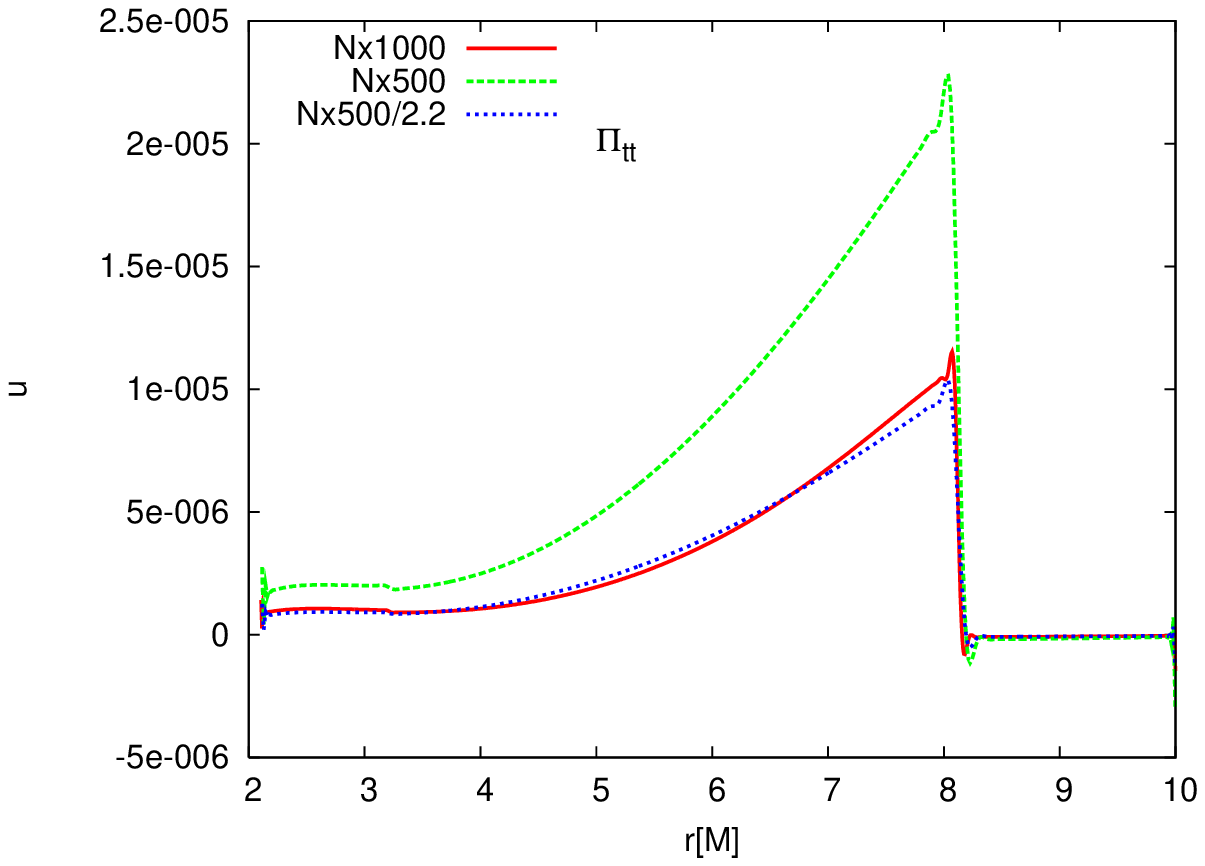}\\
\includegraphics[width=0.25\textwidth]{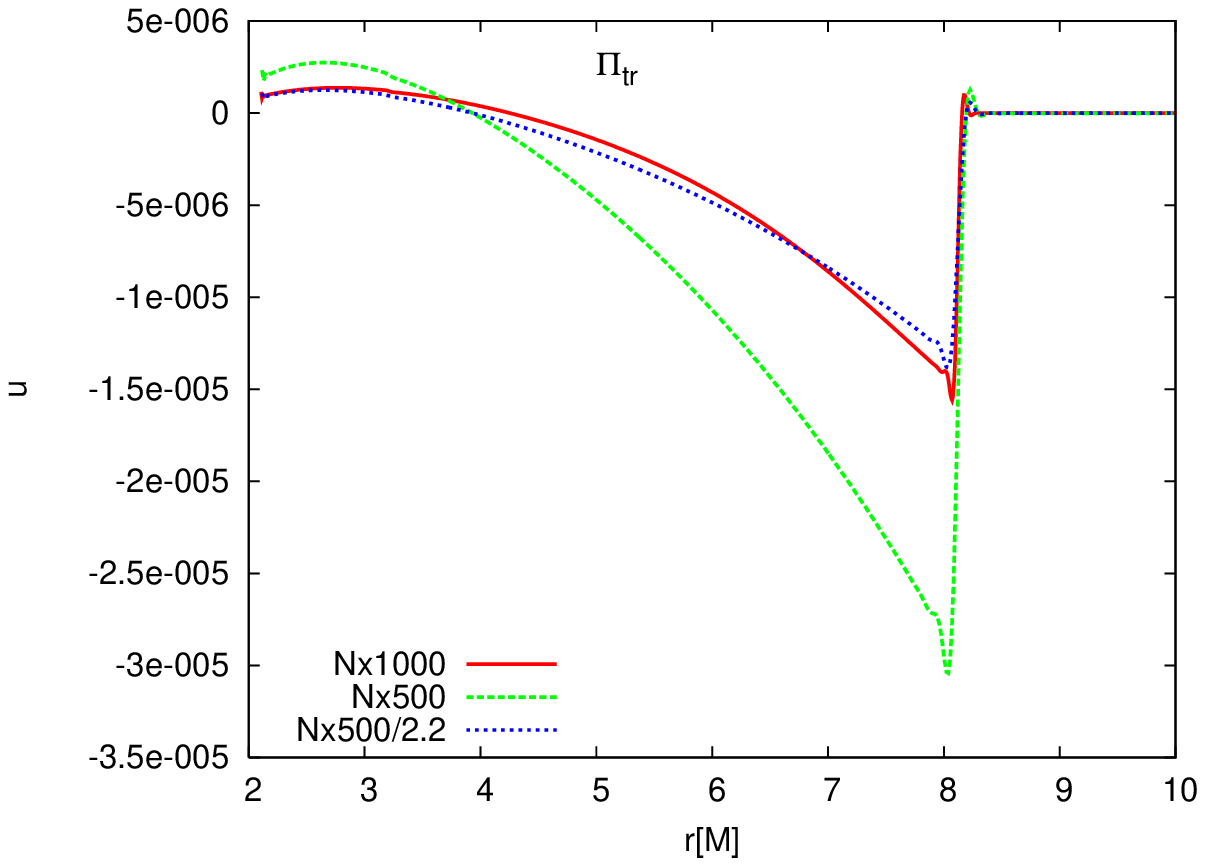}&
\includegraphics[width=0.25\textwidth]{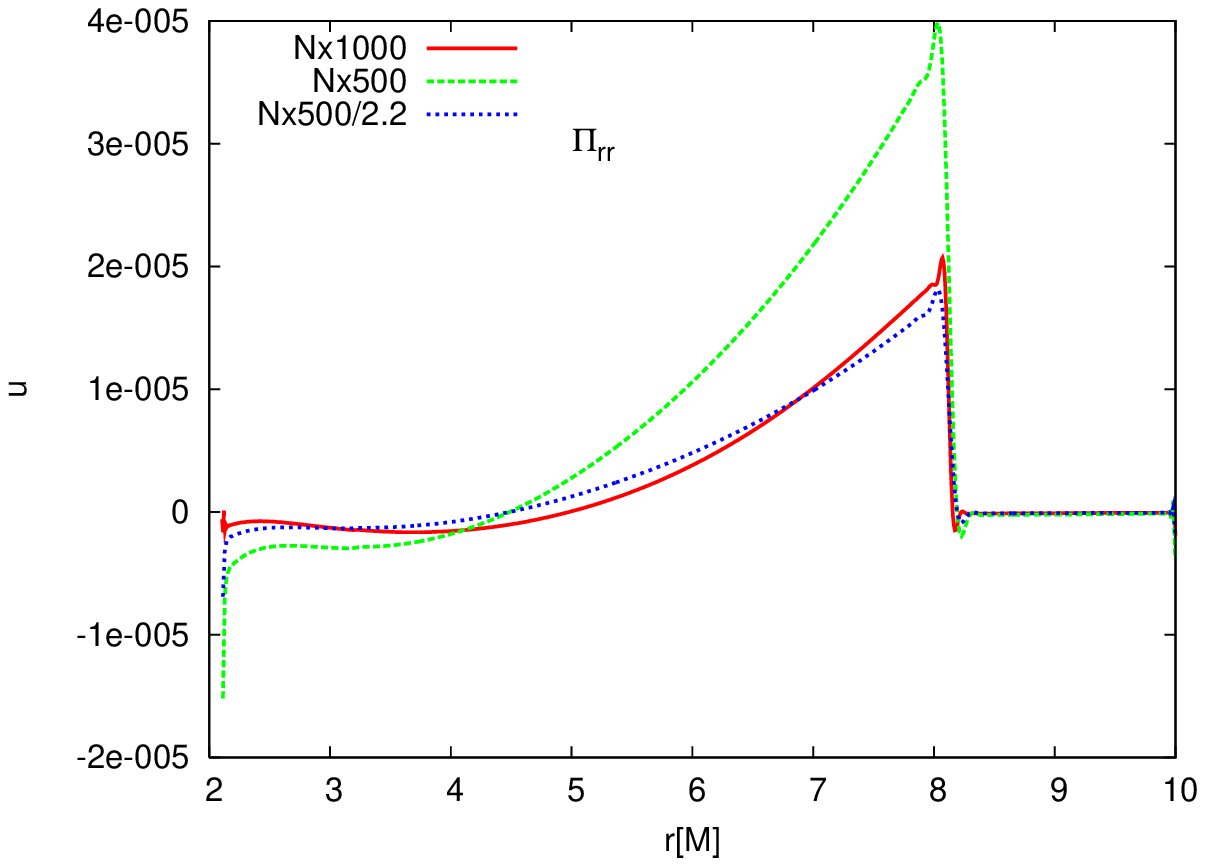}&
\includegraphics[width=0.25\textwidth]{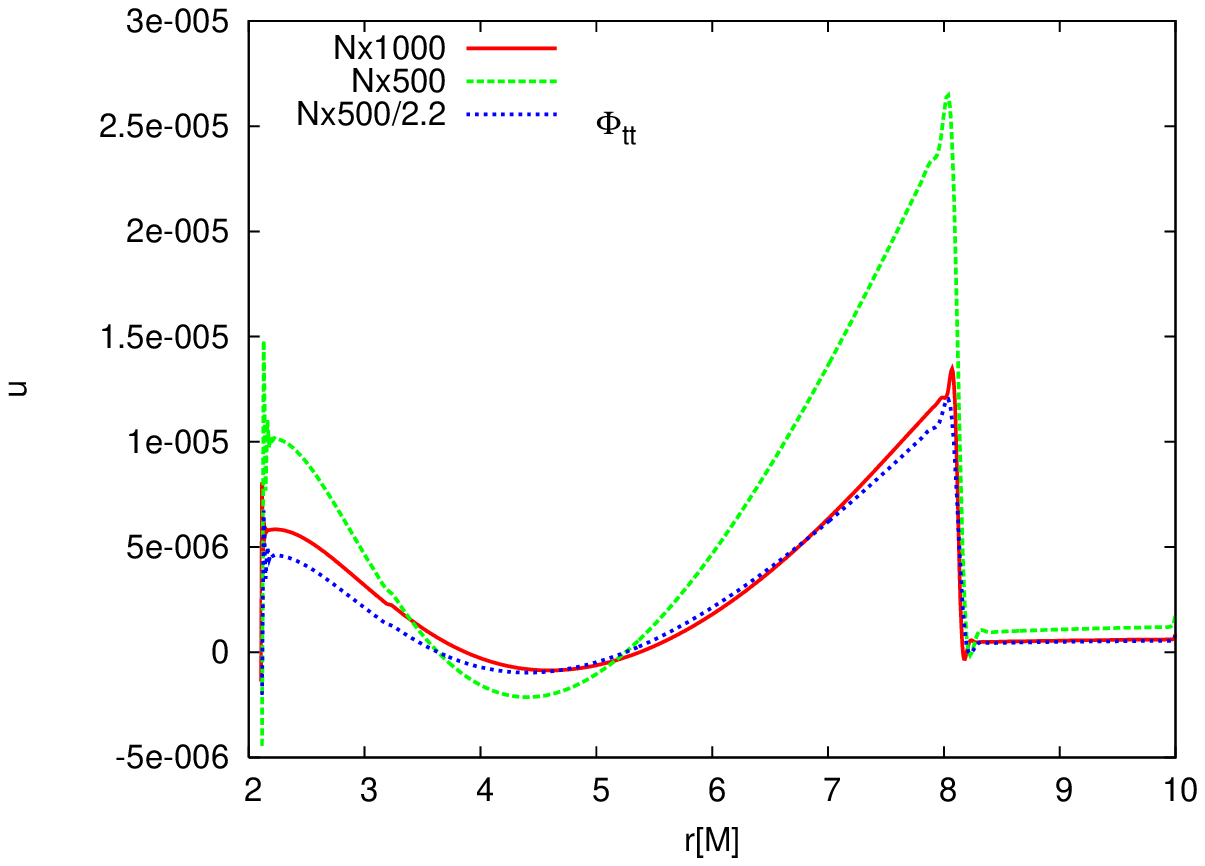}&
\includegraphics[width=0.25\textwidth]{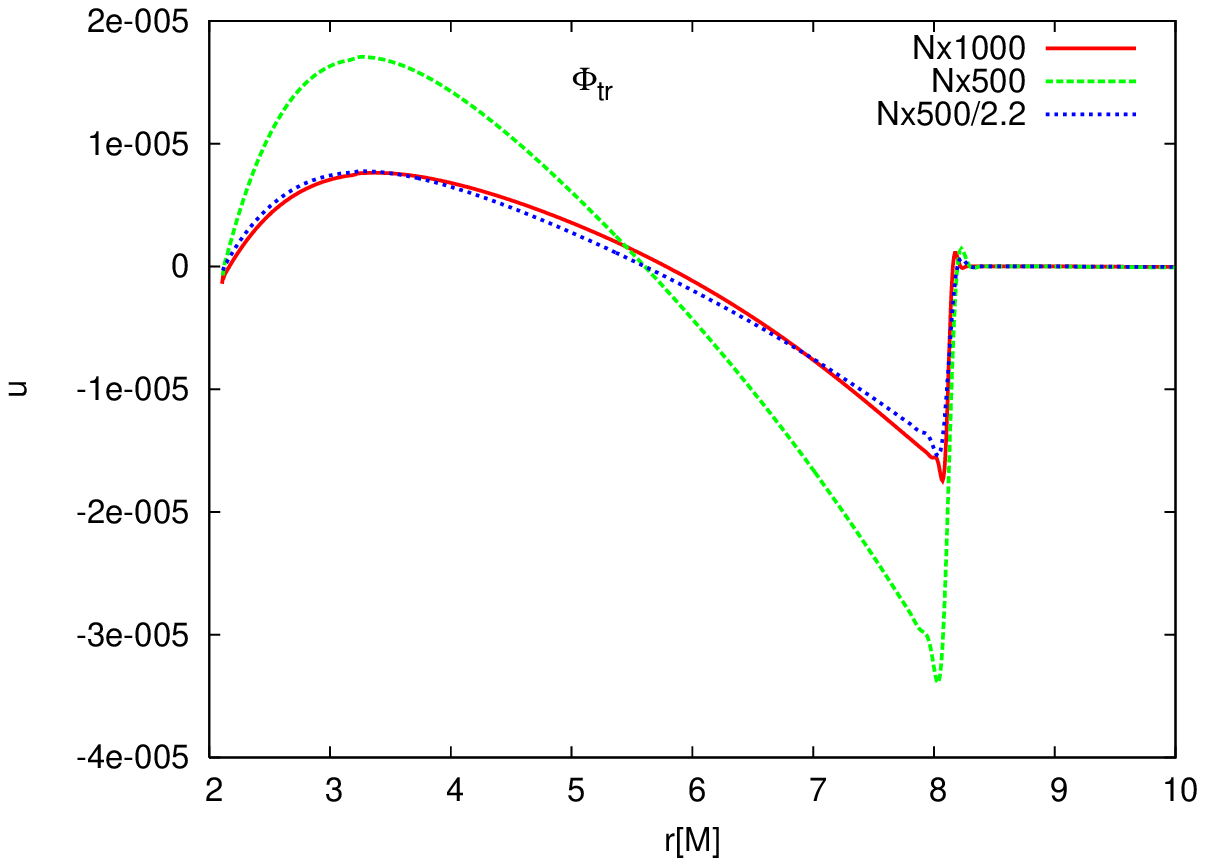}\\
\includegraphics[width=0.25\textwidth]{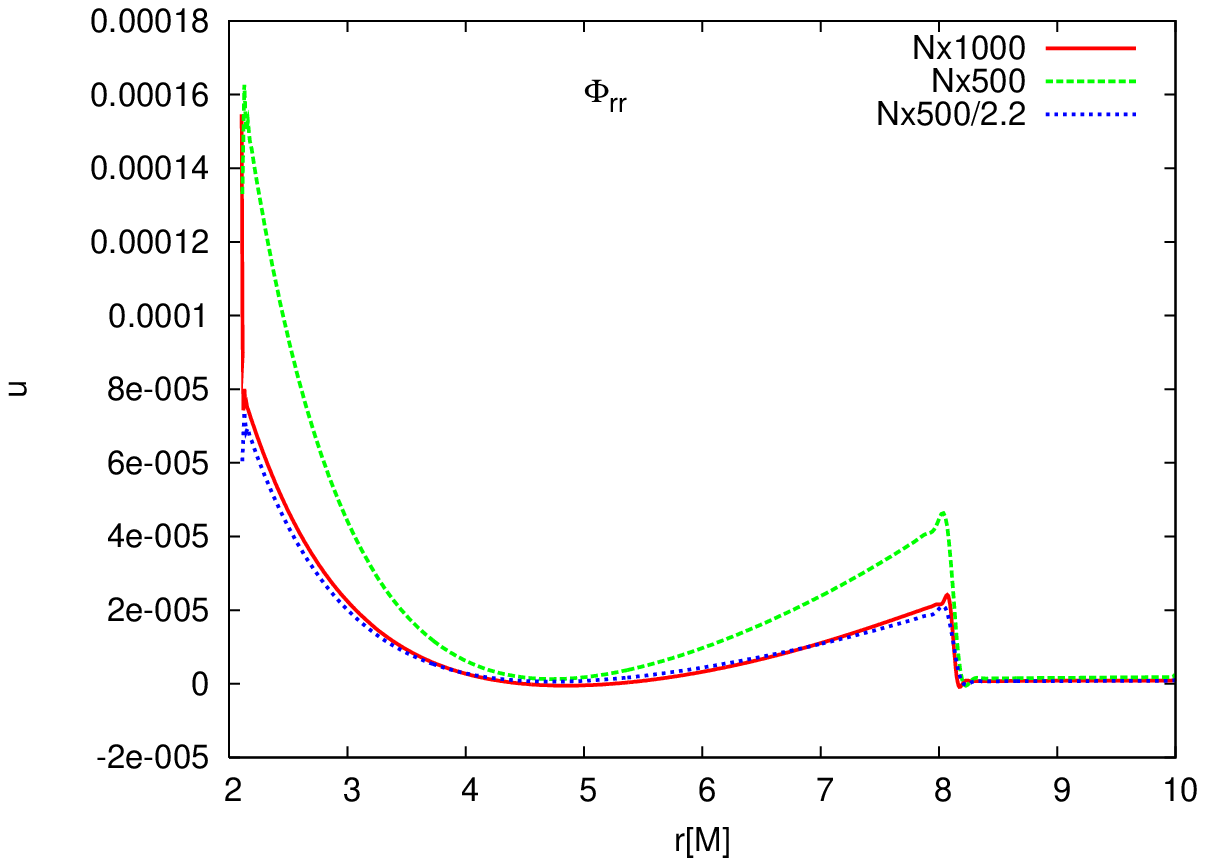}&
\includegraphics[width=0.25\textwidth]{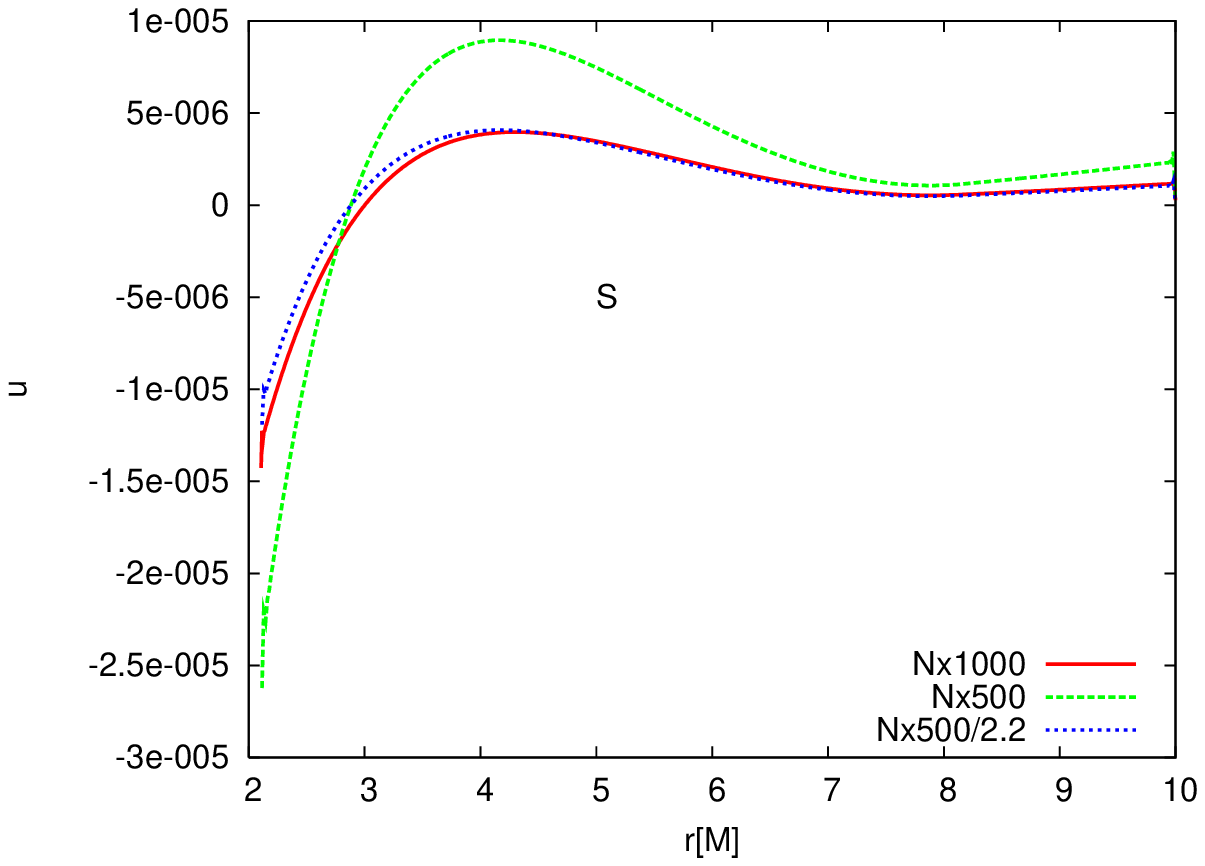}&
\includegraphics[width=0.25\textwidth]{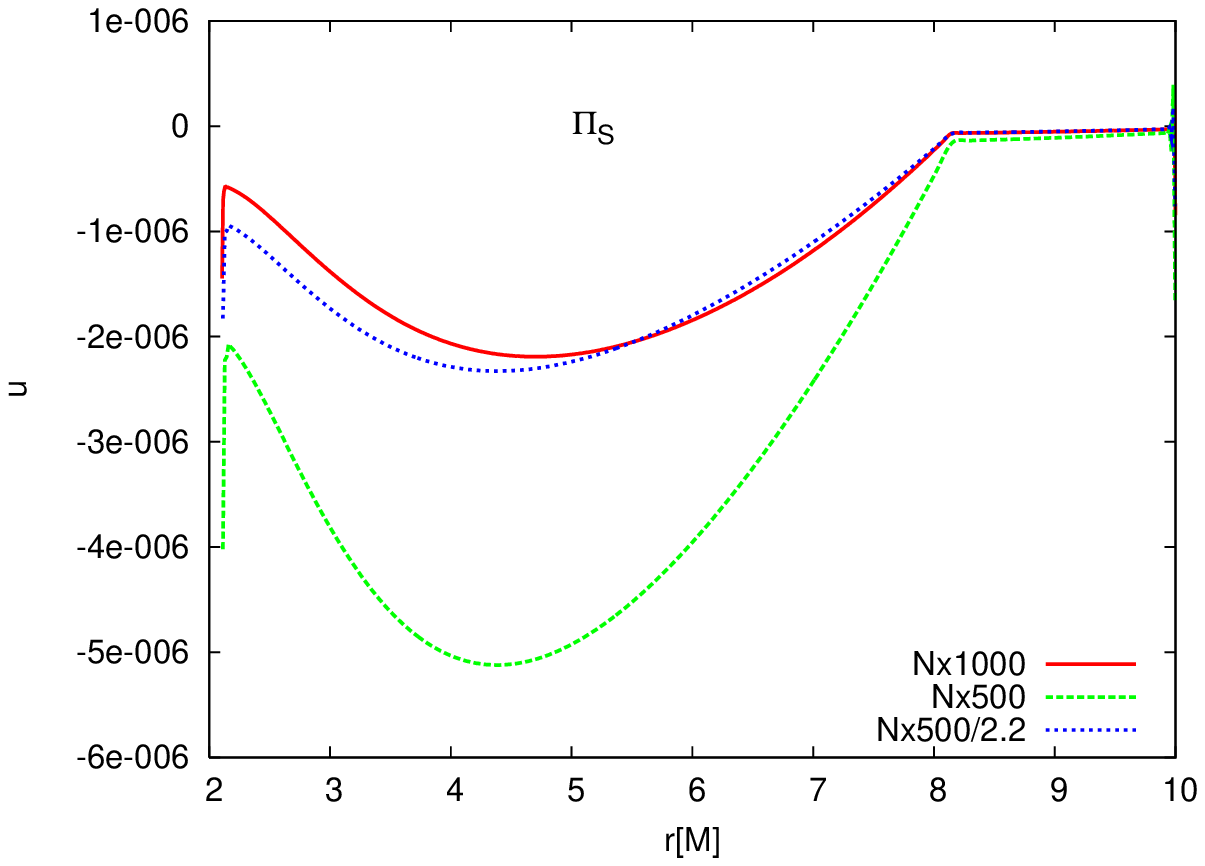}&
\includegraphics[width=0.25\textwidth]{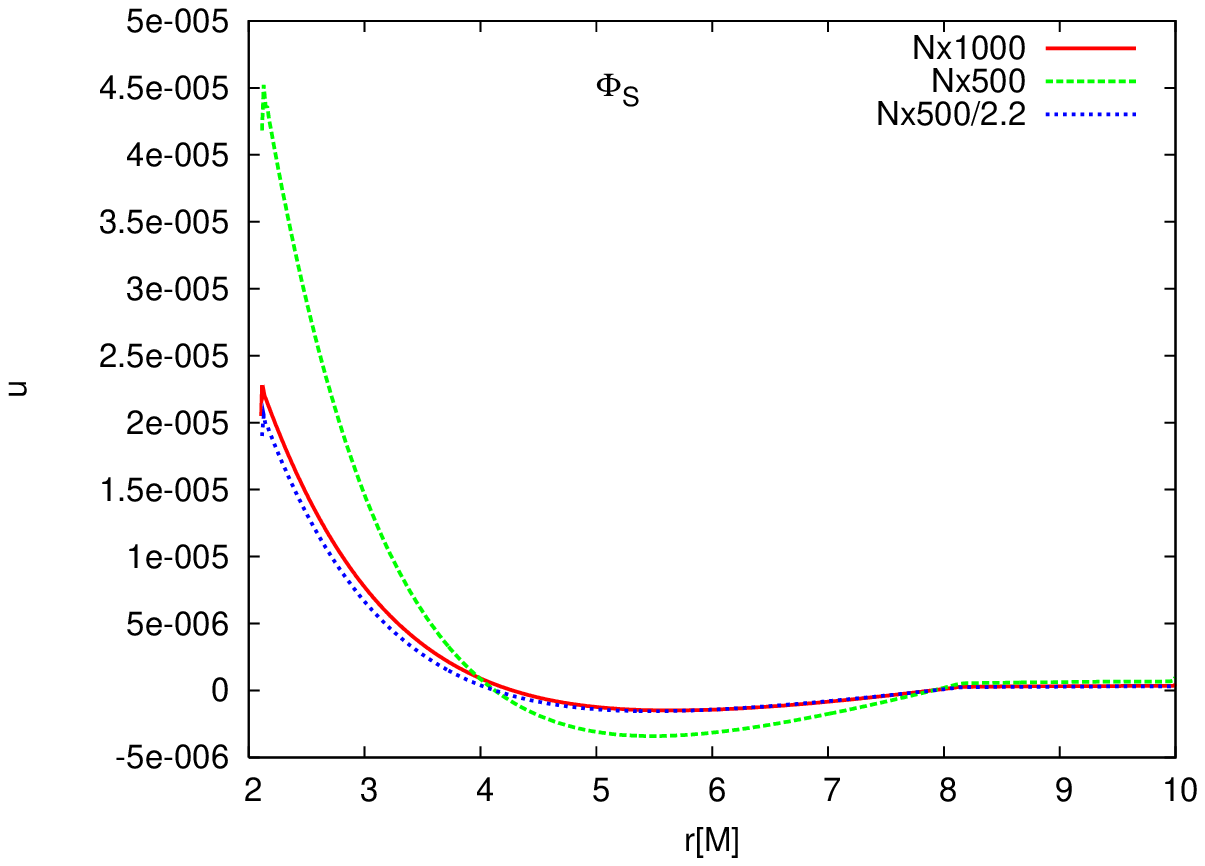}
\end{tabular}
\caption{The numerical error for different variables at $t=9.96M$. Results for $N_x=1000$ and $N_x=500$ are compared. Here highest polynomial $P_1$ is used. First order convergence behavior which corresponds to the factor 2.2 is apparent. Isotropic coordinate is used.}\label{fig12}
\end{figure*}
The Schwarzschild metric in isotropic coordinate takes the form (c.f., Eq.~(1.60) of \cite{Shapiro2010})
\begin{align}
ds^2=&-(\frac{1-\frac{M}{2r}}{1+\frac{M}{2r}})^2dt^2\nonumber\\
&+(1+\frac{M}{2r})^4[dr^2+r^2(d\theta^2+\sin^2\theta d\phi^2)].\label{isometric}
\end{align}
Correspondingly the source function is
\begin{align}
&H_t=H_\phi=0,\\
&H_r=\frac{\frac{2}{r}}{(1+\tfrac{M}{2r})(1-\tfrac{M}{2r})},\\
&H_\theta=\cot\theta.
\end{align}
We plug these source functions into our dynamical system (\ref{problem_eq}). Note this isotropic coordinate is singular at the event horizon $r=2M$. So we chose our computational domain $2.1M<r<10M$. At the boundaries, we apply the Dirichlet boundary condition based on the given metric form (\ref{isometric}).

\begin{figure}
\begin{tabular}{cc}
\includegraphics[width=0.25\textwidth]{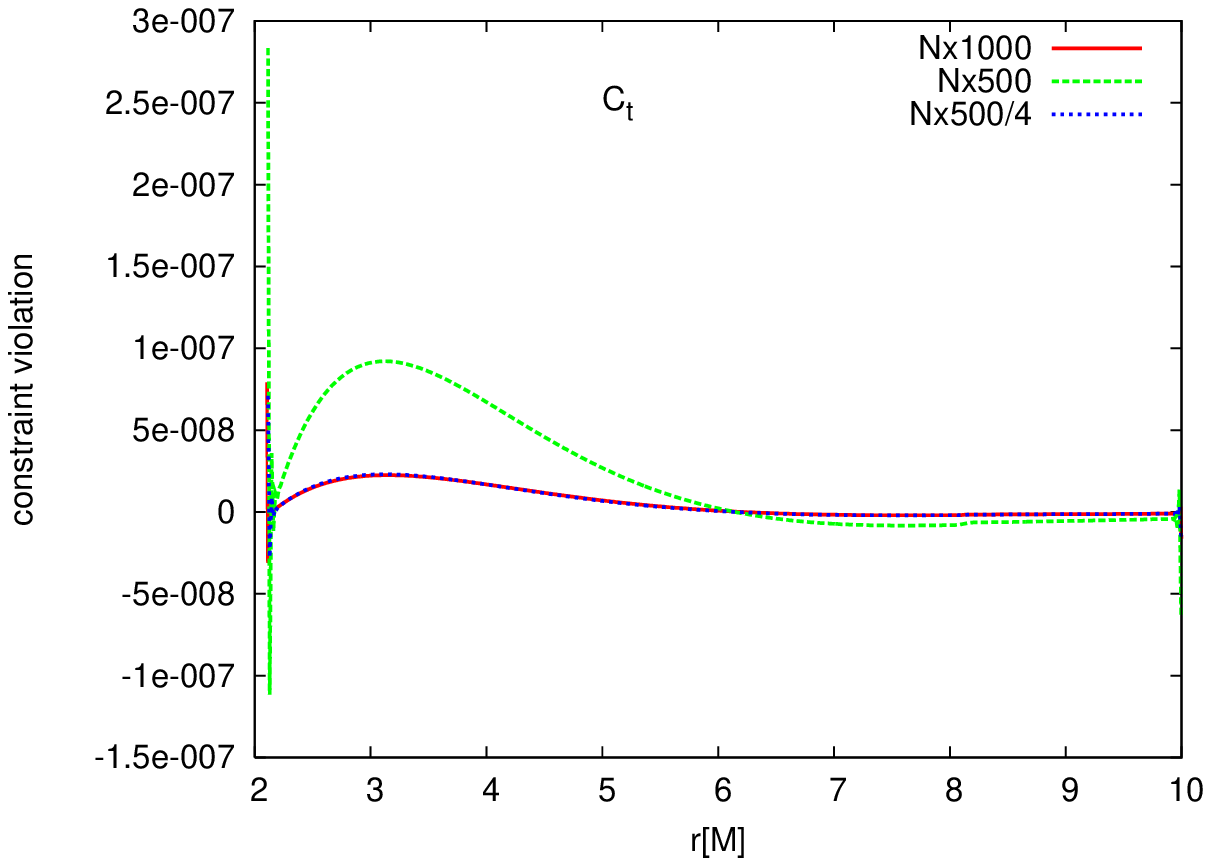}&
\includegraphics[width=0.25\textwidth]{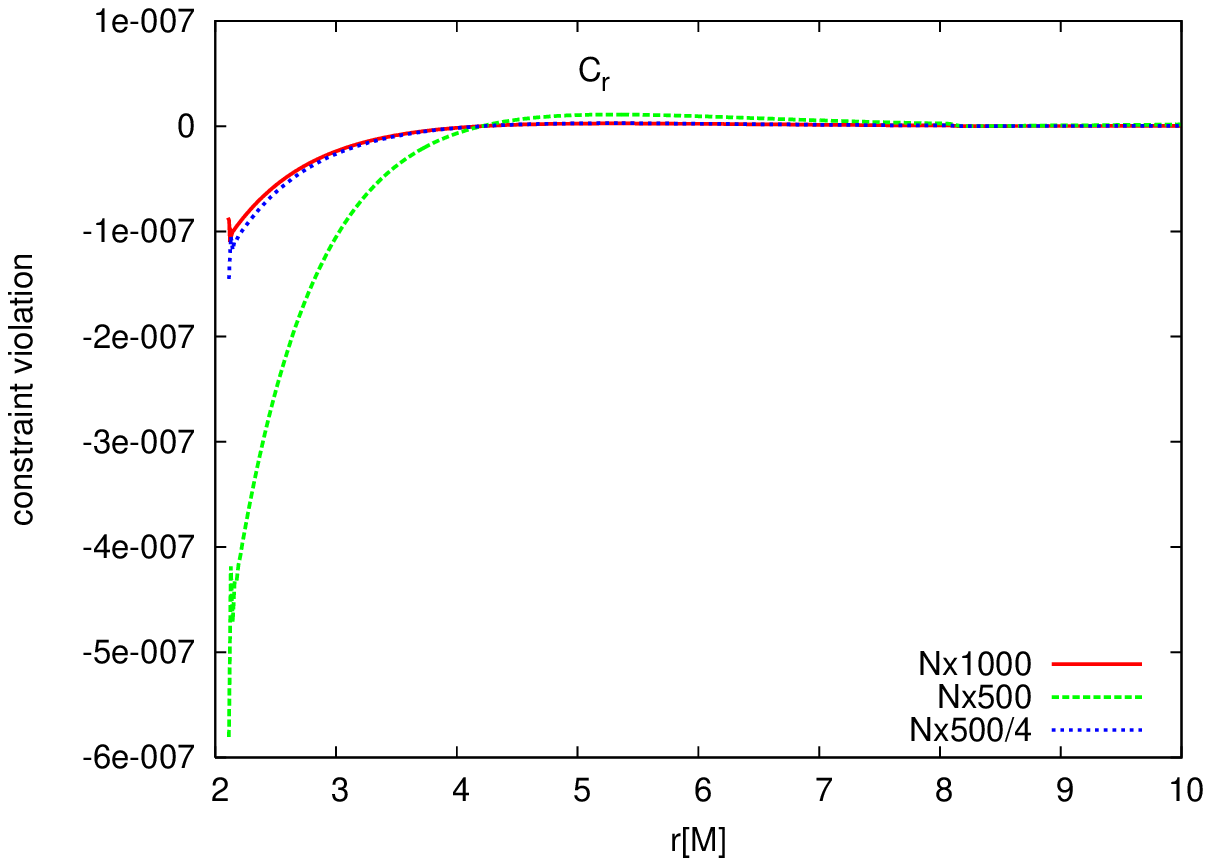}
\end{tabular}
\caption{The constraint violation $C_{t,r}$ corresponding to the Fig.~\ref{fig12}. Isotropic coordinate is used.}\label{fig13}
\end{figure}

\begin{figure}
\begin{tabular}{c}
\includegraphics[width=0.5\textwidth]{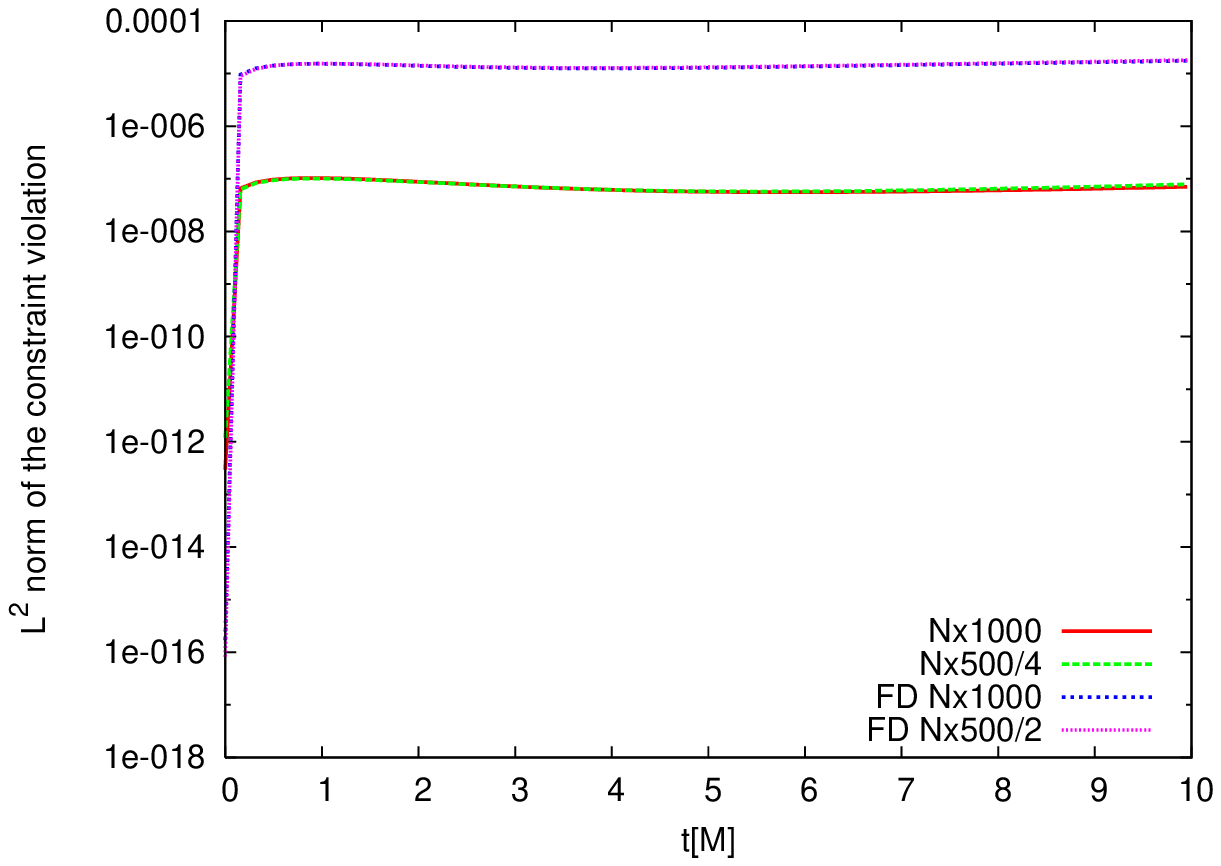}
\end{tabular}
\caption{$L^2$ norm of the constraint violation $\sqrt{\int (C^2_t+C_r^2)dx}$. Isotropic coordinate is used. FD means `Finite Difference'.}\label{fig14}
\end{figure}

Compared to the Kerr-Schild coordinate case, there is no shock formation in the current isotropic coordinate case. So the resulted configurations as shown in the Fig.~\ref{fig12} are smooth compared to the results shown in the Fig.~\ref{fig5}. Correspondingly the convergence behavior is better. The constraint violation convergence behavior is similar to the Kerr-Schild coordinate case, as shown in the Figs.~\ref{fig13} and \ref{fig14}. The long term stability behavior is similar to the results shown in the Fig.~\ref{fig10}. So we do not plot them any more here.
\subsection{Schwarzschild black hole in Painleve-Gullstrand-like (PG) coordinate}
\begin{figure*}
\begin{tabular}{cccc}
\includegraphics[width=0.25\textwidth]{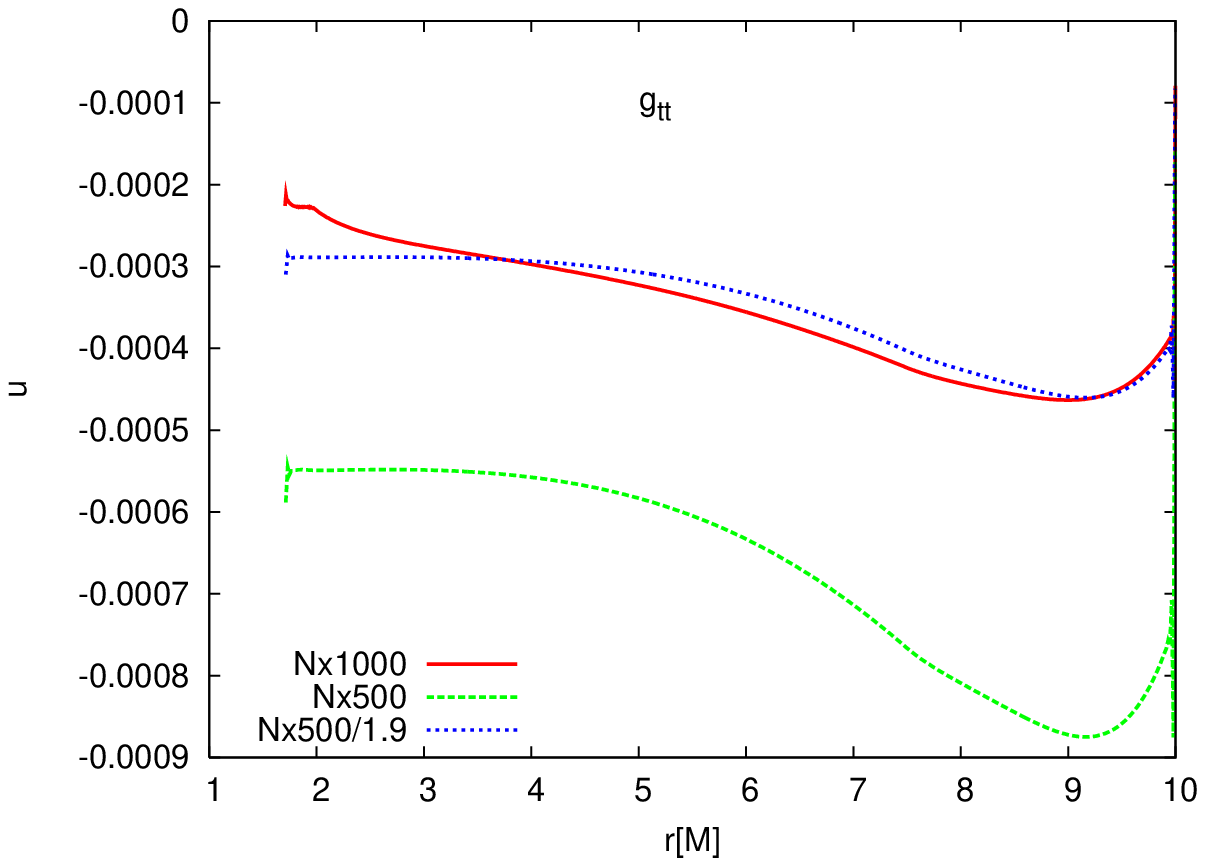}&
\includegraphics[width=0.25\textwidth]{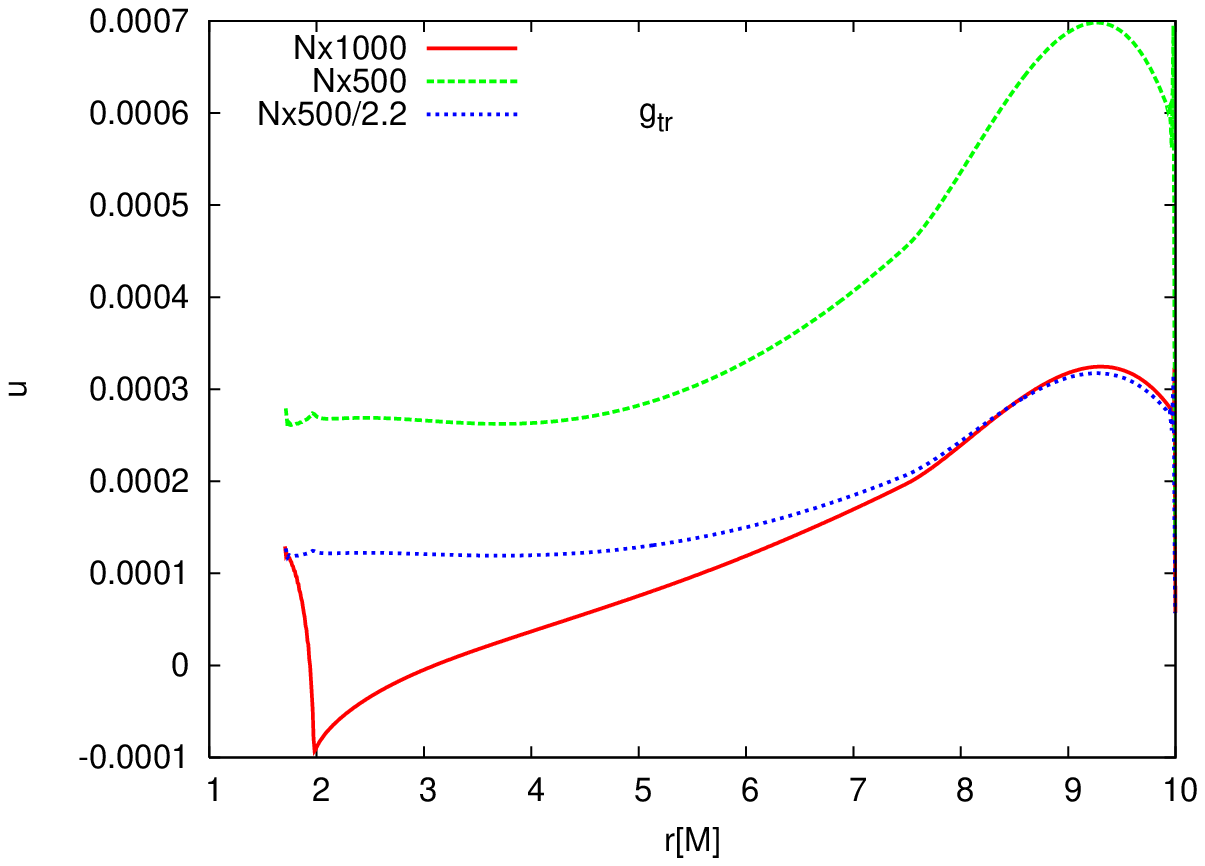}&
\includegraphics[width=0.25\textwidth]{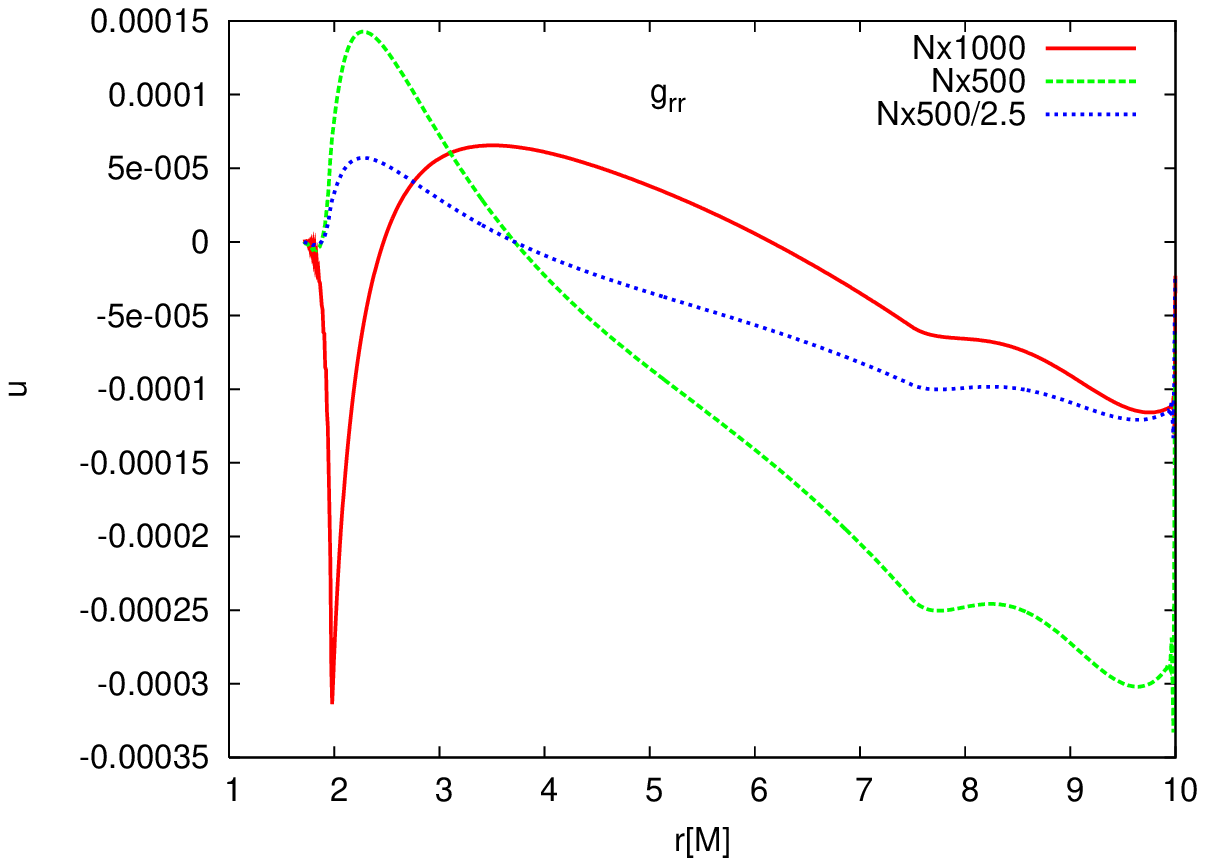}&
\includegraphics[width=0.25\textwidth]{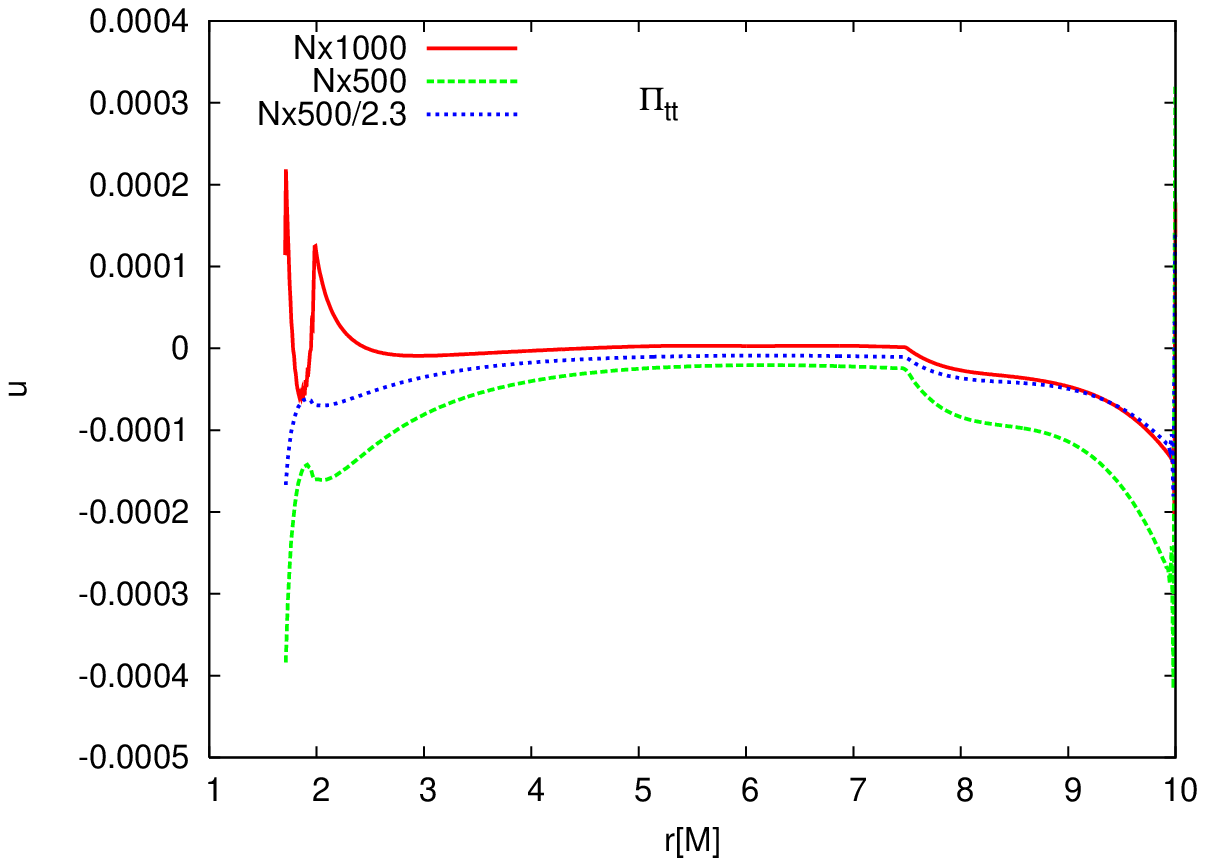}\\
\includegraphics[width=0.25\textwidth]{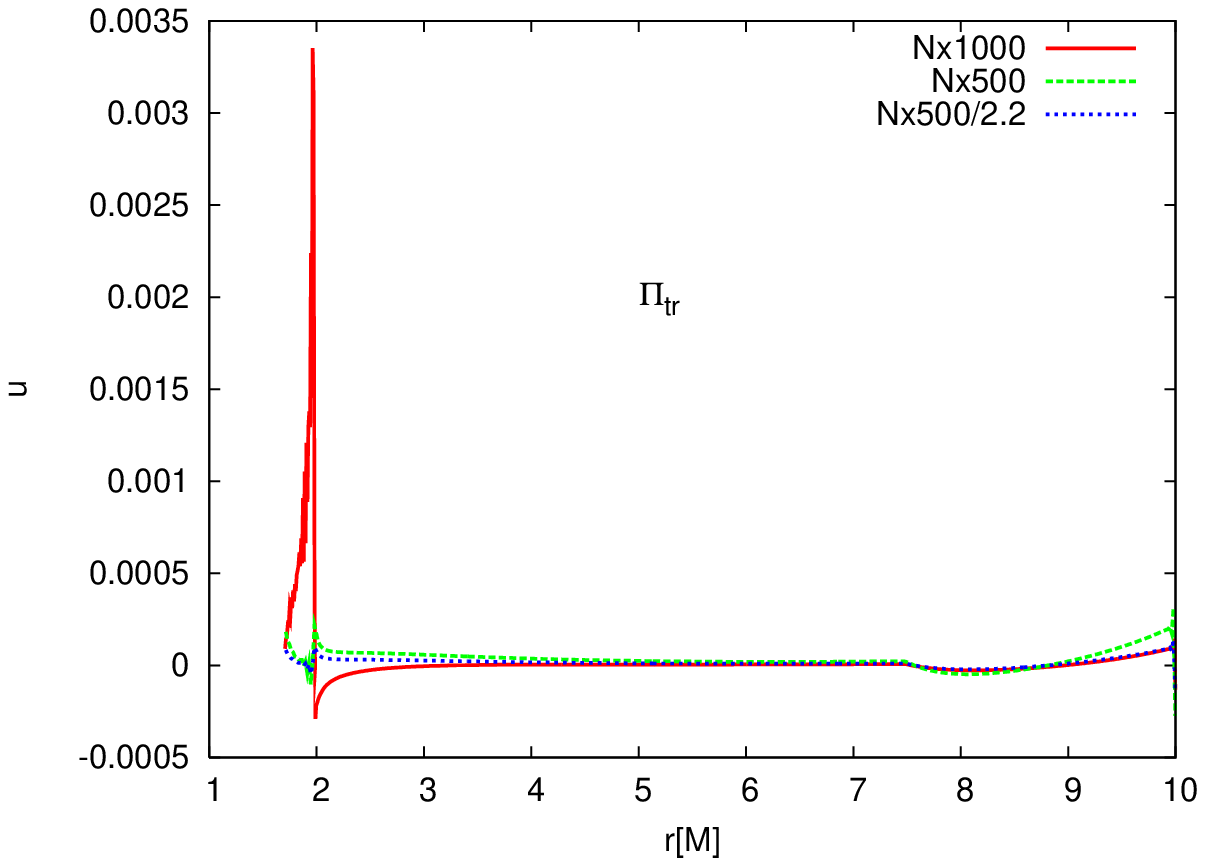}&
\includegraphics[width=0.25\textwidth]{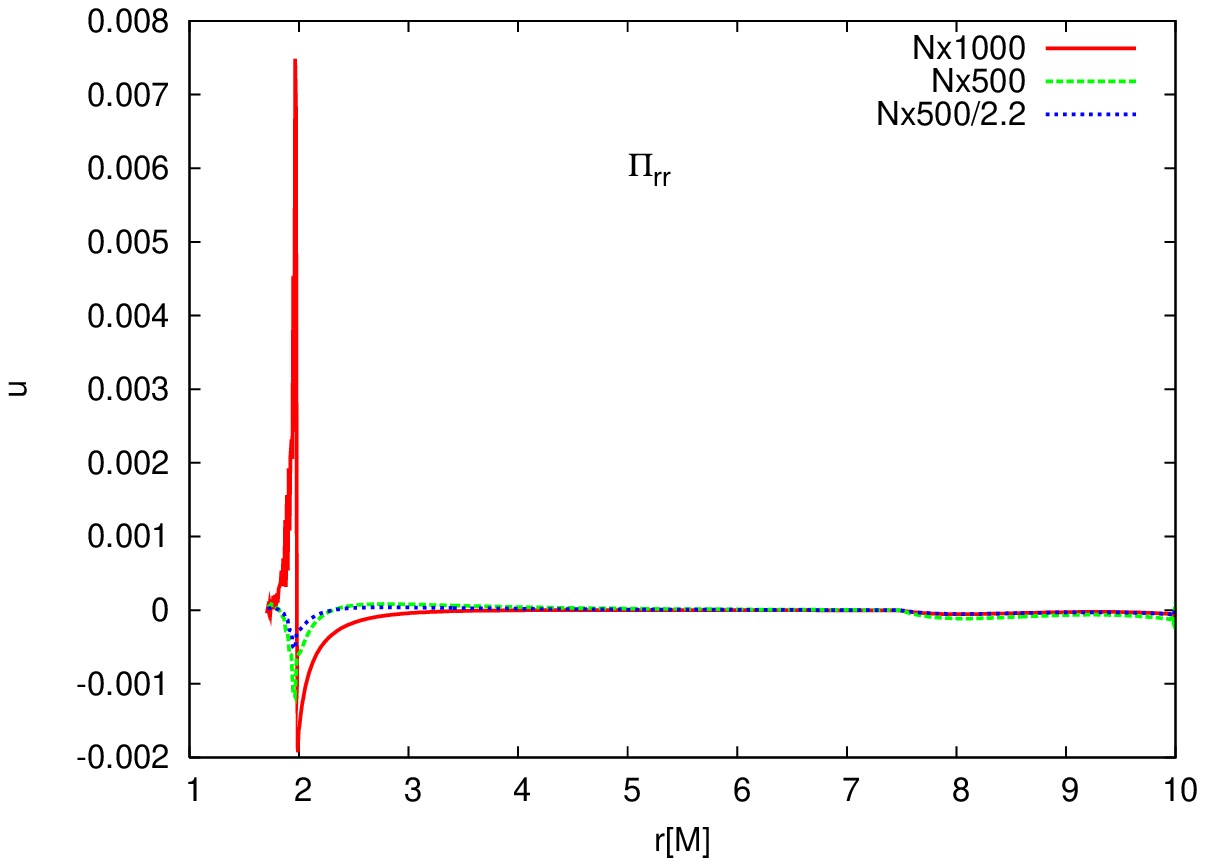}&
\includegraphics[width=0.25\textwidth]{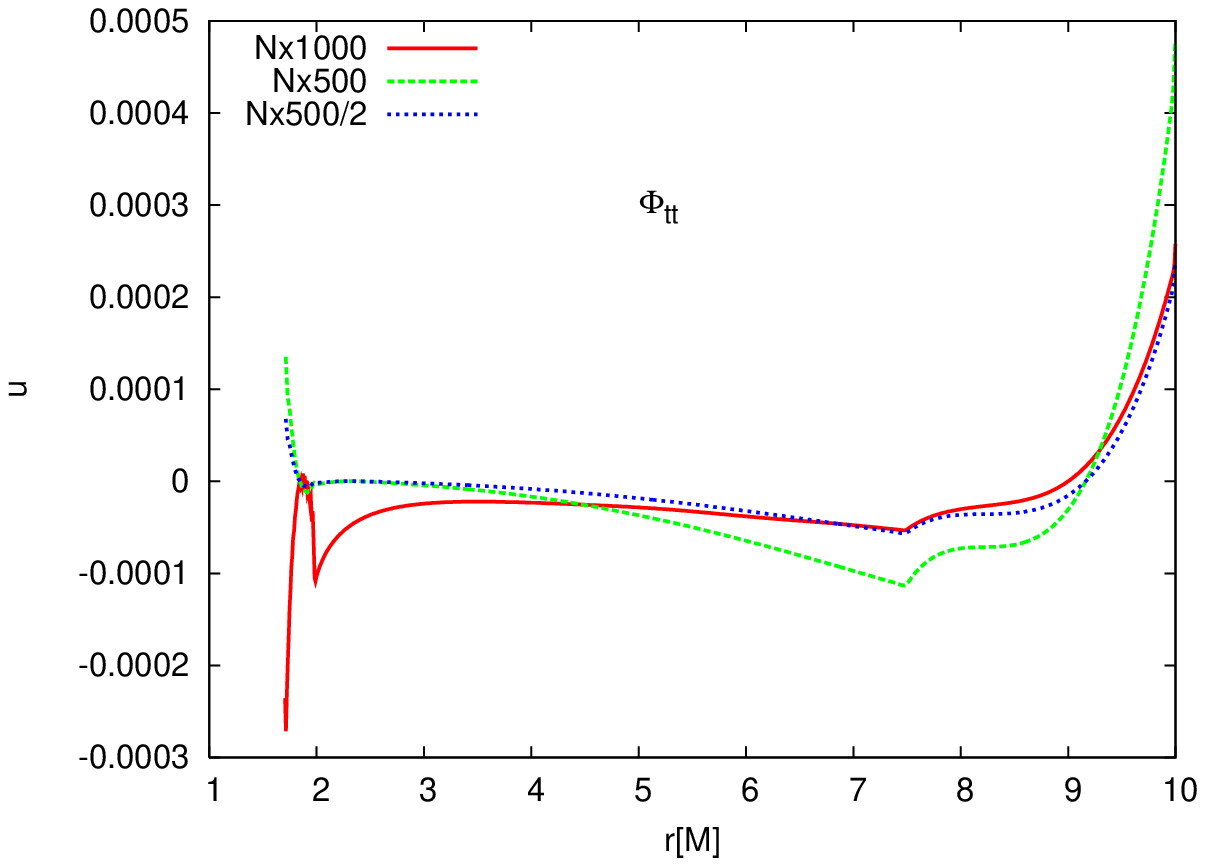}&
\includegraphics[width=0.25\textwidth]{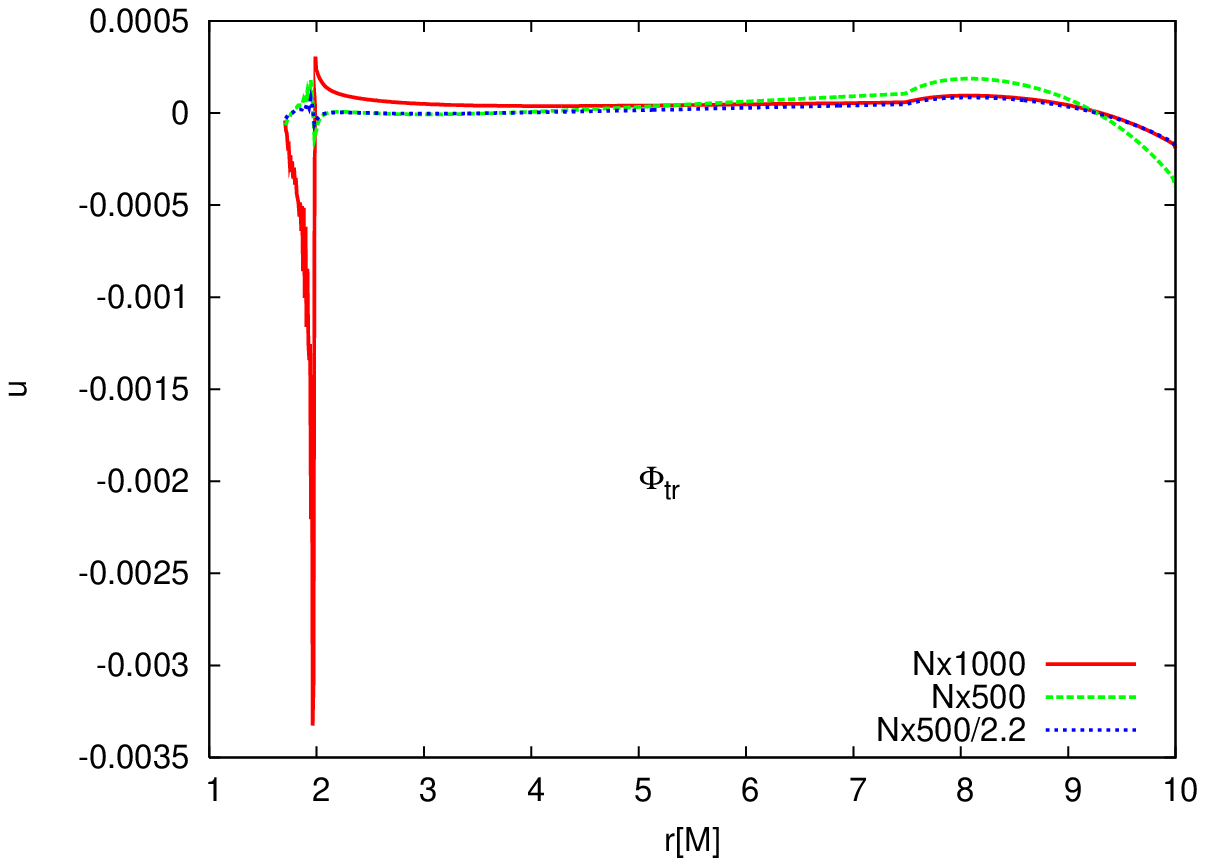}\\
\includegraphics[width=0.25\textwidth]{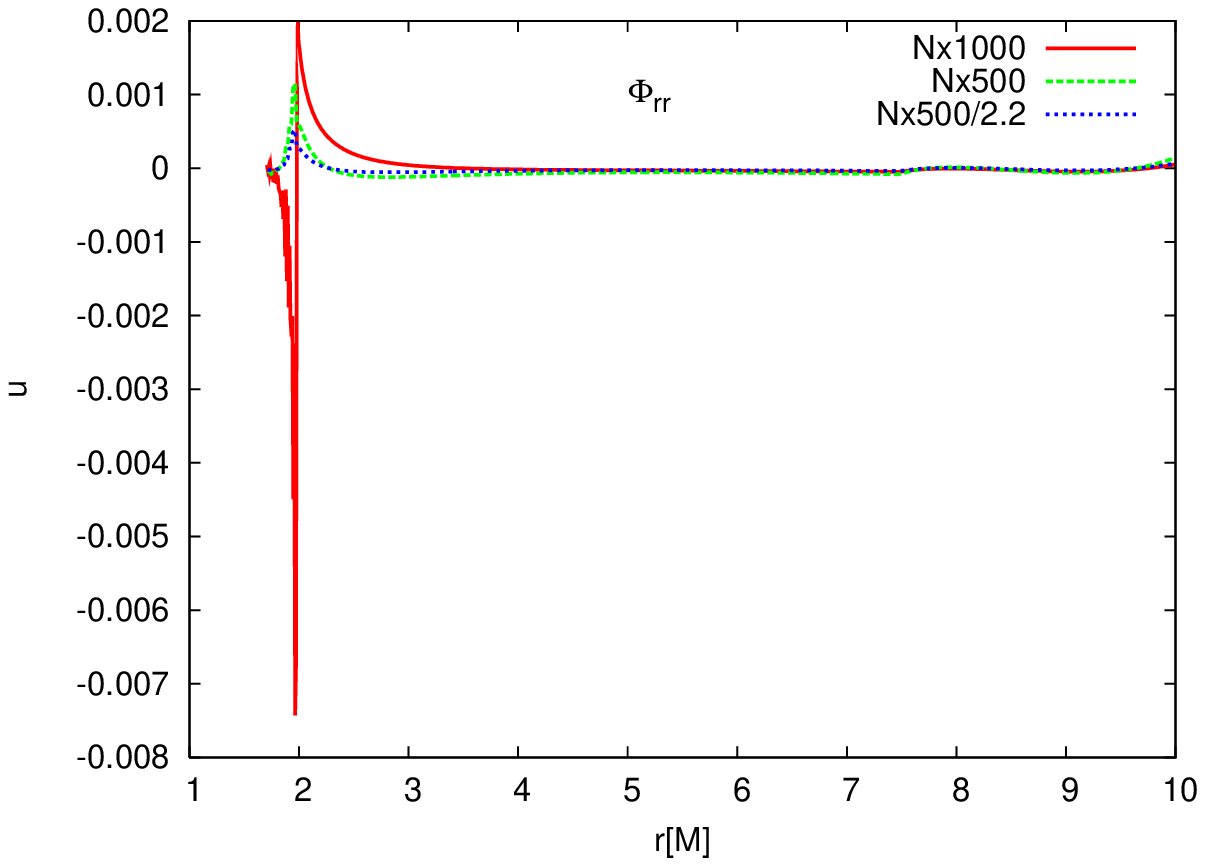}&
\includegraphics[width=0.25\textwidth]{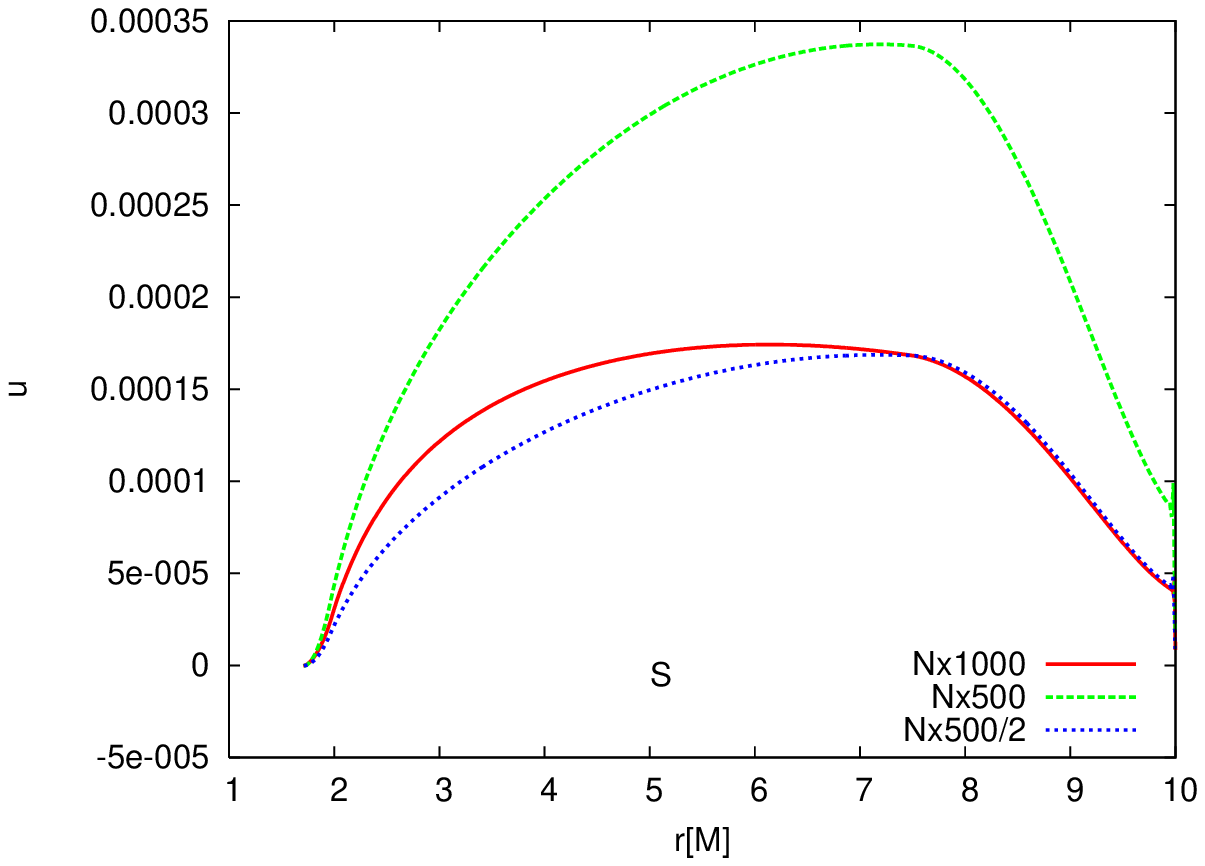}&
\includegraphics[width=0.25\textwidth]{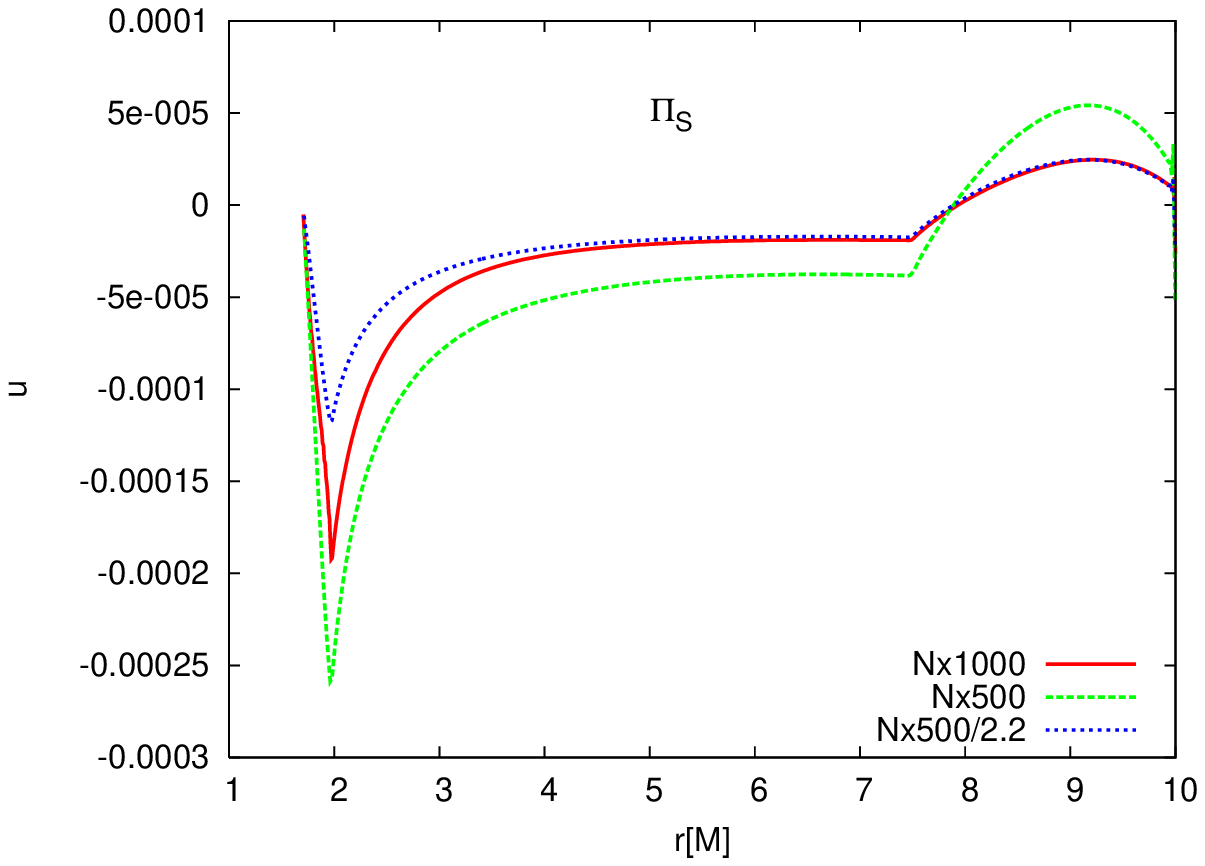}&
\includegraphics[width=0.25\textwidth]{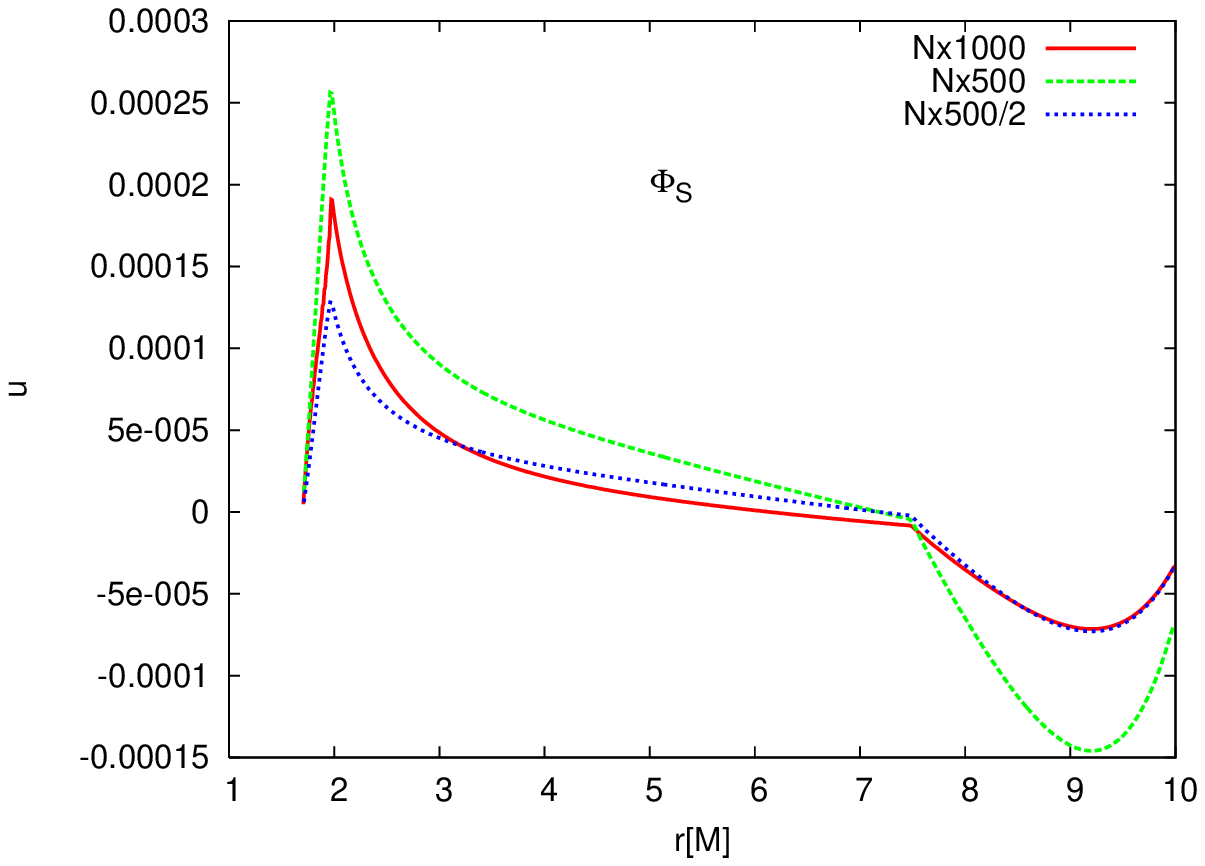}
\end{tabular}
\caption{The numerical error for different variables at $t=9.96M$. Results for $N_x=1000$ and $N_x=500$ are compared. Here highest polynomial $P_1$ is used. First order convergence behavior which corresponds to the factor 2.2 is apparent. PG coordinate is used.}\label{fig15}
\end{figure*}

\begin{figure}
\begin{tabular}{cc}
\includegraphics[width=0.25\textwidth]{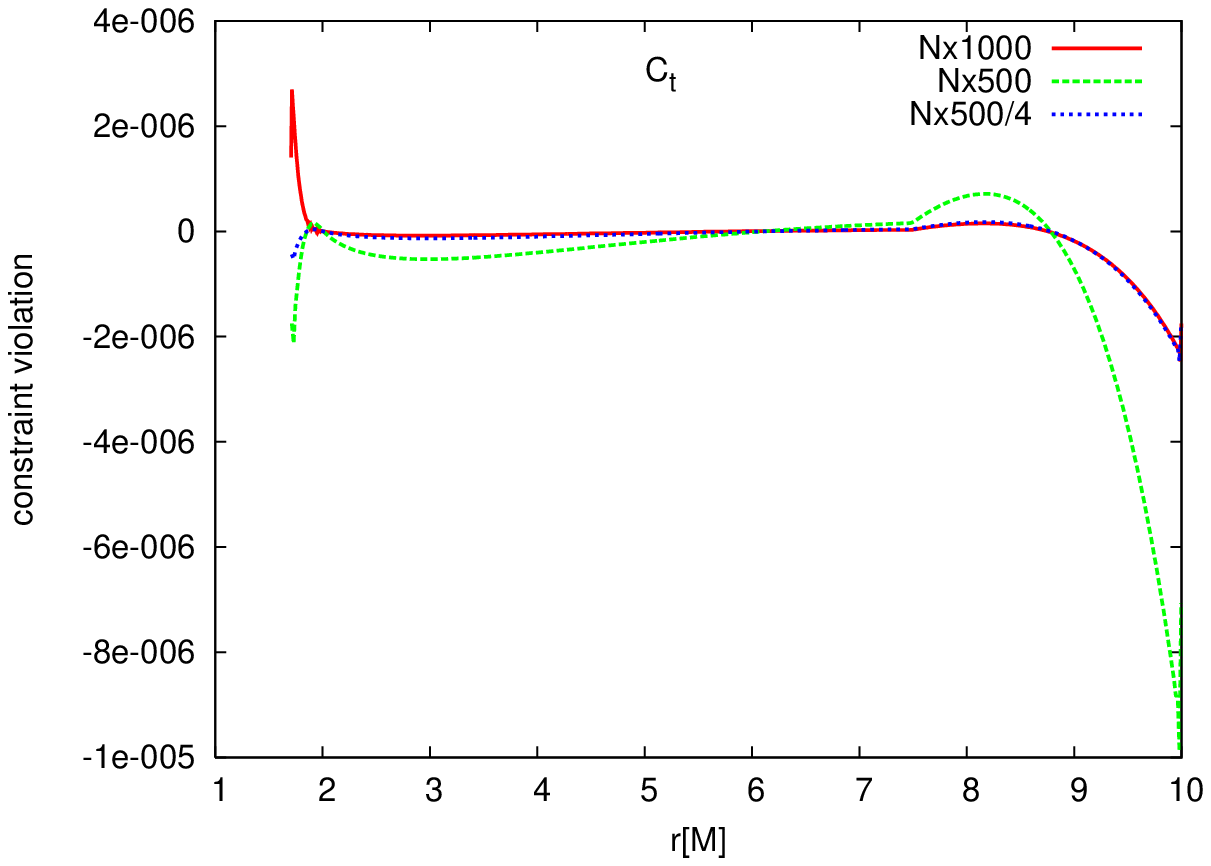}&
\includegraphics[width=0.25\textwidth]{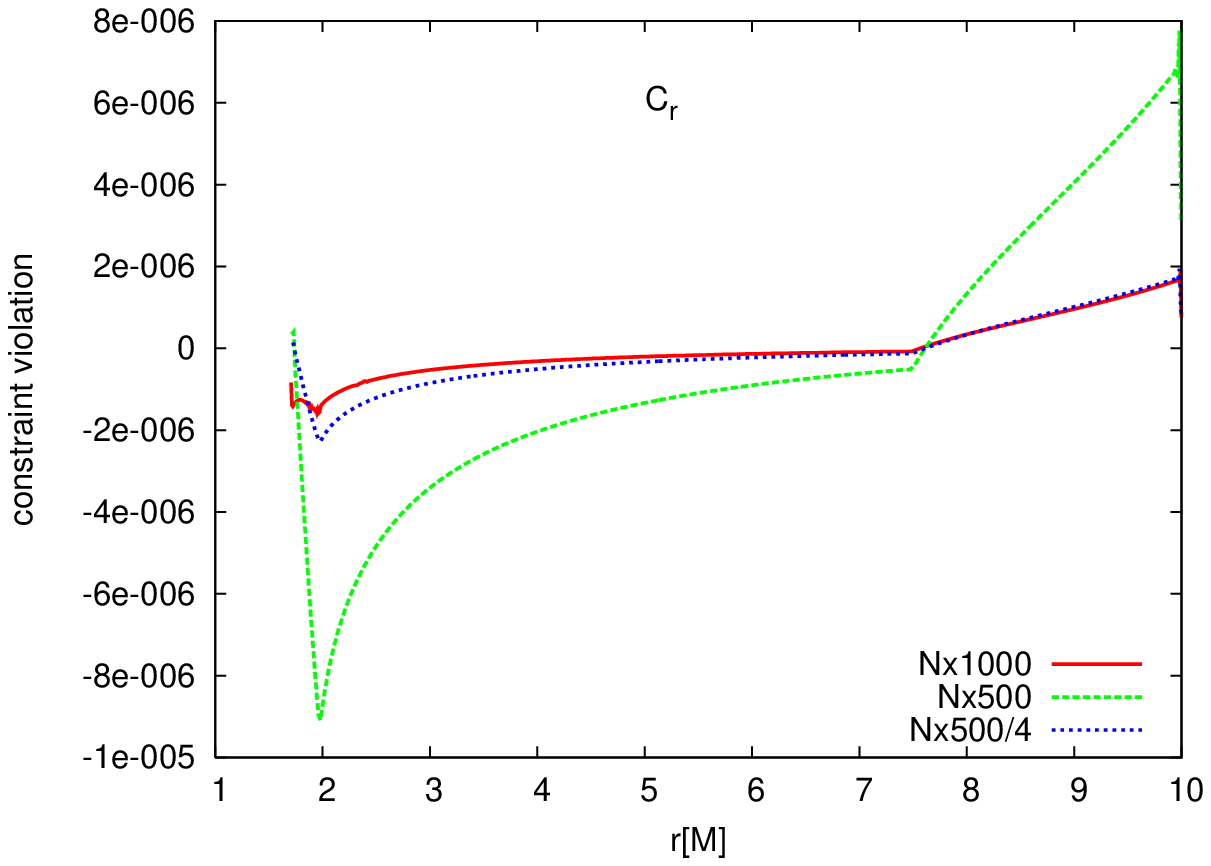}
\end{tabular}
\caption{The constraint violation $C_{t,r}$ corresponding to the Fig.~\ref{fig15}. PG coordinate is used.}\label{fig16}
\end{figure}

\begin{figure}
\begin{tabular}{c}
\includegraphics[width=0.5\textwidth]{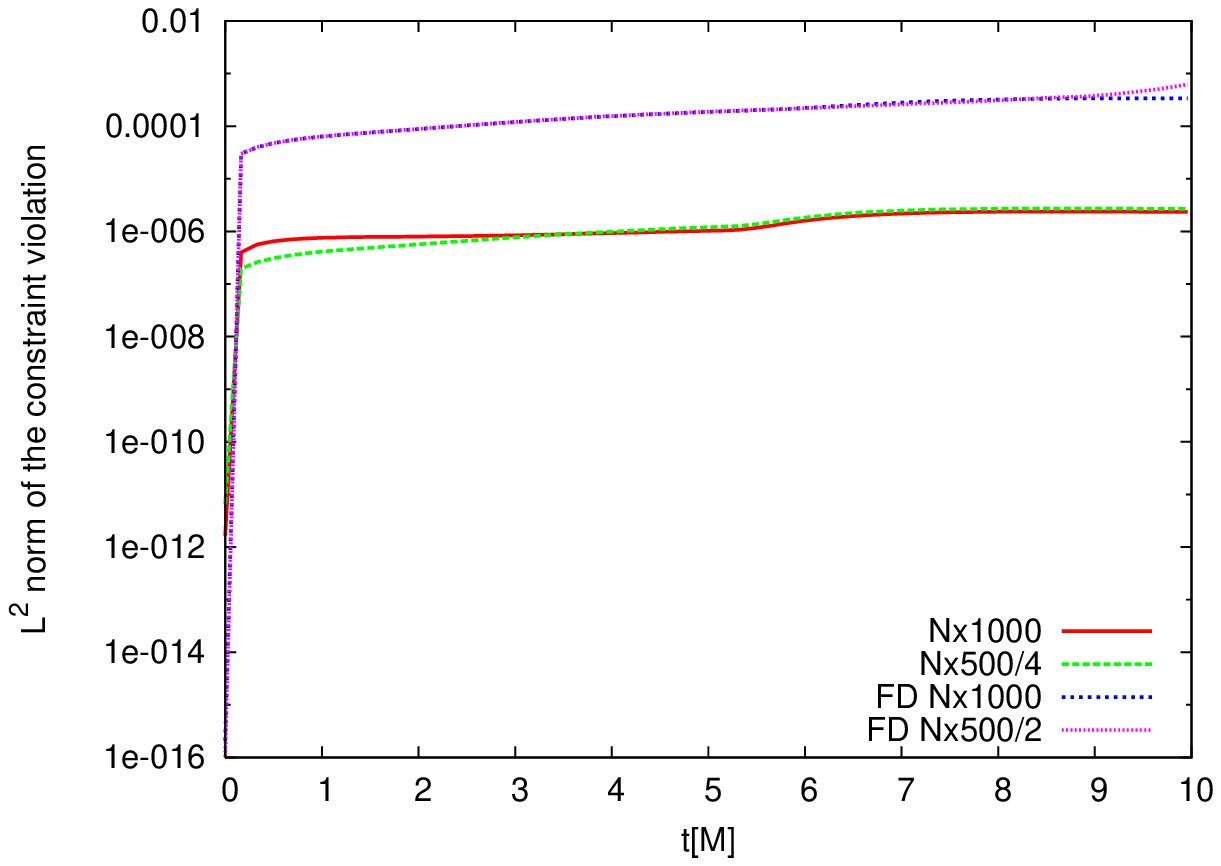}
\end{tabular}
\caption{$L^2$ norm of the constraint violation $\sqrt{\int (C^2_t+C_r^2)dx}$. PG coordinate is used. FD means `Finite Difference'.}\label{fig17}
\end{figure}
The Schwarzschild metric in Painleve-Gullstrand-like (PG) coordinate takes the form (c.f., Eq.~(52) of \cite{0264-9381-22-2-014})
\begin{align}
ds^2=&-(1-\frac{2M}{r})dt^2-2\sqrt{\frac{2M}{r}}dtdr\nonumber\\
&+dr^2+r^2d\theta^2+r^2\sin^2\theta d\phi^2,\label{pgmetric}
\end{align}
Correspondingly the source function is
\begin{align}
&H_t=-\sqrt{\frac{M}{2r}}\frac{1}{r}\frac{2M+r}{r},\\
&H_r=\frac{1}{r}\frac{M+2r}{r},\\
&H_\theta=\cot\theta,\\
&H_\phi=0.
\end{align}
We plug these source functions into our dynamical system (\ref{problem_eq}). Like the Kerr-Schild coordinate, the PG coordinate is also horizon penetrating. So we chose our computational domain $1.7M<r<10M$. At the boundaries, we apply the Dirichlet boundary condition based on the given metric form (\ref{pgmetric}).

Compared to the Kerr-Schild coordinate case, there are even more shocks appear in the current PG coordinate case. In the Kerr-Schild coordinate case only $\Phi_{tr}$ forms shock. In the current PG coordinate case, $\Pi_{tr}$, $\Pi_{rr}$, $\Phi_{tr}$ and $\Phi_{rr}$ form shocks near $r=2M$. So the resulted configurations as shown in the Fig.~\ref{fig15} are more mess compared to the results shown in the Fig.~\ref{fig5}. Correspondingly the convergence behavior is worse. The constraint violation convergence behavior is similar to the Kerr-Schild coordinate case, as shown in the Figs.~\ref{fig13} and \ref{fig14}. The long term stability behavior is similar to the results shown in the Fig.~\ref{fig10}. So we do not plot them more here.

Based on this comparison among three different coordinates, we can conclude gauge choices affect the shock formation quite a bit in the generalized harmonic gauge formulation. Assuming spectral method can not treat shock properly, we suspect this is the possible reason that it is very hard for spectral code to find a gauge driver for binary black hole simulation \cite{PhysRevD.80.084019,PhysRevD.80.124010}.

\begin{figure}
\begin{tabular}{c}
\includegraphics[width=0.5\textwidth]{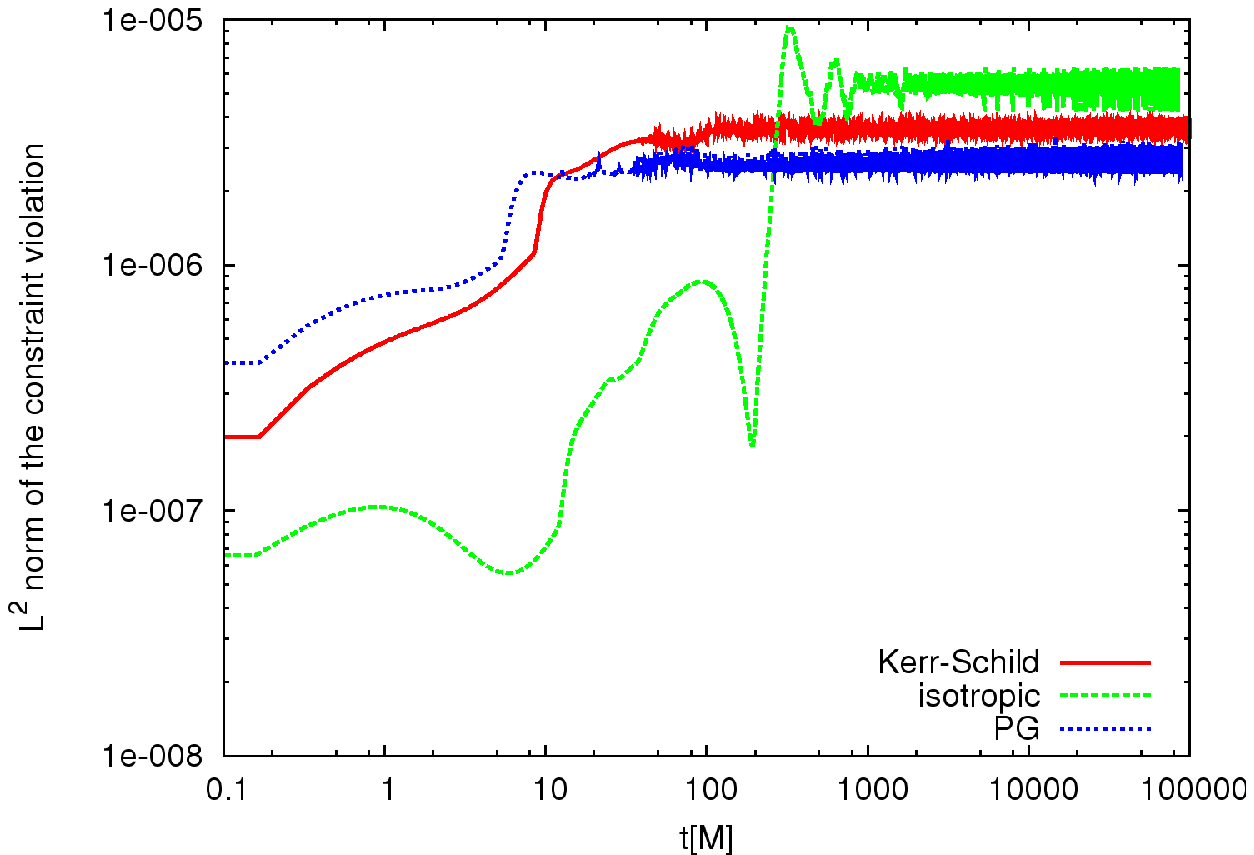}
\end{tabular}
\caption{Long term evolution behavior comparison for Kerr-Schild, isotropic and Painleve-Gullstrand-like coordinates. The $L^2$ norm of the constraint violation $\sqrt{\int (C^2_t+C_r^2)dx}$ is plotted. The default parameters setting is used.}\label{fig18}
\end{figure}

Finally, we compare the long term evolution behavior for the three different coordinates. In particular, the long term constraint violation is plotted in the Fig.~\ref{fig18}. Here the default setting $\gamma_0=-\gamma_1=\gamma_2=1$, $\gamma_4=\gamma_5=\frac{1}{2}$ together with limiter is used. At beginning, the constraint violation for isotropic coordinate is smallest, and the one for PG coordinate is largest. Interestingly, the constraint violation for isotropic coordinate becomes the largest one while the one for PG coordinate becomes smallest one. In isotropic coordinate the shift vector $\beta^i=0$ which results in zero speed freedom (c.f. the Appendix.~\ref{App::I}). These zero speed freedom may make some numerical error pile up. Consequently the isotropic coordinate results in larger constraint violation at later evolution time. Similar behavior has been found by us before in the comparison of BSSN formulation and Z4c formulation \cite{CaoHil12,PhysRevD.88.084057}.

\section{Conclusion and discussion}\label{Sec::V}
Finite difference and pseudo-spectral methods have been well studied for Einstein equations. As the third category method of solving partial differential equations numerically, finite element method has been paid few attention. But the finite element method may combine the advantages of both finite difference method and spectral method. Especially the finite element method may admit the robust property and good parallel scalability of finite difference method and the high convergence property of spectral method.

In \cite{PhysRevD.91.044033} we have investigated the finite element method to solve the constraint equations in numerical relativity for initial data. The robustness of the finite element method has been seen there. In the current paper we move on to investigate the finite element method to solve the evolution equations in numerical relativity. For the first time the local discontinuous Galerkin method is proposed for Einstein equations. In order to simplify the problem for the first study, we focus our attention on the spherical symmetric spacetime. Correspondingly a new formulation for such a problem is proposed. Based on our formulation, detail numerical algorithms including the numerical limiter and boundary conditions are designed.

Numerically we have tested our numerical scheme against the Schwarzschild black hole spacetime. In particular, three different gauge conditions including the Kerr-Schild coordinate, the isotropic coordinate and the Painleve-Gullstrand-like coordinate are tested. All of them show stable and well convergent results. Our test results not only indicate that finite element method is robust for solving Einstein equations, but also show some interesting points. Firstly compared to finite difference method, the finite element method is more accurate which implies the good convergence advantage of spectral method. In particular, our test results show that the constraint violation for the finite element method is more than one order smaller than that of finite difference method. All three gauge condition cases show the same result.

Secondly, we find that the finite element method can capture the shock formation in the numerical solution automatically. Physically we believe such shock comes from numerical error. But if such numerical error is not treated well, it may grow and destroy the numerical simulation. Comparing the results of finite difference and finite element methods, we find the numerical dissipation of finite difference method trying to smooth out the shock directly. And consequently high frequency noise develop and less accurate numerical solution is resulted. In contrast, the finite element method capture the shock straightforwardly and it can evolve with such kind of shock. Consequently the numerical error is controlled in the correspondingly level. One related issue is the linear degenerate property of the Einstein equation formulation. In order to make the system linear degenerate parameters setting $\gamma_1=-1$ and $\gamma_2=0$ is needed. As we known, shock formation will not happen for linear degenerate systems. And our numerical test results do show this consequence. But the $\gamma_2=0$ setting makes the system does not admit full constraint damping property. And consequently the evolution is not stable. In contrast, constraint damping setting is more important than the linear degenerate setting for the stable evolution. And correspondingly the shock formation due to the numerical error can not be avoided in some sense. The situation is similar to three dimensional case. Based on our finding, we suspect most gauge drivers within the generalized harmonic gauge formulation may result in shock formation due to numerical error. And these shock formation may kill the numerical solution through spectral method. We plan to investigate this problem in the near future. Actually in order to make the finite element method robust enough in numerical relativity, feasible gauge driver is important. That will be a key problem of our future study.

In the current work, we only considered simple boundary conditions. As indicated in our previous work \cite{PhysRevD.88.084057}, constraint preserving boundary condition can improve the numerical accuracy quite a bit. So another important problem we need to investigate in the near future is extending our numerical scheme to including constraint preserving boundary condition.
\acknowledgments
We are thankful to Prof. Yun-Kau Lau and Prof. Chi-Wang Shu for discussions about finite element method. This work was supported by the NSFC (No.~11690023 and No.~11622546).

\appendix

\section{characteristic structure of the dynamical equation (\ref{problem_eq})}\label{App::I}
We have characteristic variables
\begin{align}
&u^0_{AB}=g_{AB},\\
&u^{-}_{AB}=\gamma_2g_{AB}-\Pi_{AB}+\sqrt{\gamma^{rr}}\Phi_{AB},\\
&u^{+}_{AB}=-\gamma_2g_{AB}+\Pi_{AB}+\sqrt{\gamma^{rr}}\Phi_{AB},\\
&u^0_{S}=S,\\
&u^{-}_{S}=\gamma_2S-\Pi_{S}+\sqrt{\gamma^{rr}}\Phi_{S},\\
&u^{+}_{S}=-\gamma_2S+\Pi_{S}+\sqrt{\gamma^{rr}}\Phi_{S}.
\end{align}
And the characteristic variables $u^0$, $u^+$ and $u^-$ admit characteristic speed $-(1+\gamma_1)\beta^r$, $-\beta^r-\alpha\sqrt{\gamma^{rr}}$ and $-\beta^r+\alpha\sqrt{\gamma^{rr}}$ respectively.

In order to make the system linearly degenerate, we need $\gamma_1=-1$ and $\gamma_2=0$. But the constraint violation may be damped by $\gamma_{0,2}$ through $e^{-\gamma_{0,2} t}$. We concern constraint damping more than the linearly degenerate. So we follow previous works such as \cite{0264-9381-23-16-S09,PhysRevD.93.063006} to set $\gamma_0=\gamma_2=1$. And more we follow \cite{PhysRevD.93.063006} to set $\gamma_4=\gamma_5=\tfrac{1}{2}$.

\section{Useful relations for spherical symmetric spacetime}\label{App::II}
Based on the metric form (\ref{metric_sphsymm}) the Christoffel symbols can be written as
\begin{align}
&\Gamma_{ABI}=\Gamma_{AIB}=\Gamma_{IAB}=\Gamma_{A\theta\phi}=\Gamma_{A\phi\theta}=0,\\
&\Gamma_{\theta A\phi}=\Gamma_{\phi A\theta}=\Gamma_{\theta\phi A}=\Gamma_{\phi\theta A}=0,\\
&\Gamma_{\theta\theta\theta}=\Gamma_{\theta\theta\phi}=\Gamma_{\theta\phi\theta}=\Gamma_{\phi\theta\theta}
=\Gamma_{\phi\phi\theta}=\Gamma_{\phi\phi\phi}=0,\\
&\Gamma_{A\theta\theta}=-re^{2S}[\gamma^r{}_A(r\Phi_S+1)+n_A(r\Pi_S-n^r)],\\
&\Gamma_{A\phi\phi}=-r\sin^2\theta e^{2S}[\gamma^r{}_A(r\Phi_S+1)+n_A(r\Pi_S-n^r)],\\
&\Gamma_{\theta A\theta}=re^{2S}[\gamma^r{}_A(r\Phi_S+1)+n_A(r\Pi_S-n^r)],\\
&\Gamma_{\phi A\phi}=r\sin^2\theta e^{2S}[\gamma^r{}_A(r\Phi_S+1)+n_A(r\Pi_S-n^r)],\\
&\Gamma_{\theta\theta A}=re^{2S}[\gamma^r{}_A(r\Phi_S+1)+n_A(r\Pi_S-n^r)],\\
&\Gamma_{\phi\phi A}=r\sin^2\theta e^{2S}[\gamma^r{}_A(r\Phi_S+1)+n_A(r\Pi_S-n^r)],\\
&\Gamma_{ABC}=\gamma^r{}_{(B}\Phi_{C)A}
-\tfrac{1}{2}\gamma^r{}_A\Phi_{BC}+n_{(B}\Pi_{C)A}
-\tfrac{1}{2}n_A\Pi_{BC}\,,\\
&\Gamma_{\theta\phi\phi}=-r^2\sin\theta\cos\theta e^{2S},\\
&\Gamma_{\phi\theta\phi}=r^2\sin\theta\cos\theta e^{2S},\\
&\Gamma_A=g^{BC}\Gamma_{ABC}-\frac{2}{r}[\gamma^r{}_A(r\Phi_S+1)+n_A(r\Pi_S-n^r)],\\
&\Gamma_\theta=-\cot\theta,\\
&\Gamma_\phi=0.
\end{align}

\section{useful relations for Legendre polynomials}\label{App::III}
The Legendre polynomials we adopted in the current work take form
\begin{align}
&P_0=1,\\
&P_1=x_l,\\
&P_2=x_l^2-\tfrac{1}{3},\\
&P_3=x_l^3-\tfrac{3}{5}x_l,\\
&P_4=x_l^4-\tfrac{6}{7}x^2_l+\tfrac{3}{35},\\
&P_5=x_l^5-\tfrac{10}{9}x_l^3+\tfrac{5}{21}x_l,\\
&P_6=x_l^6-\tfrac{15}{11}x_l^4+\tfrac{5}{11}x_l^2-\tfrac{5}{231}.
\end{align}
And their norms are respectively
\begin{align}
&\langle P_0|P_0\rangle=2,\\
&\langle P_1|P_1\rangle=\tfrac{2}{3},\\
&\langle P_2|P_2\rangle=\tfrac{8}{45},\\
&\langle P_3|P_3\rangle=\tfrac{8}{175},\\
&\langle P_4|P_4\rangle=\tfrac{128}{11025},\\
&\langle P_5|P_5\rangle=\tfrac{128}{43695},\\
&\langle P_6|P_6\rangle=\tfrac{512}{693693}.
\end{align}
The inner product is defined in the Eq.~(\ref{Eq::innerprod}).

\bibliographystyle{apsrev}
\bibliography{refs}

\end{document}